\documentclass[12pt,a4paper]{article}
\usepackage[T2A]{fontenc}
\usepackage[cp1251]{inputenc}
\usepackage{cite}
\usepackage{mathtext}
\usepackage{amssymb,amsthm,amsmath}
\usepackage{array}
\usepackage[dvips]{epsfig}
\usepackage[russian, english]{babel}
\usepackage[title,titletoc]{appendix}

\begin{document}
\author{\bf Yu.A. Markov$^{1,\,2}\!\,$\thanks{e-mail:markov@icc.ru}\, and
M.A. Markova$^{1}\!\,$\thanks{e-mail:markova@icc.ru}}
\title{Path integral representation\\ for inverse third order wave operator\\
within the Duffin-Kemmer-Petiau formalism.\!\! II}
%
%%%%%%%%%%%%%%%%%%%%%%%%%%%%%%%%%%%%%%%%%%%%%%%%%%%%%%%%%%
%
\date{\it\normalsize
%\begin{center}
\begin{itemize}
\item[]$^{1}$Matrosov Institute for System Dynamics and Control Theory SB RAS
Irkutsk, Russia
\vspace{-0.3cm}
\item[]$^{2}$Department of Physics, Tomsk State University, Tomsk, Russia
\end{itemize}}
\thispagestyle{empty}
\maketitle{}
%
%%%%%%%%%%%%%%%%%%% Making dual numbered equations %%%%%%%%%%%%%%%%%%%%%%
%
\def\theequation{\arabic{section}.\arabic{equation}}
\vspace{-0.6cm}
\[
{\bf Abstract}
\]

%\vspace{-0.3cm}

{
\noindent
This paper is an immediate continuation of the first part of our paper \cite{part_I}. Here, in a para-Grassmann algebra we introduce a noncommutative, associative star product $*$ (the Moyal product), which is a direct generalization of the star product in the algebra of Grassmann numbers. Isomorphism between the algebra of para-Grassmann numbers of order 2 equipped with the star product and with the algebra of creation and annihilation operators $a_{n}^{\pm}$ obeying the para-Fermi statistics of the same order is established. Two independent approaches to the calculation of the Moyal product $*$ are considered. It is shown that in calculating the matrix elements in the basis of parafermion coherent states of various expressions it should be taken into account constantly that we work in the so-called Ohnuki and Kamefuchi's generalized state-vector space $\mathfrak{U}_{\;G}$, whose state vectors  include para-Grassmann numbers $\xi_{k}$ in their definition, instead of the standard state-vector space $\mathfrak{U}$ (the Fock space). Otherwise, the wide array of contradictions arises. An immediate consequence of using the extended state-vector space $\mathfrak{U}_{\;G}$ is a necessity to consider the quadratic Casimir operators $\hat{C}_{2}$ and $\hat{C}_{2}^{\prime}$ of the orthogonal groups $SO(2M)$ and $SO(2M + 1)$, correspondingly. The action rules of the Casimir operators on the state vectors, an explicit form of their matrix elements are defined and a more general connection between the Harish-Chandra operator $\hat{\omega}^2$ and the Geyer operator $a_{0}^{2}$ is obtained. The notions of the triple star product, the star exponent and the Moyal bracket are introduced. It is shown that the use of the notions of the star product $*$ and the generalized state-vector space $\mathfrak{U}_{\;G}$ allows one to make a form of the matrix element of the contribution into the generalized Hamilton operator ${\cal H}$, which is linear in the covariant derivative $\hat{D}_{\mu}$, more compact and visual.
}
{}

%{\sl PACS:} 12.38.Mh, 24.85.+p, 11.15.Kc

\newpage

%%%%%%%%%%%%%%%%%%%%%%%% section 1 %%%%%%%%%%%%%%%%%%%%%%%%%%%%

\section{Introduction}
\setcounter{equation}{0}
\label{section_1}

In the second part of our paper, we consider more closely those questions that we have already touched on in the first part \cite{part_I} (to be refereed to as ``Part I'' throughout this text). The expression obtained in section 10 of Part I for the matrix element $\langle\hspace{0.02cm}(k)^{\prime}_{p}\hspace{0.02cm}|\hspace{0.02cm}
[\hspace{0.03cm}\chi,\hat{\cal L}(z,\hat{D})\hspace{0.01cm}]\hspace{0.01cm}|\hspace{0.02cm}
(k - 1)_{x}\hspace{0.02cm}\rangle$ linear in the covariant derivative $\hat{D}_{\mu}$ has a somewhat cumbersome and tangled form. Here, in the second part we would like to make a form of this matrix element more compact and visual. For this purpose, in a para-Grassmann algebra we enter noncommutative and associative star product $*$ (the Moyal product), which in fact represents an integral convolution  of a certain type of two para-Grassmann--valued functions. This is a direct generalization of a similar product in Grassmann algebra (see, for example, Bayen {\it et al.} \cite{bayen_1978}, Tyutin \cite{tyutin_2001}, Smilga \cite{smilga_2002}, Hirshfeld and Henselder \cite{hirshfeld_2002}, Daoud \cite{daoud_2003}). In our case this product arises naturally in calculating the matrix elements of complicated operator expressions in the basis of parafermion coherent states. Moreover, the star product enables us to have a better understanding of a connection between the algebra of creation and annihilation operators $a_{k}^{\pm}$ of para-Fermi particles and the para-Grassmann algebra equipped with the product $*$ making them in fact isomorphic as it takes place for the usual fermionic operators and Grassmann variables \cite{daoud_2003}.\\
\indent For calculating the $*$ operation we will use two different approaches. The first of them was suggested in the paper by Omote and Kamefuchi \cite{omote_1979}. The idea of this approach consists in reducing the integration over the para-Grassmann variables to calculating certain operator expressions averaged over the vacuum state. Here, ultimately, the calculation boils down to shift of the annihilation operators $a_{k}^{-}$ to the right until the vacuum conditions can be employed. Here, the rearrangement rules are defined by the algebra of operators $a_{k}^{\pm}$ obeying the para-Fermi statistics of the special order $p$. Another way of heuristic character is to separate by hand from various contributions to the integral only those which give nontrivial result after an integration. For separating these contributions we make use of a simple fact that the integral with respect to para-Grassmann variable $\mu$ of order $p$ is not vanishing only when the integrand contains this para-Grassmann variable exactly to the power of $p$ \cite{ohnuki_1980}. Two independent approaches to calculating the product $*$ allows one to verify independently the results of calculations which becomes more complicated when the order $p$ increases.\\
\indent One of the most interesting features of the approach developed here is the necessity of introduction instead of  the standard Fock space $\mathfrak{U}$ an enlarged Fock space $\mathfrak{U}_{\,G}$, whose state vectors can also include the para-Grassmann numbers $\xi_k$. The need for such a generalization was first noted by Ohnuki and Kamefuchi \cite{ohnuki_1980} and it is connected with that the para-Fermi operators $a_{k}^{\pm}$ and the para-Grassmann numbers $\xi_k$ are not (anti)commutative for parastatistics of order $p\geq 2$. As a consequence, the parafermion coherent state (in the form which we use throughout this and in the preceding works) could not be represented in the form of an expansion in states with a certain number of para-Fermi particles. We have already discussed briefly this fact in section 11 of Part I. Careful consideration of this circumstance is required at all stages of calculations, or otherwise this leads to contradictions of various kinds.\\
\indent One of the important elements of the Lie algebra $\mathfrak{so}(2M)$ and $\mathfrak{so}(2M + 1)$ are the quadratic Casimir operators $\hat{C}_2$ and $\hat{C}_{2}^{\prime}$, correspondingly. The need to deal with these operators have arisen naturally within the framework of Ohnuki and Kamefuchi's generalized state-vector space. In our specific case $p = 2$ these operators are closely related to elements of the centre of the Duffin-Kemmer-Petiau algebra as they were defined by Harish-Chandra \cite{harish-chandra_1947}. In our derivation we use the representation of the quadratic Casimir operators in terms of the creation and annihilation operators $a_{k}^{\pm}$ of parafermions from the paper by Omote {\it et al.} \cite{omote_1976} and Bracken and Green \cite{bracken_1972} (see also Gould and Paldus \cite{gould_1986}).\\
\indent The paper is organized as follows. Sections 2 and 3 are devoted to the establishment of the relation between the functions $\Omega\hspace{0.03cm}(\bar{\xi}^{\,\prime},\xi)$ and $\widetilde{\Omega}\hspace{0.03cm} (\bar{\xi}^{\,\prime},\xi)$, which are the matrix elements of the operator $a_{0}$ of the algebra ${\mathfrak so}$(2M + 2) and the Geyer operator $a_{0}^{2}$ in the basis of parafermion coherent states. In section 4 an important notion of the star product $*$ within the framework of the algebra of para-Grassmann numbers is introduced. Isomorphism between the algebra of creation and annihilation operators $a_{k}^{\pm}$ obeying para-Fermi statistics of order $p = 2$ and the algebra of para-Grassmann numbers $\xi_{k}$ and $\hat{\xi}_{k}^{\prime}$ of the same order, in which a product is defined by the operation $*$\hspace{0.01cm}, is proved. In section 5 the triple star product $\Omega* \Omega*\Omega$ of the function $\Omega = \Omega\hspace{0.03cm}(\bar{\xi}^{\,\prime},\xi)$ and the star exponential $\exp_{*}\bigl(-i\hspace{0.02cm}\frac{2\pi}{3}\,\Omega\bigr)$ are considered. In this section we have concluded ultimately that it is impossible to present the function $\widetilde{\Omega}\hspace{0.03cm} (\bar{\xi}^{\,\prime},\xi)$ as the star product of two functions $\Omega\hspace{0.03cm}(\bar{\xi}^{\,\prime},\xi)$, i.e. $\widetilde{\Omega} \neq \Omega\!\hspace{0.04cm} *\hspace{0.01cm}\Omega$. In section 6 to overcome a contradiction of the previous section, an analysis of the connection between the Harish-Chandra operator $\hat{\omega}^{2}$ and the Geyer operator $a_{0}^{2}$ is performed again and a more exact relation between these operators (in comparison with a similar relation obtained in section 8 of Part I) is derived. In section 7 the quadratic Casimir operators $\hat{C}_{2}$ and $\hat{C}_{2}^{\prime}$ of the groups $SO(2M)$ and $SO(2M + 1)$ are taken into consideration that enables one, in particular, to make a form of the connection between the operators $\hat{\omega}^{2}$ and $a_{0}^{2}$ more compact and explicit. The action rules of the Casimir operators on the state vectors, an explicit form of the matrix elements of these operators and their representations through the operator $\hat{\Lambda} \equiv \sum_{k=1}^2 \{\hspace{0.02cm}a_{k}^{+}, a_{k}^{-} \}$ are defined.\\
\indent Section 8 is devoted to discussion of the so-called generalized state-vector space $\mathfrak{U}_{\;G}$ as it was defined by Ohnuki and Kamefuchi \cite{ohnuki_1980}, whose state vectors include the para-Grassmann numbers $\xi_{k}$. In section 9 the representation for the Harish-Chandra operator $\hat{\omega}^{2}$ in terms of the operator 
$\hat{\Lambda}$ is written out. In the same section it is proved that by using the refined connection between the operators $\hat{\omega}^{2}$ and $a_{0}^{2}$, we correctly reproduce the star product of three functions $\Omega$, namely $\Omega * \Omega * \Omega = \Omega$. In section 10 the expressions for the commutators $[\hspace{0.02cm}\hat{A}, a^{\pm}_{k}\hspace{0.02cm}]$ and $[\hspace{0.02cm}\hat{A},[\hspace{0.03cm}a_{0},a^{\pm}_{k}\hspace{0.02cm}]\hspace{0.03cm}]$, where $\hat{A}\equiv\alpha\exp\bigl(-i\hspace{0.02cm}\frac{2\pi}{3}\,a^{\phantom{+}}_{0}\bigr)$, are calculated. It is shown that the expressions obtained in calculating their matrix elements in the basis of the parafermion coherent states are contradictory with one another. It is pointed out that a reason of this contradiction is the use of trilinear commutation relation including two operators $a_{0}$ and one of the operators $a_{k}^{\pm}$ in the form as it was defined in the original paper by Geyer \cite{geyer_1968}. The most general form of this trilinear relation including the Casimir operator $\hat{C}_{2}^{\prime}$ is derived that allows one to obtain finally the consistent expressions for the commutators $[\hspace{0.02cm}\hat{A}, a^{\pm}_{k}\hspace{0.02cm}]$ and $[\hspace{0.02cm}\hat{A},[\hspace{0.03cm}a_{0},a^{\pm}_{k}\hspace{0.02cm}]\hspace{0.03cm}]$. In section 11 the final form of the matrix element $\langle\hspace{0.02cm}(k)^{\prime}_{p}\hspace{0.02cm}|\hspace{0.02cm}
\hat{A}\hspace{0.02cm}\hat{\eta}_{\mu}(q)\hat{D}_{\mu}\hspace{0.01cm}|\hspace{0.02cm}(k - 1)_{x}\hspace{0.02cm}\rangle$ linear in the covariant derivative $\hat{D}_{\mu}$, is written out. In the concluding section 12 the key points of our work are specified and the remained unsolved problems are briefly discussed. \\
\indent In Appendix A all of the necessary formulae of integration with respect to a para-Grassmann variable $\mu$ of order 2 are listed. In Appendix B the formulae of differentiations with respect to the para-Grassmann variable used in the text of our paper are given. In Appendix C the derivation of the formula (68) given in the paper by Harish-Chandra \cite{harish-chandra_1947} is considered again and its corrected expression is obtained. Appendix D is devoted to the proof of the relation $[\hspace{0.03cm}a_{0},\hat{\Lambda}\hspace{0.02cm}] = 0$. In particular, it is shown that in contrast to the relation $\{\hspace{0.01cm}a_{0},\hat{\Lambda}\} = 8\hspace{0.02cm}a_{0}$, this operator relation does not fall into a sum of two independent relations $[\hspace{0.03cm}a_{0},\hat{\Lambda}_{i}\hspace{0.02cm}] = 0$, where $\hat{\Lambda}_i \equiv \{\hspace{0.02cm}a_{i}^{+}, a_{i}^{-}\},\, i = 1, 2$.

%%%%%%%%%%%%%%%%%%%%%%%% section 2 %%%%%%%%%%%%%%%%%%%%%%%%%%%%

\section{\bf Connection between the functions $\widetilde{\Omega}\hspace{0.03cm} (\bar{\xi}^{\,\prime},\xi)$ and $\Omega\hspace{0.03cm}(\bar{\xi}^{\,\prime},\xi)$}
\setcounter{equation}{0}
\label{section_2}

In sections 5 and 7 of the first part of our paper \cite{part_I} we have derived matrix elements of the operators $a_0$ and $a_0^2$. For convenience of further references, we present these expressions ones again
\begin{equation}
\langle\hspace{0.02cm}\bar{\xi}^{\,\prime}\hspace{0.02cm}|\,
a_{0}|\,\xi\hspace{0.02cm}\rangle
=
\Omega\,\langle\hspace{0.02cm}\bar{\xi}^{\,\prime}\hspace{0.02cm}|\,\xi\hspace{0.02cm}\rangle,
\label{eq:2q}
\end{equation}
where
\begin{equation}
\begin{split}
\Omega \equiv \Omega\hspace{0.03cm}(\bar{\xi}^{\,\prime}, \xi)
=
-\frac{1}{2}\,\biggl\{\!
&\biggl(\displaystyle\frac{1}{2}\;[\hspace{0.03cm}\bar{\xi}^{\,\prime}_{1}, \bar{\xi}^{\,\prime}_{2}\hspace{0.03cm}]\biggr)\!
\biggl(\,\displaystyle\frac{1}{2}\;[\hspace{0.03cm}\xi^{\phantom{\prime}}_{1}, \xi^{\phantom{\prime}}_{2}\hspace{0.03cm}]\biggr)
+
\biggl(\displaystyle\frac{1}{2}\;[\hspace{0.03cm}\bar{\xi}^{\,\prime}_{1}, \xi^{\phantom{\prime}}_{2}\hspace{0.03cm}]\biggr)\!
\biggl(\displaystyle\frac{1}{2}\;[\hspace{0.03cm}\bar{\xi}^{\,\prime}_{2}, \xi^{\phantom{\prime}}_{1}\hspace{0.03cm}]\biggr) - \\[1ex]
-\, &\biggl(\displaystyle\frac{1}{2}\;[\hspace{0.03cm}\bar{\xi}^{\,\prime}_{1}, \xi^{\phantom{\prime}}_{1}\hspace{0.03cm}]\biggr)\!
\biggl(\displaystyle\frac{1}{2}\;[\hspace{0.03cm}\bar{\xi}^{\,\prime}_{2}, \xi^{\phantom{\prime}}_{2}\hspace{0.03cm}]\biggr)
+
2\biggl(\displaystyle\frac{1}{2}\;[\hspace{0.03cm}\bar{\xi}^{\,\prime}_{1}, \xi^{\phantom{\prime}}_{1}\hspace{0.03cm}]
+
\displaystyle\frac{1}{2}\;[\hspace{0.03cm}\bar{\xi}^{\,\prime}_{2}, \xi^{\phantom{\prime}}_{2}\hspace{0.03cm}] - 1\biggr)\!
\biggr\}
\end{split}
\label{eq:2w}
\end{equation}
and
\begin{equation}
\langle\hspace{0.02cm}\bar{\xi}^{\,\prime}\hspace{0.02cm}|\,
a^{2}_{0}|\,\xi\hspace{0.02cm}\rangle = \widetilde{\Omega} \,
\langle\hspace{0.02cm}\bar{\xi}^{\,\prime}\hspace{0.02cm}|\,\xi\hspace{0.02cm}\rangle,
\label{eq:2e}
\end{equation}
with
\begin{equation}
\widetilde{\Omega} = \widetilde{\Omega}\hspace{0.03cm}(\bar{\xi}^{\,\prime},\xi) =
\label{eq:2r}
\end{equation}
\[
= 1 - \sum\limits^{2}_{k = 1}
\biggl[\biggl(\displaystyle\frac{1}{2}\;[\hspace{0.03cm}\bar{\xi}^{\,\prime}_{k}, \xi^{\phantom{\prime}}_{k}\hspace{0.02cm}]\biggr)^{\!\!2\,} -
\displaystyle\frac{1}{2}\;[\hspace{0.03cm}\bar{\xi}^{\,\prime}_{k}, \xi^{\phantom{\prime}}_{k}\hspace{0.02cm}]
+ 1\biggr]\,
+ 2\hspace{0.02cm}\prod\limits^{2}_{k = 1}
\biggl[\biggl(\displaystyle\frac{1}{2}\;[\hspace{0.03cm}\bar{\xi}^{\,\prime}_{k}, \xi^{\phantom{\prime}}_{k}\hspace{0.02cm}]\biggr)^{\!\!2\,} -
\displaystyle\frac{1}{2}\;[\hspace{0.03cm}\bar{\xi}^{\,\prime}_{k}, \xi^{\phantom{\prime}}_{k}\hspace{0.02cm}] + 1\biggr].
\]
In this section and in the subsequent three sections we would like to establish a connection between the functions $\widetilde{\Omega}$ and $\Omega$ and thereby to clarify whether the operator $a_{0}^{2}$ is the square of the operator $a_{0}$. For this purpose, we use the insertion of resolution of the identity operator
\[
\iint\!|\,\mu\hspace{0.02cm}\rangle\hspace{0.02cm} \langle\hspace{0.02cm}\bar{\mu}\hspace{0.02cm}|\,
{\rm e}^{\,-\textstyle\frac{\!1}{2}\,[\hspace{0.03cm}\bar{\mu},\mu\hspace{0.03cm}]}
\hspace{0.03cm}(d\mu)_{2}\hspace{0.03cm}(d\bar{\mu})_{2} = \hat{1}
\]
and an explicit form of the overlap function
\begin{equation}
\langle\hspace{0.02cm}\bar{\xi}^{\,\prime}\hspace{0.02cm}|\,\xi\hspace{0.02cm}\rangle
=
{\rm e}^{\,\textstyle\hspace{0.02cm}\frac{\!1}{2}\,
[\hspace{0.02cm}\bar{\xi}^{\,\prime}\hspace{0.02cm},\xi^{\phantom{\prime}\!}
\hspace{0.02cm}\hspace{0.02cm}]},
\label{eq:2t}
\end{equation}
where $\bar{\mu},\,\mu$ are para-Grassmann numbers,
\[
(d\mu)_{2}\equiv d^{\hspace{0.02cm}2\!}\mu_{\hspace{0.02cm}2}\hspace{0.03cm} d^{\hspace{0.02cm}2\!}\mu_{1}, \quad
(d\bar{\mu})_{2}\equiv d^{\hspace{0.02cm}2}\bar{\mu}_{1}\hspace{0.03cm} d^{\hspace{0.02cm}2}\bar{\mu}_{\hspace{0.02cm}2}
\]
are the measure of integration and as usual, for the sake of brevity we make use of the notations
\[
[\hspace{0.03cm}\bar{\mu},\mu\hspace{0.03cm}] = \sum\limits_{k = 1}^{2}\;
[\hspace{0.03cm}\bar{\mu}_{k},\mu_{k}\hspace{0.03cm}]
\]
and so on. Then, by virtue of the definitions (\ref{eq:2q}) and (\ref{eq:2t}) we get
\[
\langle\hspace{0.02cm}\bar{\xi}^{\,\prime}\hspace{0.02cm}|\,
a^{2}_{0}|\,\xi\hspace{0.02cm}\rangle =\!
\iint\langle\hspace{0.02cm}\bar{\xi}^{\,\prime}\hspace{0.02cm}|\,
a_{0}|\,\mu\hspace{0.02cm}\rangle\hspace{0.03cm}
\langle\hspace{0.02cm}\bar{\mu}\hspace{0.02cm}|\,
a_{0}|\,\xi\hspace{0.02cm}\rangle\,
{\rm e}^{\,-\textstyle\frac{\!1}{2}\,[\hspace{0.03cm}\bar{\mu},\mu\hspace{0.03cm}]}
\hspace{0.03cm}(d\mu)_{2}\hspace{0.03cm}(d\bar{\mu})_{2} =
\]
\[
=\! \iint\!\Omega\hspace{0.03cm}(\bar{\xi}^{\,\prime},\mu)\hspace{0.03cm}
\Omega\hspace{0.03cm}(\bar{\mu}, \xi)\,
{\rm e}^{\,-\textstyle\frac{\!1}{2}\,[\hspace{0.03cm}\bar{\mu}\! -\! \bar{\xi}^{\,\prime},\mu\! -\! \xi\hspace{0.03cm}]} \hspace{0.03cm}(d\mu)_{2}\hspace{0.03cm}(d\bar{\mu})_{2}\;
{\rm e}^{\,\textstyle\hspace{0.02cm}\frac{\!1}{2}\,
[\hspace{0.02cm}\bar{\xi}^{\,\prime}\hspace{0.02cm},\xi
\hspace{0.02cm}\hspace{0.02cm}]}.
\]
Substituting  the representation (\ref{eq:2e}) into the left-hand side of this expression and cancelling the factor (\ref{eq:2t}), we obtain
\[
\widetilde{\Omega}\hspace{0.03cm}(\bar{\xi}^{\,\prime},\xi)
=
\! \iint\!\Omega\hspace{0.03cm}(\bar{\xi}^{\,\prime},\mu)\hspace{0.03cm}
\Omega\hspace{0.03cm}(\bar{\mu}, \xi)\,
{\rm e}^{\,-\textstyle\frac{\!1}{2}\,[\hspace{0.03cm}\bar{\mu}\! -\! \bar{\xi}^{\,\prime},\mu\! -\! \xi\hspace{0.03cm}]} \hspace{0.03cm}(d\mu)_{2}\hspace{0.03cm}(d\bar{\mu})_{2}.
\]
The last step is a shift of the variables of integration:
\[
\bar{\mu}\rightarrow \bar{\mu} + \bar{\xi}^{\,\prime},\quad \mu\rightarrow \mu + \xi
\]
and finally we get
\begin{equation}
\widetilde{\Omega}\hspace{0.03cm}(\bar{\xi}^{\,\prime},\xi)
=
\! \iint\!\Omega\hspace{0.03cm}(\bar{\xi}^{\,\prime},\xi + \mu)\hspace{0.03cm}
\Omega\hspace{0.03cm}(\bar{\xi}^{\,\prime} + \bar{\mu}, \xi)\,
{\rm e}^{\,-\textstyle\frac{\!1}{2}\,[\hspace{0.03cm}\bar{\mu},\mu\hspace{0.03cm}]} \hspace{0.03cm}(d\mu)_{2}\hspace{0.03cm}(d\bar{\mu})_{2}.
\label{eq:2y}
\end{equation}
By these means, it is necessary to take the integral on the right-hand side of (\ref{eq:2y}) for the function $\Omega$ defined by the expression (\ref{eq:2w}) and to compare the obtained expression with (\ref{eq:2r}).\\
\indent The first step is an expansion of the function $\Omega$ in the integrand (\ref{eq:2y}) in powers of $\mu$ and $\bar{\mu}$. From (\ref{eq:2w}) it is easy to see that these expansions have the following form:
\begin{equation}
\Omega\hspace{0.03cm}(\bar{\xi}^{\,\prime},\xi + \mu) = \Omega\hspace{0.03cm}(\bar{\xi}^{\,\prime},\xi)
-
\frac{1}{2}\,\biggl[\,\frac{\partial\hspace{0.03cm}\Omega\hspace{0.03cm}(\bar{\xi}^{\,\prime},\xi)}
{\partial\hspace{0.03cm}\xi}\hspace{0.03cm},\hspace{0.03cm} \mu\,\biggr]\, +
\label{eq:2u}
\end{equation}
\[
+\,\biggl(-\,\frac{\!1}{2^{\hspace{0.02cm}3}}\biggr)\hspace{0.01cm}
\Bigl\{\hspace{0.03cm}\stackrel{\bf (1)}{[\hspace{0.03cm}\bar{\xi}^{\,\prime}_{1}, \bar{\xi}^{\,\prime}_{2}\hspace{0.03cm}]\hspace{0.03cm}
[\hspace{0.03cm}\mu^{\phantom{\prime}}_{1}, \mu^{\phantom{\prime}}_{2}\hspace{0.03cm}]}
+
\stackrel{\bf (2)}{[\hspace{0.03cm}\bar{\xi}^{\,\prime}_{2}, \mu^{\phantom{\prime}}_{1}\hspace{0.03cm}]
\hspace{0.03cm}[\hspace{0.03cm}\bar{\xi}^{\,\prime}_{1}, \mu^{\phantom{\prime}}_{2}\hspace{0.03cm}]}
-
\stackrel{\bf (3)}{[\hspace{0.03cm}\bar{\xi}^{\,\prime}_{1}, \mu^{\phantom{\prime}}_{1}\hspace{0.03cm}]\hspace{0.03cm}
[\hspace{0.03cm}\bar{\xi}^{\,\prime}_{2}, \mu^{\phantom{\prime}}_{2}\hspace{0.03cm}]}
\Bigr\}
\]
and
\begin{equation}
\Omega\hspace{0.03cm}(\bar{\xi}^{\,\prime} + \bar{\mu},\xi) = \Omega\hspace{0.03cm}(\bar{\xi}^{\,\prime},\xi)
+
\frac{1}{2}\,\biggl[\,\bar{\mu}\hspace{0.03cm}, \frac{\partial\hspace{0.03cm}\Omega\hspace{0.03cm}(\bar{\xi}^{\,\prime},\xi)}
{\partial\hspace{0.03cm}\bar{\xi}^{\,\prime}}\hspace{0.03cm}\,\biggr]\, +
\label{eq:2i}
\end{equation}
\[
+\,\biggl(-\,\frac{\!1}{2^{\hspace{0.02cm}3}}\biggr)\hspace{0.01cm}
\Bigl\{\hspace{0.03cm}\stackrel{\bf (1)}{[\hspace{0.03cm}\bar{\mu}_{1}, \bar{\mu}_{2}\hspace{0.03cm}]\hspace{0.03cm}
[\hspace{0.03cm}\xi_{1}, \xi_{2}\hspace{0.03cm}]}
+
\stackrel{\bf (2)}{[\hspace{0.03cm}\bar{\mu}_{2}, \xi_{1}\hspace{0.03cm}]
\hspace{0.03cm}[\hspace{0.03cm}\bar{\mu}_{1}, \xi_{2}\hspace{0.03cm}]}
-
\stackrel{\bf (3)}{[\hspace{0.03cm}\bar{\mu}_{1}, \xi_{1}\hspace{0.03cm}]\hspace{0.03cm}
[\hspace{0.03cm}\bar{\mu}_{2}, \xi_{2}\hspace{0.03cm}]}
\Bigr\}.
\]
We have introduced the markers $\small{\bf (1)-(3)}$ over the terms which are quadratic in $\mu$ and $\bar{\mu}$.\\
\indent Let us consider the terms in (\ref{eq:2u}) and (\ref{eq:2i}) linear in $\mu$ and $\bar{\mu}$. Substituting them into (\ref{eq:2y}), one obtains
\[
\begin{split}
&\Omega\hspace{0.03cm}(\bar{\xi}^{\,\prime},\xi)\!\iint
\biggl(\biggl[\,\bar{\mu}\hspace{0.03cm}, \frac{\partial\hspace{0.03cm}\Omega\hspace{0.03cm}(\bar{\xi}^{\,\prime},\xi)}
{\partial\hspace{0.03cm}\bar{\xi}^{\,\prime}}\hspace{0.03cm}\,\biggr]
-
\biggl[\,\frac{\partial\hspace{0.03cm}\Omega\hspace{0.03cm}(\bar{\xi}^{\,\prime},\xi)}
{\partial\hspace{0.03cm}\xi}\hspace{0.03cm},\hspace{0.03cm} \mu\,\biggr]\hspace{0.02cm}
\biggr)\,
{\rm e}^{\,-\textstyle\frac{\!1}{2}\,[\hspace{0.03cm}\bar{\mu},\mu\hspace{0.03cm}]} \hspace{0.03cm}(d\mu)_{2}\hspace{0.03cm}(d\bar{\mu})_{2} = \\[1ex]
&\Omega\hspace{0.03cm}(\bar{\xi}^{\,\prime},\xi)\biggl(
\int\biggl[\,\bar{\mu}\hspace{0.03cm}, \frac{\partial\hspace{0.03cm}\Omega\hspace{0.03cm}(\bar{\xi}^{\,\prime},\xi)}
{\partial\hspace{0.03cm}\bar{\xi}^{\,\prime}}\hspace{0.03cm}\,\biggr]
\delta^{(2)}(\bar{\mu})\hspace{0.03cm}(d\bar{\mu})_{2}
-
\int\biggl[\,\frac{\partial\hspace{0.03cm}\Omega\hspace{0.03cm}(\bar{\xi}^{\,\prime},\xi)}
{\partial\hspace{0.03cm}\xi}\hspace{0.03cm},\hspace{0.03cm} \mu\,\biggr]
\delta^{(2)}(\mu)\hspace{0.03cm}(d\mu)_{2}\biggr) = 0.
\end{split}
\]
Here, the para-Grassmann $\delta$-functions are
\[
\delta^{(2)}(\bar{\mu}) \equiv \delta(\bar{\mu}_{1})\hspace{0.03cm}\delta(\bar{\mu}_{2}),
\quad
\delta^{(2)}(\mu) \equiv \delta(\mu_{1})\hspace{0.03cm}\delta(\mu_{2}),
\]
where $\delta(\bar{\mu}_{i})$ and $\delta(\mu_{i}),\,i = 1, 2,$ are defined by the formula \eqref{ap:A5}.\\
\indent Further we consider the contributions quadratic in $\mu$ and $\bar{\mu}$. At first we calculate a contribution with derivatives of the function $\Omega$:
\begin{equation}
-\frac{\!1}{2^{2}}\!\iint
\biggl[\,\frac{\partial\hspace{0.03cm}\Omega\hspace{0.03cm}(\bar{\xi}^{\,\prime},\xi)}
{\partial\hspace{0.03cm}\xi}\hspace{0.03cm},\hspace{0.03cm} \mu\,\biggr]\!
\biggl[\,\bar{\mu}\hspace{0.03cm}, \frac{\partial\hspace{0.03cm}\Omega\hspace{0.03cm}(\bar{\xi}^{\,\prime},\xi)}
{\partial\hspace{0.03cm}\bar{\xi}^{\,\prime}}\hspace{0.03cm}\,\biggr]\hspace{0.03cm}
{\rm e}^{\,-\textstyle\frac{\!1}{2}\,[\hspace{0.03cm}\bar{\mu},\mu\hspace{0.03cm}]} \hspace{0.03cm}(d\mu)_{2}\hspace{0.03cm}(d\bar{\mu})_{2}.
\label{eq:2o}
\end{equation}
There are two independent ways of taking the integrals of this type. The first of them was suggested in the paper by Omote and Kamefuchi \cite{omote_1979}. We take the integral (\ref{eq:2o}) by the first way and then verify the result of calculations by another way.\\
\indent Following an approach suggested by Omote and Kamefuchi, we present the exponential in the integrand (\ref{eq:2o}) in the form of the overlap function
\begin{equation}
{\rm e}^{\,\textstyle\hspace{0.02cm}-\frac{\!1}{2}\,
[\hspace{0.02cm}\bar{\mu}\hspace{0.02cm},\mu\hspace{0.02cm}\hspace{0.02cm}]}
=
\langle\hspace{0.02cm}-\bar{\mu}\hspace{0.02cm}|\,\mu\hspace{0.02cm}\rangle.
\label{eq:2p}
\end{equation}
Then, instead of (\ref{eq:2o}), we will have a chain of equalities
\begin{equation}
-\frac{\!1}{2^{\hspace{0.02cm}2}}\!\iint\biggl(
\biggl[\,\frac{\partial\hspace{0.03cm}\Omega\hspace{0.03cm}(\bar{\xi}^{\,\prime},\xi)}
{\partial\hspace{0.03cm}\xi_{1}}\hspace{0.03cm},\hspace{0.03cm} \mu_{1}\,\biggr]
+
\biggl[\,\frac{\partial\hspace{0.03cm}\Omega\hspace{0.03cm}(\bar{\xi}^{\,\prime},\xi)}
{\partial\hspace{0.03cm}\xi_{2}}\hspace{0.03cm},\hspace{0.03cm} \mu_{2}\,\biggr]
\biggr)
\langle\hspace{0.02cm}-\bar{\mu}\hspace{0.02cm}|\,\mu\hspace{0.02cm}\rangle
\hspace{0.03cm}(d\mu)_{2}
\biggl[\,\bar{\mu}\hspace{0.03cm}, \frac{\partial\hspace{0.03cm}\Omega\hspace{0.03cm}(\bar{\xi}^{\,\prime},\xi)}
{\partial\hspace{0.03cm}\bar{\xi}^{\,\prime}}\hspace{0.03cm}\,\biggr]
\hspace{0.03cm}(d\bar{\mu})_{2} =
\hspace{0.4cm}
\label{eq:2a}
\end{equation}
\[
-\frac{\!1}{2^{\hspace{0.02cm}2}}\!\iint\langle\hspace{0.02cm}-\bar{\mu}\hspace{0.02cm}
|\hspace{0.03cm} \biggl(
\biggl[\,\frac{\partial\hspace{0.03cm}\Omega\hspace{0.03cm}(\bar{\xi}^{\,\prime},\xi)}
{\partial\hspace{0.03cm}\xi_{1}}\hspace{0.03cm},\hspace{0.03cm} a^{-}_{1}\,\biggr]
+
\biggl[\,\frac{\partial\hspace{0.03cm}\Omega\hspace{0.03cm}(\bar{\xi}^{\,\prime},\xi)}
{\partial\hspace{0.03cm}\xi_{2}}\hspace{0.03cm},\hspace{0.03cm} a^{-}_{2}\,\biggr]
\biggr)|\,\mu\hspace{0.02cm}\rangle
\hspace{0.03cm}(d\mu)_{2}
\biggl[\,\bar{\mu}\hspace{0.03cm}, \frac{\partial\hspace{0.03cm}\Omega\hspace{0.03cm}(\bar{\xi}^{\,\prime},\xi)}
{\partial\hspace{0.03cm}\bar{\xi}^{\,\prime}}\hspace{0.03cm}\,\biggr]
\hspace{0.03cm}(d\bar{\mu})_{2} =
\hspace{1.1cm}
\]
\[
-\frac{\!1}{2^{\hspace{0.02cm}4}}\!\iint\langle\hspace{0.02cm}-\bar{\mu}\hspace{0.02cm}
|\hspace{0.03cm} \biggl(
\biggl[\,\frac{\partial\hspace{0.03cm}\Omega\hspace{0.03cm}(\bar{\xi}^{\,\prime},\xi)}
{\partial\hspace{0.03cm}\xi_{1}}\hspace{0.03cm},\hspace{0.03cm} a^{-}_{1}\,\biggr]
+
\biggl[\,\frac{\partial\hspace{0.03cm}\Omega\hspace{0.03cm}(\bar{\xi}^{\,\prime},\xi)}
{\partial\hspace{0.03cm}\xi_{2}}\hspace{0.03cm},\hspace{0.03cm} a^{-}_{2}\,\biggr]
\biggr) (a^{+}_{1})^{2} (a^{+}_{2})^{2} |\,0\hspace{0.02cm}\rangle
\hspace{0.03cm}(d\mu)_{2}
\biggl[\,\bar{\mu}\hspace{0.03cm}, \frac{\partial\hspace{0.03cm}\Omega\hspace{0.03cm}(\bar{\xi}^{\,\prime},\xi)}
{\partial\hspace{0.03cm}\bar{\xi}^{\,\prime}}\hspace{0.03cm}\,\biggr]
\hspace{0.03cm}(d\bar{\mu})_{2}.
\]
Here, at the last step we have formally used the definition of $\delta$-function \eqref{ap:A5} as applied to the creation operator $a_{k}^{+}$:
\[
\int\!|\,\mu\hspace{0.02cm}\rangle \hspace{0.03cm}(d\mu)_{2} =\!
\int\!{\rm e}^{\,\textstyle\hspace{0.02cm}-\frac{\!1}{2}\,
[\hspace{0.02cm}\mu\hspace{0.02cm},a^{+}\hspace{0.02cm}\hspace{0.02cm}]}
|\,0\hspace{0.02cm}\rangle \hspace{0.03cm}(d\mu)_{2}
=
\frac{\!1}{2^{\hspace{0.02cm}2}}\, (a^{+}_{1})^{2} (a^{+}_{2})^{2} |\,0\hspace{0.02cm}\rangle.
\]
The last matrix element in (\ref{eq:2a}) is calculated when the operators $a_{1}^{-}$ and $a_{2}^{-}$ are shifted, with the use of (I.C.3) and (I.C.4), to the right until the vacuum conditions can be employed. The result of calculations is
\begin{equation}
\begin{split}
&\langle\hspace{0.02cm}-\bar{\mu}\hspace{0.02cm}
|\hspace{0.03cm} \biggl(
\biggl[\,\frac{\partial\hspace{0.03cm}\Omega\hspace{0.03cm}(\bar{\xi}^{\,\prime},\xi)}
{\partial\hspace{0.03cm}\xi_{1}}\hspace{0.03cm},\hspace{0.03cm} a^{-}_{1}\,\biggr]
+
\biggl[\,\frac{\partial\hspace{0.03cm}\Omega\hspace{0.03cm}(\bar{\xi}^{\,\prime},\xi)}
{\partial\hspace{0.03cm}\xi_{2}}\hspace{0.03cm},\hspace{0.03cm} a^{-}_{2}\,\biggr]
\biggr) (a^{+}_{1})^{2} (a^{+}_{2})^{2} |\,0\hspace{0.02cm}\rangle =\\[1ex]
&= 2^{\hspace{0.02cm}2}\biggl(\delta(\bar{\mu}_{2})\hspace{0.03cm}
\biggl\{\,\frac{\partial\hspace{0.03cm}\Omega\hspace{0.03cm}(\bar{\xi}^{\,\prime},\xi)}
{\partial\hspace{0.03cm}\xi_{1}}\hspace{0.03cm},\hspace{0.03cm} \bar{\mu}_{1}\biggr\}
+\hspace{0.03cm}
\delta(\bar{\mu}_{1})\hspace{0.03cm}
\biggl\{\,\frac{\partial\hspace{0.03cm}\Omega\hspace{0.03cm}(\bar{\xi}^{\,\prime},\xi)}
{\partial\hspace{0.03cm}\xi_{2}}\hspace{0.03cm},\hspace{0.03cm} \bar{\mu}_{2}\biggr\}
\biggr).
\end{split}
\label{eq:2s}
\end{equation}
The remaining integral with respect to $(d\bar{\mu})_{2}$ in (\ref{eq:2s}) is easily calculated by using the formula \eqref{ap:A4}
\begin{equation}
-\frac{\!1}{2^{2}}\!\int\hspace{0.03cm}\biggl[\,\bar{\mu}\hspace{0.03cm}, \frac{\partial\hspace{0.03cm}\Omega\hspace{0.03cm}(\bar{\xi}^{\,\prime},\xi)}
{\partial\hspace{0.03cm}\bar{\xi}^{\,\prime}}\hspace{0.03cm}\,\biggr]\!\hspace{0.03cm}
\biggl(\delta(\bar{\mu}_{2})\hspace{0.03cm}
\biggl\{\,\frac{\partial\hspace{0.03cm}\Omega\hspace{0.03cm}(\bar{\xi}^{\,\prime},\xi)}
{\partial\hspace{0.03cm}\xi_{1}}\hspace{0.03cm},\hspace{0.03cm} \bar{\mu}_{1}\biggr\}
+\hspace{0.03cm}
\delta(\bar{\mu}_{1})\hspace{0.03cm}
\biggl\{\,\frac{\partial\hspace{0.03cm}\Omega\hspace{0.03cm}(\bar{\xi}^{\,\prime},\xi)}
{\partial\hspace{0.03cm}\xi_{2}}\hspace{0.03cm},\hspace{0.03cm} \bar{\mu}_{2}\biggr\}
\biggr)\hspace{0.03cm}
d^{\hspace{0.02cm}2}\bar{\mu}_{1}\hspace{0.03cm} d^{\hspace{0.02cm}2}\bar{\mu}_{2} =
\label{eq:2d}
\end{equation}
\[
\begin{split}
&-\frac{\!1}{2^{\hspace{0.02cm}2}}\,\biggl(\int\hspace{0.03cm}\biggl[\,\bar{\mu}_{1}\hspace{0.03cm}, \frac{\partial\hspace{0.03cm}\Omega\hspace{0.03cm}(\bar{\xi}^{\,\prime},\xi)}
{\partial\hspace{0.03cm}\bar{\xi}^{\,\prime}_{1}}\hspace{0.03cm}\,\biggr]\!\hspace{0.03cm}
\biggl\{\,\frac{\partial\hspace{0.03cm}\Omega\hspace{0.03cm}(\bar{\xi}^{\,\prime},\xi)}
{\partial\hspace{0.03cm}\xi^{\phantom{\prime}}_{1}}\hspace{0.03cm},\hspace{0.03cm} \bar{\mu}_{1}\biggr\}
\hspace{0.04cm}d^{\hspace{0.02cm}2}\bar{\mu}_{1}
+
\int\hspace{0.03cm}\biggl[\,\bar{\mu}_{2}\hspace{0.03cm}, \frac{\partial\hspace{0.03cm}\Omega\hspace{0.03cm}(\bar{\xi}^{\,\prime},\xi)}
{\partial\hspace{0.03cm}\bar{\xi}^{\,\prime}_{2}}\hspace{0.03cm}\,\biggr]\!\hspace{0.03cm}
\biggl\{\,\frac{\partial\hspace{0.03cm}\Omega\hspace{0.03cm}(\bar{\xi}^{\,\prime},\xi)}
{\partial\hspace{0.03cm}\xi^{\phantom{\prime}}_{2}}\hspace{0.03cm},\hspace{0.03cm} \bar{\mu}_{2}\biggr\}\hspace{0.04cm}d^{\hspace{0.02cm}2}\bar{\mu}_{2}
\biggr) = \\[1.5ex]
&-\frac{1}{2}\,\biggl(\hspace{0.03cm}\biggl[\,\frac{\partial\hspace{0.03cm}
\Omega\hspace{0.03cm}(\bar{\xi}^{\,\prime},\xi)}
{\partial\hspace{0.03cm}\xi^{\phantom{\prime}}_{1}}\hspace{0.03cm}, \frac{\partial\hspace{0.03cm}\Omega\hspace{0.03cm}(\bar{\xi}^{\,\prime},\xi)}
{\partial\hspace{0.03cm}\bar{\xi}^{\,\prime}_{1}}\hspace{0.03cm}\,\biggr]
+
\biggl[\,\frac{\partial\hspace{0.03cm}
\Omega\hspace{0.03cm}(\bar{\xi}^{\,\prime},\xi)}
{\partial\hspace{0.03cm}\xi^{\phantom{\prime}}_{2}}\hspace{0.03cm}, \frac{\partial\hspace{0.03cm}\Omega\hspace{0.03cm}(\bar{\xi}^{\,\prime},\xi)}
{\partial\hspace{0.03cm}\bar{\xi}^{\,\prime}_{2}}\hspace{0.03cm}\,\biggr]
\biggr)
\equiv
-\frac{\!1}{2}\,\biggl[\,\frac{\partial\hspace{0.03cm}
\Omega\hspace{0.03cm}(\bar{\xi}^{\,\prime},\xi)}
{\partial\hspace{0.03cm}\xi}\hspace{0.03cm}, \frac{\partial\hspace{0.03cm}\Omega\hspace{0.03cm}(\bar{\xi}^{\,\prime},\xi)}
{\partial\hspace{0.03cm}\bar{\xi}^{\,\prime}}\hspace{0.03cm}\,\biggr].
\end{split}
\]
Therefore, we can already write out the first two terms in the expansion of convolution integral (\ref{eq:2y})
\begin{equation}
\widetilde{\Omega}\hspace{0.03cm}(\bar{\xi}^{\,\prime},\xi)
=
[\Omega\hspace{0.03cm}(\bar{\xi}^{\,\prime},\xi)]^{\hspace{0.04cm}2}
+
\biggl(-\frac{\!1}{2}\biggr)\biggl[\,\frac{\partial\hspace{0.03cm}
\Omega\hspace{0.03cm}(\bar{\xi}^{\,\prime},\xi)}
{\partial\hspace{0.03cm}\xi}\hspace{0.03cm}, \frac{\partial\hspace{0.03cm}\Omega\hspace{0.03cm}(\bar{\xi}^{\,\prime},\xi)}
{\partial\hspace{0.03cm}\bar{\xi}^{\,\prime}}\hspace{0.03cm}\,\biggr]
+\, \ldots\,.
\label{eq:2f}
\end{equation}
\indent Let us analyse the contributions cubic in $\mu$ and $\bar{\mu}$. We take for example the term with derivative from the expansion (\ref{eq:2u}) and the contribution ${\bf (1)}$ from (\ref{eq:2i}). Then, instead of (\ref{eq:2p}), we have
\[
\frac{\!1}{2^{4}}\,[\hspace{0.03cm}\xi_{1}, \xi_{2}\hspace{0.03cm}]\!\iint
\biggl[\,\frac{\partial\hspace{0.03cm}\Omega\hspace{0.03cm}(\bar{\xi}^{\,\prime},\xi)}
{\partial\hspace{0.03cm}\xi}\hspace{0.03cm},\hspace{0.03cm} \mu\,\biggr]
[\hspace{0.03cm}\bar{\mu}_{1}, \bar{\mu}_{2}\hspace{0.03cm}]\,
{\rm e}^{\,-\textstyle\frac{\!1}{2}\,[\hspace{0.03cm}\bar{\mu},\mu\hspace{0.03cm}]} \hspace{0.03cm}(d\mu)_{2}\hspace{0.03cm}(d\bar{\mu})_{2} =
\]
\vspace{0.01cm}
\[
\frac{\!1}{2^{4}}\,[\hspace{0.03cm}\xi_{1}, \xi_{2}\hspace{0.03cm}]\!\iint\!
\biggl(\!\delta(\bar{\mu}_{2})\hspace{0.03cm}
\biggl\{\,\frac{\partial\hspace{0.03cm}\Omega\hspace{0.03cm}(\bar{\xi}^{\,\prime},\xi)}
{\partial\hspace{0.03cm}\xi_{1}}\hspace{0.03cm},\hspace{0.03cm} \bar{\mu}_{1}\biggr\}
+\hspace{0.03cm}
\delta(\bar{\mu}_{1})\hspace{0.03cm}
\biggl\{\,\frac{\partial\hspace{0.03cm}\Omega\hspace{0.03cm}(\bar{\xi}^{\,\prime},\xi)}
{\partial\hspace{0.03cm}\xi_{2}}\hspace{0.03cm},\hspace{0.03cm} \bar{\mu}_{2}\biggr\}
\!\biggr)
[\hspace{0.03cm}\bar{\mu}_{1}, \bar{\mu}_{2}\hspace{0.03cm}]\hspace{0.05cm}
d^{\hspace{0.02cm}2}\bar{\mu}_{1}\hspace{0.03cm} d^{\hspace{0.02cm}2}\bar{\mu}_{2} = 0.
\]
Here, we have used the results of the previous calculations, Eqs.\,(\ref{eq:2a}) and (\ref{eq:2s}). Similarly, we can verify that in (\ref{eq:2y}) the remaining contributions cubic in $\mu$ and $\bar{\mu}$ also vanish.

%%%%%%%%%%%%%%%%%%%%%%%% section 3 %%%%%%%%%%%%%%%%%%%%%%%%%%%%

\section{\bf Contributions of the fourth order in $\mu$ and $\bar{\mu}$}
\setcounter{equation}{0}
\label{section_3}

Now we turn to the consideration of the remaining contributions of fourth order in $\mu$ and $\bar{\mu}$. Let us consider the ``diagonal'' contribution ${\bf (1)-(1)}$, where the first one
${\bf (1)}$ designates a term with a mark ${\bf (1)}$ in (\ref{eq:2u}) and the second one
${\bf (1)}$ denotes a similar term in (\ref{eq:2i}):
\begin{equation}
\frac{\!1}{2^{\hspace{0.02cm}6}}\,[\hspace{0.03cm}\bar{\xi}^{\,\prime}_{1},
\bar{\xi}^{\,\prime}_{2}\hspace{0.03cm}]\hspace{0.03cm}
[\hspace{0.03cm}\xi^{\phantom{\prime}}_{1}, \xi^{\phantom{\prime}}_{2}\hspace{0.03cm}]
\iint[\hspace{0.03cm}\mu_{1}, \mu_{2}\hspace{0.03cm}]
[\hspace{0.03cm}\bar{\mu}_{1}, \bar{\mu}_{2}\hspace{0.03cm}]\,
{\rm e}^{\,-\textstyle\frac{\!1}{2}\,[\hspace{0.03cm}\bar{\mu},\mu\hspace{0.03cm}]} \hspace{0.03cm}(d\mu)_{2}\hspace{0.03cm}(d\bar{\mu})_{2}.
\label{eq:3q}
\end{equation}
The use of the representation (\ref{eq:2p}), (\ref{eq:2a}) leads to necessity of the calculation of the matrix element
\[
\langle\hspace{0.02cm}-\bar{\mu}\hspace{0.02cm}
|\hspace{0.03cm}[\hspace{0.03cm}a^{-}_{1}, a^{-}_{2}\hspace{0.03cm}]
\hspace{0.02cm}(a^{+}_{2})^{2} (a^{+}_{1})^{2} |\,0\hspace{0.02cm}\rangle.
\]
In view of the commutation rules (I.B.4)\,--\,(I.B.6), we obtain
\[
[\hspace{0.03cm}a^{-}_{1}, a^{-}_{2}\hspace{0.03cm}]
\hspace{0.02cm}(a^{+}_{2})^{2} (a^{+}_{1})^{2}
=
\]
\[
(a^{+}_{2})^{2} (a^{+}_{1})^{2}\hspace{0.02cm}
[\hspace{0.03cm}a^{-}_{1}, a^{-}_{2}\hspace{0.03cm}]
-
2\hspace{0.03cm}(a^{+}_{2})^{2}a^{-}_{2}a^{+}_{1} -
2\hspace{0.03cm}a^{+}_{1}a^{+}_{2}a^{-}_{1}a^{+}_{1} -
2\hspace{0.03cm}a^{+}_{2}a^{+}_{1}a^{+}_{1}a^{-}_{1}
+ 4\hspace{0.03cm}a^{+}_{2}a^{+}_{1}
- 2\hspace{0.03cm}(a^{+}_{2})^{2}a^{+}_{1}a^{-}_{2},
\]
and therefore the required matrix element equals
\[
\langle\hspace{0.02cm}-\bar{\mu}\hspace{0.02cm}
|\hspace{0.03cm}[\hspace{0.03cm}a^{-}_{1}, a^{-}_{2}\hspace{0.03cm}]
\hspace{0.02cm}(a^{+}_{2})^{2} (a^{+}_{1})^{2} |\,0\hspace{0.02cm}\rangle
= -4\hspace{0.03cm}[\hspace{0.03cm}\bar{\mu}_{1}, \bar{\mu}_{2}\hspace{0.03cm}],
\]
and the integral (\ref{eq:3q}) transforms to
\[
-\frac{\!1}{2^{\hspace{0.03cm}6}}\,[\hspace{0.03cm}\bar{\xi}^{\,\prime}_{1},
\bar{\xi}^{\,\prime}_{2}\hspace{0.03cm}]\hspace{0.03cm}
[\hspace{0.03cm}\xi^{\phantom{\prime}}_{1}, \xi^{\phantom{\prime}}_{2}\hspace{0.03cm}]
\iint
[\hspace{0.03cm}\bar{\mu}_{1}, \bar{\mu}_{2}\hspace{0.03cm}]^{\hspace{0.03cm}2}
\hspace{0.03cm}d^{2}\bar{\mu}_{1}\hspace{0.04cm}d^{2}\bar{\mu}_{2}
=
-\frac{\!1}{2^{\hspace{0.03cm}3}}\,[\hspace{0.03cm}\bar{\xi}^{\,\prime}_{1},
\bar{\xi}^{\,\prime}_{2}\hspace{0.03cm}]\hspace{0.03cm}
[\hspace{0.03cm}\xi^{\phantom{\prime}}_{1}, \xi^{\phantom{\prime}}_{2}\hspace{0.03cm}]
\int\!\delta(\bar{\mu}_{2})\hspace{0.03cm}d^{2}\bar{\mu}_{2}.
\]
Here, we have used the integration formula \eqref{ap:A3} and the definition of $\delta$-function, Eq.\,\eqref{ap:A5}. As a result, the desired contribution of fourth order ${\bf (1)-(1)}$ is equal to
\begin{equation}
{\bf (1)\!-\!(1)}: \quad -\frac{\!1}{2^{\hspace{0.03cm}3}}\,
[\hspace{0.03cm}\bar{\xi}^{\,\prime}_{1},\bar{\xi}^{\,\prime}_{2}\hspace{0.03cm}]
\hspace{0.03cm}
[\hspace{0.03cm}\xi^{\phantom{\prime}}_{1}, \xi^{\phantom{\prime}}_{2}\hspace{0.03cm}].
\label{eq:3w}
\end{equation}
\indent Let us consider the second ``diagonal'' contribution of fourth order to the convolution (\ref{eq:2y}), namely 
${\bf (2)-(2)}$ one:
\begin{equation}
\frac{\!1}{2^{\hspace{0.02cm}6}}\iint
[\hspace{0.03cm}\bar{\xi}^{\,\prime}_{2}, \mu_{1}\hspace{0.03cm}]
\hspace{0.03cm}
[\hspace{0.03cm}\bar{\xi}^{\,\prime}_{1}, \mu_{2}\hspace{0.03cm}]
\hspace{0.03cm}
[\hspace{0.03cm}\bar{\mu}_{2}, \xi^{\phantom{\prime}}_{1}\hspace{0.03cm}]
\hspace{0.03cm}
[\hspace{0.03cm}\bar{\mu}_{1}, \xi^{\phantom{\prime}}_{2}\hspace{0.03cm}]\,
{\rm e}^{\,-\textstyle\frac{\!1}{2}\,[\hspace{0.03cm}\bar{\mu},\mu\hspace{0.03cm}]} \hspace{0.03cm}(d\mu)_{2}\hspace{0.03cm}(d\bar{\mu})_{2}.
\label{eq:3e}
\end{equation}
Here, we are faced with an analysis of matrix element of the following type:
\begin{equation}
\langle\hspace{0.02cm}-\bar{\mu}\hspace{0.02cm}
|\hspace{0.03cm}[\hspace{0.03cm}\bar{\xi}^{\prime}_{2}, a^{-}_{1}\hspace{0.03cm}]
[\hspace{0.03cm}\bar{\xi}^{\prime}_{1}, a^{-}_{2}\hspace{0.03cm}]
\hspace{0.02cm}(a^{+}_{2})^{2} (a^{+}_{1})^{2} |\,0\hspace{0.02cm}\rangle.
\label{eq:3r}
\end{equation}
By using the commutation rules (I.C.3) and (I.C.4), after somewhat cumbersome calculations we find
\[
\begin{split}
&\hspace{3cm}[\hspace{0.03cm}\bar{\xi}^{\prime}_{2}, a^{-}_{1}\hspace{0.03cm}]
[\hspace{0.03cm}\bar{\xi}^{\prime}_{1}, a^{-}_{2}\hspace{0.03cm}]
\hspace{0.02cm}(a^{+}_{2})^{2} (a^{+}_{1})^{2}
=
(a^{+}_{2})^{2} (a^{+}_{1})^{2}\hspace{0.02cm}
[\hspace{0.03cm}\bar{\xi}^{\prime}_{2}, a^{-}_{1}\hspace{0.03cm}]
[\hspace{0.03cm}\bar{\xi}^{\prime}_{1}, a^{-}_{2}\hspace{0.03cm}]
\,+\\[1ex]
&+
2\hspace{0.03cm}(\hspace{0.03cm}\bar{\xi}^{\prime}_{2}\hspace{0.03cm}a^{+}_{1} (a^{+}_{2})^{2}\hspace{0.03cm}
[\hspace{0.03cm}\bar{\xi}^{\prime}_{1}, a^{-}_{2}\hspace{0.03cm}]
+
\hspace{0.03cm}a^{+}_{1} \hspace{0.03cm}\bar{\xi}^{\prime}_{2} (a^{+}_{2})^{2}\hspace{0.03cm}
[\hspace{0.03cm}\bar{\xi}^{\prime}_{1}, a^{-}_{2}\hspace{0.03cm}]
+
\hspace{0.03cm}(a^{+}_{1})^{2}\hspace{0.03cm}\bar{\xi}^{\prime}_{1}
\hspace{0.03cm}a^{+}_{2} \hspace{0.03cm}
[\hspace{0.03cm}\bar{\xi}^{\prime}_{2}, a^{-}_{1}\hspace{0.03cm}]
+
\hspace{0.03cm}(a^{+}_{1})^{2}\hspace{0.03cm}a^{+}_{2}
\hspace{0.03cm}\bar{\xi}^{\prime}_{1}\hspace{0.03cm}
[\hspace{0.03cm}\bar{\xi}^{\prime}_{2}, a^{-}_{1}\hspace{0.03cm}])
\,+\\[1ex]
&\hspace{3cm}+4\hspace{0.03cm}(\hspace{0.03cm}\bar{\xi}^{\prime}_{2}
\hspace{0.03cm}a^{+}_{1} \bar{\xi}^{\prime}_{1}\hspace{0.03cm}a^{+}_{2}
+
\hspace{0.03cm}a^{+}_{1}\hspace{0.03cm}\bar{\xi}^{\prime}_{2} \,\bar{\xi}^{\prime}_{1}\hspace{0.03cm}a^{+}_{2}
+
\hspace{0.03cm}\bar{\xi}^{\prime}_{2}\hspace{0.03cm}a^{+}_{1} a^{+}_{2}\hspace{0.03cm}\bar{\xi}^{\prime}_{1}
+
\hspace{0.03cm}a^{+}_{1}\hspace{0.03cm}\bar{\xi}^{\prime}_{2} \,a^{+}_{2}\hspace{0.03cm}\bar{\xi}^{\prime}_{1}).
\end{split}
\]
Substituting this expression into (\ref{eq:3r}), we obtain
\[
\langle\hspace{0.02cm}-\bar{\mu}\hspace{0.02cm}
|\hspace{0.03cm}[\hspace{0.03cm}\bar{\xi}^{\prime}_{2}, a^{-}_{1}\hspace{0.03cm}]
[\hspace{0.03cm}\bar{\xi}^{\prime}_{1}, a^{-}_{2}\hspace{0.03cm}]
\hspace{0.02cm}(a^{+}_{2})^{2} (a^{+}_{1})^{2} |\,0\hspace{0.02cm}\rangle
=
\]
\[
\hspace{0.1cm}= 4\hspace{0.03cm}(\hspace{0.03cm}\bar{\xi}^{\prime}_{2}\hspace{0.03cm}
\bar{\mu}^{\phantom{\prime}}_{1} \hspace{0.03cm}\bar{\xi}^{\prime}_{1}\hspace{0.03cm}\bar{\mu}^{\phantom{\prime}}_{2}
+
\bar{\mu}^{\phantom{\prime}}_{1} \hspace{0.03cm}\bar{\xi}^{\prime}_{2} \,\bar{\xi}^{\prime}_{1}\hspace{0.03cm}\bar{\mu}^{\phantom{\prime}}_{2}
+
\bar{\xi}^{\prime}_{2}\hspace{0.03cm}\bar{\mu}^{\phantom{\prime}}_{1}\hspace{0.03cm}
\bar{\mu}^{\phantom{\prime}}_{2}\hspace{0.03cm}\bar{\xi}^{\prime}_{1}
+
\bar{\mu}^{\phantom{\prime}}_{1}\hspace{0.03cm}\bar{\xi}^{\prime}_{2} \,\bar{\mu}^{\phantom{\prime}}_{2}\hspace{0.03cm}\bar{\xi}^{\prime}_{1})
\equiv 4\{\bar{\mu}^{\phantom{\prime}}_{1},\bar{\xi}^{\prime}_{2}\}
\{\bar{\mu}^{\phantom{\prime}}_{2},\bar{\xi}^{\prime}_{1}\}.
\]
In this case, the integral (\ref{eq:3e}) is reduced to a product of two independent integrals:
\[
\frac{\!1}{2^{\hspace{0.02cm}6}}\,\biggl(\int
[\hspace{0.03cm}\bar{\mu}_{1}, \xi^{\phantom{\prime}}_{2}\hspace{0.03cm}]
\{\bar{\mu}_{1},\bar{\xi}^{\prime}_{2}\}\hspace{0.04cm}d^{2}\bar{\mu}_{1}\biggr)\!
\biggl(\int[\hspace{0.03cm}\bar{\mu}_{2}, \xi^{\phantom{\prime}}_{1}\hspace{0.03cm}]
\{\bar{\mu}_{2},\bar{\xi}^{\prime}_{1}\}\hspace{0.04cm}d^{2}\bar{\mu}_{2}\biggr) =
\frac{\!1}{2^{\hspace{0.02cm}4}}\,
[\hspace{0.03cm}\bar{\xi}^{\,\prime}_{1}, \xi^{\phantom{\prime}}_{1}\hspace{0.03cm}]
\hspace{0.03cm}
[\hspace{0.03cm}\bar{\xi}^{\,\prime}_{2}, \xi^{\phantom{\prime}}_{2}\hspace{0.03cm}].
\]
Here, we have used the integration formula \eqref{ap:A4}. As a result, for the contribution ${\bf (2)-(2)}$ we derive
\begin{equation}
{\bf (2)\!-\!(2)}: \quad \frac{\!1}{2^{\hspace{0.02cm}4}}\,
[\hspace{0.03cm}\bar{\xi}^{\,\prime}_{1}, \xi^{\phantom{\prime}}_{1}\hspace{0.03cm}]
\hspace{0.03cm}
[\hspace{0.03cm}\bar{\xi}^{\,\prime}_{2}, \xi^{\phantom{\prime}}_{2}\hspace{0.03cm}].
\label{eq:3t}
\end{equation}
From the expression (\ref{eq:2u}) and (\ref{eq:2i}) it is easy to see that the third ``diagonal'' contribution ${\bf (3)-(3)}$ follows from previous one by a simple replacement: $\bar{\xi}_{1}^{\prime} \rightleftarrows \bar{\xi}_{2}^{\prime},\,\xi_1 \rightleftarrows \xi_2$, consequently we can immediately write
\begin{equation}
{\bf (3)\!-\!(3)}: \quad \frac{\!1}{2^{\hspace{0.02cm}4}}\,
[\hspace{0.03cm}\bar{\xi}^{\,\prime}_{1}, \xi^{\phantom{\prime}}_{1}\hspace{0.03cm}]
\hspace{0.03cm}
[\hspace{0.03cm}\bar{\xi}^{\,\prime}_{2}, \xi^{\phantom{\prime}}_{2}\hspace{0.03cm}].
\label{eq:3y}
\end{equation}
\indent Let us now examine the fourth order ``off-diagonal'' contributions. The calculations completely similar to the previous ones result in the following expressions:
\begin{equation}
\begin{split}
&{\bf (1)\!-\!(2)} = {\bf (2)\!-\!(1)}: \quad -\frac{\!1}{2^{\hspace{0.03cm}4}}\,
[\hspace{0.03cm}\bar{\xi}^{\,\prime}_{1},\bar{\xi}^{\,\prime}_{2}\hspace{0.03cm}]
\hspace{0.03cm}
[\hspace{0.03cm}\xi^{\phantom{\prime}}_{1}, \xi^{\phantom{\prime}}_{2}\hspace{0.03cm}],
\\[1ex]
&{\bf (1)\!-\!(3)} = {\bf (3)\!-\!(1)}: \quad -\frac{\!1}{2^{\hspace{0.03cm}4}}\,
[\hspace{0.03cm}\bar{\xi}^{\,\prime}_{1},\bar{\xi}^{\,\prime}_{2}\hspace{0.03cm}]
\hspace{0.03cm}
[\hspace{0.03cm}\xi^{\phantom{\prime}}_{1}, \xi^{\phantom{\prime}}_{2}\hspace{0.03cm}],
\\[1ex]
&{\bf (2)\!-\!(3)}  = {\bf (3)\!-\!(2)}: \quad -\frac{\!1}{2^{\hspace{0.02cm}4}}\,
[\hspace{0.03cm}\bar{\xi}^{\,\prime}_{1}, \xi^{\phantom{\prime}}_{2}\hspace{0.03cm}]
\hspace{0.03cm}
[\hspace{0.03cm}\bar{\xi}^{\,\prime}_{2}, \xi^{\phantom{\prime}}_{1}\hspace{0.03cm}].
\end{split}
\label{eq:3u}
\end{equation}
Substituting the obtained terms of expansion (\ref{eq:2f}), (\ref{eq:3w}), (\ref{eq:3t}), (\ref{eq:3y}) and (\ref{eq:3u}) into (\ref{eq:2y}) and collecting similar terms, we finally obtain
\begin{equation}
\widetilde{\Omega}\hspace{0.03cm}(\bar{\xi}^{\,\prime},\xi)
=
[\hspace{0.03cm}\Omega\hspace{0.03cm}(\bar{\xi}^{\,\prime},\xi)]^{\hspace{0.04cm}2}
+
\biggl(-\frac{1}{2}\biggr)\biggl[\,\frac{\partial\hspace{0.03cm}
\Omega\hspace{0.03cm}(\bar{\xi}^{\,\prime},\xi)}
{\partial\hspace{0.03cm}\xi}\hspace{0.03cm}, \frac{\partial\hspace{0.03cm}\Omega\hspace{0.03cm}(\bar{\xi}^{\,\prime},\xi)}
{\partial\hspace{0.03cm}\bar{\xi}^{\,\prime}}\hspace{0.03cm}\,\biggr]
+
\biggl(-\frac{\!3}{2^{\hspace{0.03cm}3}}\biggr)
[\hspace{0.03cm}\bar{\xi}^{\,\prime}_{1},\bar{\xi}^{\,\prime}_{2}\hspace{0.03cm}]
\hspace{0.03cm}
[\hspace{0.03cm}\xi^{\phantom{\prime}}_{1}, \xi^{\phantom{\prime}}_{2}\hspace{0.03cm}]
\,+
\label{eq:3i}
\end{equation}
\[
+\,\frac{\!1}{2^{\hspace{0.02cm}3}}\,
[\hspace{0.03cm}\bar{\xi}^{\,\prime}_{1}, \xi^{\phantom{\prime}}_{1}\hspace{0.03cm}]
\hspace{0.03cm}
[\hspace{0.03cm}\bar{\xi}^{\,\prime}_{2}, \xi^{\phantom{\prime}}_{2}\hspace{0.03cm}]
+
\biggl(-\frac{\!1}{2^{\hspace{0.02cm}3}}\biggr)
[\hspace{0.03cm}\bar{\xi}^{\,\prime}_{1}, \xi^{\phantom{\prime}}_{2}\hspace{0.03cm}]
\hspace{0.03cm}
[\hspace{0.03cm}\bar{\xi}^{\,\prime}_{2}, \xi^{\phantom{\prime}}_{1}\hspace{0.03cm}].
\]
This expression can be rewritten in terms of derivatives of the function $\Omega$. Making use of the differentiation formulae (I.C.9)\,--\,(I.C.12) and an explicit form of the function (\ref{eq:2w}), we find instead of (\ref{eq:3i})
\[
\widetilde{\Omega}\hspace{0.03cm}(\bar{\xi}^{\,\prime},\xi)
=
[\hspace{0.03cm}\Omega\hspace{0.03cm}(\bar{\xi}^{\,\prime},\xi)]^{\hspace{0.04cm}2}
+
\biggl(-\frac{1}{2}\biggr)\biggl[\,\frac{\partial\hspace{0.03cm}
\Omega\hspace{0.03cm}(\bar{\xi}^{\,\prime},\xi)}
{\partial\hspace{0.03cm}\xi}\hspace{0.03cm}, \frac{\partial\hspace{0.03cm}\Omega\hspace{0.03cm}(\bar{\xi}^{\,\prime},\xi)}
{\partial\hspace{0.03cm}\bar{\xi}^{\,\prime}}\hspace{0.03cm}\,\biggr]
+
\biggl(-\frac{\!3}{2^{\hspace{0.02cm}3}}\biggr)
\biggl(\frac{\partial^{2}\hspace{0.03cm}\Omega\hspace{0.03cm}(\bar{\xi}^{\,\prime},\xi)}
{\partial\hspace{0.03cm}\xi_{1}\hspace{0.04cm}\partial\hspace{0.03cm}\xi_{2}}
\cdot
\frac{\partial^{2}\hspace{0.03cm}\Omega\hspace{0.03cm}(\bar{\xi}^{\,\prime},\xi)}
{\partial\hspace{0.03cm}\bar{\xi}^{\,\prime}_{1}\hspace{0.04cm}
\partial\hspace{0.03cm}\bar{\xi}^{\,\prime}_{2}}\biggr)
\,+
\]
\[
+\,\frac{\!1}{2^{\hspace{0.02cm}3}}\,
\biggl(\frac{\partial^{2}\hspace{0.03cm}\Omega\hspace{0.03cm}(\bar{\xi}^{\,\prime},\xi)}
{\partial\hspace{0.03cm}\bar{\xi}^{\,\prime}_{1}\hspace{0.04cm}
\partial\hspace{0.03cm}\xi^{\phantom{\prime}}_{1}} - 2 \biggr)
\biggl(\frac{\partial^{2}\hspace{0.03cm}\Omega\hspace{0.03cm}(\bar{\xi}^{\,\prime},\xi)}
{\partial\hspace{0.03cm}\bar{\xi}^{\,\prime}_{2}\hspace{0.04cm}
\partial\hspace{0.03cm}\xi^{\phantom{\prime}}_{2}} - 2 \biggr)
+
\biggl(-\frac{\!1}{2^{\hspace{0.02cm}3}}\biggr)
\biggl(\frac{\partial^{2}\hspace{0.03cm}\Omega\hspace{0.03cm}(\bar{\xi}^{\,\prime},\xi)}
{\partial\hspace{0.03cm}\bar{\xi}^{\,\prime}_{1}\hspace{0.04cm}
\partial\hspace{0.03cm}\xi^{\phantom{\prime}}_{2}}
\cdot
\frac{\partial^{2}\hspace{0.03cm}\Omega\hspace{0.03cm}(\bar{\xi}^{\,\prime},\xi)}
{\partial\hspace{0.03cm}\bar{\xi}^{\,\prime}_{2}\hspace{0.04cm}
\partial\hspace{0.03cm}\xi^{\phantom{\prime}}_{1}}\biggr).
\]
\indent In closing this section, we would like to consider another more direct way of taking the integrals with the para-Grassmann variables. As an example, let us examine the contribution ${\bf (1)-(1)}$, i.e. the expression (\ref{eq:3q}). Instead of representation (\ref{eq:2p}) we use the expansion of exponential function
\begin{equation}
\begin{split}
&{\rm e}^{\,\textstyle\hspace{0.02cm}-\frac{\!1}{2}\,
[\hspace{0.02cm}\bar{\mu}\hspace{0.02cm},\mu\hspace{0.02cm}\hspace{0.02cm}]}
=
1- \frac{\!1}{2}\,[\hspace{0.02cm}\bar{\mu}\hspace{0.02cm},\mu\hspace{0.02cm}\hspace{0.02cm}]
+
\frac{\!1}{2!}\,\biggl(\frac{\!1}{2}\biggr)^{\!\!2}
[\hspace{0.02cm}\bar{\mu}\hspace{0.02cm},
\mu\hspace{0.02cm}\hspace{0.02cm}]^{\hspace{0.04cm}2}
-
\frac{\!1}{3!}\,\biggl(\frac{\!1}{2}\biggr)^{\!\!3}
[\hspace{0.02cm}\bar{\mu}\hspace{0.02cm},
\mu\hspace{0.02cm}\hspace{0.02cm}]^{\hspace{0.04cm}3}
+
\frac{\!1}{4!}\,\biggl(\frac{\!1}{2}\biggr)^{\!\!4}
[\hspace{0.02cm}\bar{\mu}\hspace{0.02cm},
\mu\hspace{0.02cm}\hspace{0.02cm}]^{\hspace{0.04cm}4}
= \\[1ex]
&= 1- \frac{\!1}{2}\,\bigl(\hspace{0.02cm}[\hspace{0.02cm}\bar{\mu}_{1}\hspace{0.02cm},
\mu_{1}\hspace{0.02cm}\hspace{0.02cm}] + [\hspace{0.02cm}\bar{\mu}_{2}\hspace{0.02cm},
\mu_{2}\hspace{0.02cm}\hspace{0.02cm}]\hspace{0.02cm}\bigr)
+
\frac{\!1}{2!}\,\biggl(\frac{\!1}{2}\biggr)^{\!\!2}
\bigl(\hspace{0.02cm}[\hspace{0.02cm}\bar{\mu}_{1}\hspace{0.02cm},
\mu_{1}\hspace{0.02cm}\hspace{0.02cm}]^{\hspace{0.04cm}2}
+
2\hspace{0.03cm}[\hspace{0.02cm}\bar{\mu}_{1}\hspace{0.02cm},
\mu_{1}\hspace{0.02cm}\hspace{0.02cm}]\hspace{0.03cm}
[\hspace{0.02cm}\bar{\mu}_{2}\hspace{0.02cm},
\mu_{2}\hspace{0.02cm}\hspace{0.02cm}]
+
[\hspace{0.02cm}\bar{\mu}_{2}\hspace{0.02cm},
\mu_{2}\hspace{0.02cm}\hspace{0.02cm}]^{\hspace{0.04cm}2}\hspace{0.02cm}\bigr)
\,- \\[1ex]
&-\,\frac{\!1}{3!}\,\biggl(\frac{\!1}{2}\biggr)^{\!\!3}
\bigl(\hspace{0.01cm}3\hspace{0.05cm}[\hspace{0.02cm}\bar{\mu}_{1}\hspace{0.02cm},
\mu_{1}\hspace{0.02cm}\hspace{0.02cm}]^{\hspace{0.04cm}2}
[\hspace{0.02cm}\bar{\mu}_{2}\hspace{0.02cm},
\mu_{2}\hspace{0.02cm}\hspace{0.02cm}]
+
3\hspace{0.05cm}[\hspace{0.02cm}\bar{\mu}_{1}\hspace{0.02cm},
\mu_{1}\hspace{0.02cm}\hspace{0.02cm}]\hspace{0.03cm}
[\hspace{0.02cm}\bar{\mu}_{2}\hspace{0.02cm},
\mu_{2}\hspace{0.02cm}\hspace{0.02cm}]^{\hspace{0.04cm}2}\hspace{0.02cm}\bigr)
+
\frac{\!1}{4!}\,\biggl(\frac{\!1}{2}\biggr)^{\!\!4}
6\hspace{0.04cm}[\hspace{0.02cm}\bar{\mu}_{1}\hspace{0.02cm},
\mu_{1}\hspace{0.02cm}\hspace{0.02cm}]^{\hspace{0.04cm}2\,}
[\hspace{0.02cm}\bar{\mu}_{2}\hspace{0.02cm},
\mu_{2}\hspace{0.02cm}\hspace{0.02cm}]^{\hspace{0.04cm}2}.
\end{split}
\label{eq:3o}
\end{equation}
As it is not difficult to see from the integration formulae \eqref{ap:A1}\,--\,\eqref{ap:A4}, a nontrivial contribution to the expression (\ref{eq:3q}) for $p = 2$ gives us only one term from the right-hand side (\ref{eq:3o}), namely
\[
\frac{\!1}{2^{2}}\,[\hspace{0.02cm}\bar{\mu}_{1}\hspace{0.02cm},
\mu_{1}\hspace{0.02cm}\hspace{0.02cm}]\hspace{0.03cm}
[\hspace{0.02cm}\bar{\mu}_{2}\hspace{0.02cm},
\mu_{2}\hspace{0.02cm}\hspace{0.02cm}],
\]
and therefore by using subsequently \eqref{ap:A3}, \eqref{ap:A4} and \eqref{ap:A5}, we derive
\[
\iint[\hspace{0.03cm}\mu_{1}, \mu_{2}\hspace{0.03cm}]
[\hspace{0.03cm}\bar{\mu}_{1}, \bar{\mu}_{2}\hspace{0.03cm}]\,
{\rm e}^{\,-\textstyle\frac{\!1}{2}\,[\hspace{0.03cm}\bar{\mu},\mu\hspace{0.03cm}]} \hspace{0.03cm}(d\mu)_{2}\hspace{0.03cm}(d\bar{\mu})_{2}
=
\frac{\!1}{2^{2}}\!\iint[\hspace{0.03cm}\mu_{1}, \mu_{2}\hspace{0.03cm}]
[\hspace{0.03cm}\bar{\mu}_{1}, \bar{\mu}_{2}\hspace{0.03cm}]\,
[\hspace{0.02cm}\bar{\mu}_{1}\hspace{0.02cm},
\mu_{1}\hspace{0.02cm}\hspace{0.02cm}]\hspace{0.03cm}
[\hspace{0.02cm}\bar{\mu}_{2}\hspace{0.02cm},
\mu_{2}\hspace{0.02cm}\hspace{0.02cm}] \hspace{0.03cm}(d\mu)_{2}\hspace{0.03cm}(d\bar{\mu})_{2}
=
\]
\[
= \frac{\!1}{2^{2}}\,2\hspace{0.02cm}i^{2}\!\!\iint[\hspace{0.03cm}\bar{\mu}_{1}, \bar{\mu}_{2}\hspace{0.03cm}]
[\hspace{0.02cm}\bar{\mu}_{1}\hspace{0.02cm},
\mu_{1}\hspace{0.02cm}\hspace{0.02cm}]\hspace{0.03cm}
\{\hspace{0.02cm}\mu_{1},\bar{\mu}_{2}\hspace{0.02cm}\}
\hspace{0.03cm}d^{\hspace{0.03cm}2}\mu_{1}\hspace{0.03cm}(d\bar{\mu})_{2}
=
\frac{\!1}{2^{2}}\,(2\hspace{0.03cm}i^{2})^{2}\!\!\iint[\hspace{0.03cm}\bar{\mu}_{1}, \bar{\mu}_{2}\hspace{0.03cm}]
[\hspace{0.03cm}\bar{\mu}_{2}, \bar{\mu}_{1}\hspace{0.03cm}]\hspace{0.03cm}
d^{\hspace{0.03cm}2}\bar{\mu}_{1}\hspace{0.03cm}d^{\hspace{0.03cm}2}\bar{\mu}_{2}
=
\]
\[
= -\frac{\!1}{2^{\hspace{0.03cm}2}}\,2^{\hspace{0.03cm}5}\!
\int\!\delta(\bar{\mu}_{2})\hspace{0.03cm}d^{\hspace{0.03cm}2}\bar{\mu}_{2}
= - 2^{3}.
\]
Substituting the obtained number into (\ref{eq:3q}), we reproduce (\ref{eq:3w}).\\
\indent Let us verify more nontrivial contribution (\ref{eq:2o}). At the beginning we write in more detail the integrand in (\ref{eq:2o}):
\begin{equation}
\biggl[\,\frac{\partial\hspace{0.03cm}\Omega}
{\partial\hspace{0.03cm}\xi}\hspace{0.03cm},\hspace{0.03cm} \mu\,\biggr]\!
\biggl[\,\bar{\mu}\hspace{0.03cm}, \frac{\partial\hspace{0.03cm}\Omega}
{\partial\hspace{0.03cm}\bar{\xi}^{\,\prime}}\hspace{0.03cm}\,\biggr]
=
\biggl[\,\frac{\partial\hspace{0.03cm}\Omega}
{\partial\hspace{0.03cm}\xi_{1}}\hspace{0.03cm},\hspace{0.03cm} \mu_{1}\,\biggr]\!
\biggl[\,\bar{\mu}_{1}\hspace{0.03cm}, \frac{\partial\hspace{0.03cm}\Omega}
{\partial\hspace{0.03cm}\bar{\xi}^{\,\prime}_{1}}\hspace{0.03cm}\,\biggr]
+
\biggl[\,\frac{\partial\hspace{0.03cm}\Omega}
{\partial\hspace{0.03cm}\xi_{2}}\hspace{0.03cm},\hspace{0.03cm} \mu_{2}\,\biggr]\!
\biggl[\,\bar{\mu}_{2}\hspace{0.03cm}, \frac{\partial\hspace{0.03cm}\Omega}
{\partial\hspace{0.03cm}\bar{\xi}^{\,\prime}_{2}}\hspace{0.03cm}\,\biggr]
\,+
\label{eq:3p}
\end{equation}
\[
+\,\biggl[\,\frac{\partial\hspace{0.03cm}\Omega}
{\partial\hspace{0.03cm}\xi_{1}}\hspace{0.03cm},\hspace{0.03cm} \mu_{1}\,\biggr]\!
\biggl[\,\bar{\mu}_{2}\hspace{0.03cm}, \frac{\partial\hspace{0.03cm}\Omega}
{\partial\hspace{0.03cm}\bar{\xi}^{\,\prime}_{2}}\hspace{0.03cm}\,\biggr]
+
\biggl[\,\frac{\partial\hspace{0.03cm}\Omega}
{\partial\hspace{0.03cm}\xi_{2}}\hspace{0.03cm},\hspace{0.03cm} \mu_{2}\,\biggr]\!
\biggl[\,\bar{\mu}_{1}\hspace{0.03cm}, \frac{\partial\hspace{0.03cm}\Omega}
{\partial\hspace{0.03cm}\bar{\xi}^{\,\prime}_{1}}\hspace{0.03cm}\,\biggr].
\]
For the first two terms on the right-hand side of (\ref{eq:3p}) in the expansion (\ref{eq:3o}) only two terms give a nontrivial contribution:
\[
-\,\frac{\!1}{2^{4}}\,
\bigl(\hspace{0.02cm}[\hspace{0.02cm}\bar{\mu}_{1}\hspace{0.02cm},
\mu_{1}\hspace{0.02cm}\hspace{0.02cm}]^{\hspace{0.04cm}2\,}
[\hspace{0.02cm}\bar{\mu}_{2}\hspace{0.02cm},
\mu_{2}\hspace{0.02cm}\hspace{0.02cm}]
+
[\hspace{0.02cm}\bar{\mu}_{1}\hspace{0.02cm},
\mu_{1}\hspace{0.02cm}\hspace{0.02cm}]\hspace{0.03cm}
[\hspace{0.02cm}\bar{\mu}_{2}\hspace{0.02cm},
\mu_{2}\hspace{0.02cm}\hspace{0.02cm}]^{\hspace{0.04cm}2}\hspace{0.02cm}\bigr).
\]
For the remaining terms in (\ref{eq:3p}) there are no such nontrivial contributions in (\ref{eq:3o}) and therefore it can be dropped. The integral with the first term on the right-hand side (\ref{eq:3p}) equals
\[
\begin{split}
&\frac{\!1}{2^{6}}\!\iint\biggl[\,\frac{\partial\hspace{0.03cm}\Omega}
{\partial\hspace{0.03cm}\xi_{1}}\hspace{0.03cm},\hspace{0.03cm} \mu_{1}\,\biggr]\!
\biggl[\,\bar{\mu}_{1}\hspace{0.03cm}, \frac{\partial\hspace{0.03cm}\Omega}
{\partial\hspace{0.03cm}\bar{\xi}^{\,\prime}_{1}}\hspace{0.03cm}\,\biggr]
[\hspace{0.02cm}\bar{\mu}_{1}\hspace{0.02cm},
\mu_{1}\hspace{0.02cm}\hspace{0.02cm}]\hspace{0.03cm}
[\hspace{0.02cm}\bar{\mu}_{2}\hspace{0.02cm},
\mu_{2}\hspace{0.02cm}\hspace{0.02cm}]^{\hspace{0.04cm}2}
\hspace{0.03cm}d^{\hspace{0.03cm}2}\mu_{2}\hspace{0.03cm}
d^{\hspace{0.03cm}2}\mu_{1}\hspace{0.03cm}(d\bar{\mu})_{2}
= \\[1ex]
&=\frac{\!1}{2^{3}}\!\iint\!\delta(\bar{\mu}_{2})\biggl[\,\frac{\partial\hspace{0.03cm}\Omega}
{\partial\hspace{0.03cm}\xi_{1}}\hspace{0.03cm},\hspace{0.03cm} \mu_{1}\,\biggr]\!
\biggl[\,\bar{\mu}_{1}\hspace{0.03cm}, \frac{\partial\hspace{0.03cm}\Omega}
{\partial\hspace{0.03cm}\bar{\xi}^{\,\prime}_{1}}\hspace{0.03cm}\,\biggr]
[\hspace{0.02cm}\bar{\mu}_{1}\hspace{0.02cm},
\mu_{1}\hspace{0.02cm}\hspace{0.02cm}]\hspace{0.03cm}
\hspace{0.03cm}d^{\hspace{0.03cm}2}\mu_{1}\hspace{0.03cm}(d\bar{\mu})_{2}
= \\[1ex]
&=\frac{i^{2}}{2^{2}}\!\iint\!\delta(\bar{\mu}_{2})\biggl[\,\bar{\mu}_{1}\hspace{0.03cm}, \frac{\partial\hspace{0.03cm}\Omega}
{\partial\hspace{0.03cm}\bar{\xi}^{\,\prime}_{1}}\hspace{0.03cm}\,\biggr]\!
\biggl\{\,\bar{\mu}_{1}\hspace{0.03cm}, \frac{\partial\hspace{0.03cm}\Omega}
{\partial\hspace{0.03cm}\xi_{1}}\hspace{0.03cm}\,\biggr\}
\hspace{0.03cm}
d^{\hspace{0.03cm}2}\bar{\mu}_{1}\hspace{0.03cm}d^{\hspace{0.03cm}2}\bar{\mu}_{2}
=
-\frac{1}{2}\,\biggl[\,\frac{\partial\hspace{0.03cm}
\Omega\hspace{0.03cm}(\bar{\xi}^{\,\prime},\xi)}
{\partial\hspace{0.03cm}\xi^{\phantom{\prime}}_{1}}\hspace{0.03cm}, \frac{\partial\hspace{0.03cm}\Omega\hspace{0.03cm}(\bar{\xi}^{\,\prime},\xi)}
{\partial\hspace{0.03cm}\bar{\xi}^{\,\prime}_{1}}\hspace{0.03cm}\,\biggr].
\end{split}
\]
The integral with the second term in (\ref{eq:3p}) is taking in a similar way and as a result, we reproduce the second term on the right-hand side (\ref{eq:2f}).

%%%%%%%%%%%%%%%%%%%%%%%% section 4 %%%%%%%%%%%%%%%%%%%%%%%%%%%%

\section{The star product}
\setcounter{equation}{0}
\label{section_4}

It remains for us to compare the expression for the convolution (\ref{eq:3i}), with the expression for the function 
$\tilde{\Omega}$, Eq.\,(\ref{eq:2r}), which follows from the matrix element of the operator $a_{0}^{2}$. But we shall defer this comparison until the following section, and here we would like to introduce an important notion of a star product $*$ in a class of functions depending on the para-Grassmann variables. For this purpose we return to the connection between the functions $\widetilde{\Omega}$ and $\Omega$ as it was defined by the convolution integral, 
Eq.\,(\ref{eq:2y}). We present the integrands as a result of action of the shift operators on $\Omega$:
\begin{equation}
\Omega\hspace{0.03cm}(\bar{\xi}^{\,\prime}\! + \bar{\mu},\xi)
=
{\rm e}^{\textstyle\frac{1}{2}\,[\hspace{0.03cm}\bar{\mu},\!\overrightarrow{\partial}\!
/\partial\bar{\xi}^{\,\prime}\hspace{0.03cm}]}\,
\Omega\hspace{0.03cm}(\bar{\xi}^{\,\prime},\xi),
\quad
\Omega\hspace{0.03cm}(\bar{\xi}^{\,\prime},\xi + \mu)
=
\Omega\hspace{0.03cm}(\bar{\xi}^{\,\prime},\xi)\,
{\rm e}^{\textstyle-\frac{1}{2}\,[\hspace{0.03cm}\mu,\!\overleftarrow{\partial}\!
/\partial\xi\hspace{0.03cm}]}\,.
\label{eq:4q}
\end{equation}
By a direct calculation, we verify that these expressions reproduce the expansions (\ref{eq:2u}) and (\ref{eq:2i}). To be specific, let us consider the first one
\[
{\rm e}^{\textstyle\frac{1}{2}\,[\hspace{0.03cm}\bar{\mu},\!\overrightarrow{\partial}\!
/\partial\bar{\xi}^{\,\prime}\hspace{0.03cm}]}\,
\Omega\hspace{0.03cm}(\bar{\xi}^{\,\prime},\xi)
=
\Omega\hspace{0.03cm}(\bar{\xi}^{\,\prime},\xi)
+
\frac{1}{2}\,\sum\limits^{2}_{l\hspace{0.03cm}=\hspace{0.03cm}1}\,
\biggl[\hspace{0.03cm}\bar{\mu}_{l}\hspace{0.03cm},\frac{\partial\hspace{0.03cm}\Omega}
{\partial\bar{\xi}^{\,\prime}_{l}\hspace{0.03cm}}\biggr]
+
\frac{1}{2!}\hspace{0.03cm}\biggl(\frac{1}{2}\biggr)^{\!\!2}
\sum\limits^{2}_{l = 1}\sum\limits^{2}_{s\hspace{0.03cm}=\hspace{0.02cm}1}\,
\biggl[\hspace{0.03cm}\bar{\mu}_{l}\hspace{0.03cm},\frac{\partial}
{\partial\bar{\xi}^{\,\prime}_{l}\hspace{0.03cm}}\biggr]\!
\biggl[\hspace{0.03cm}\bar{\mu}_{s}\hspace{0.03cm},\frac{\partial\hspace{0.03cm}\Omega}
{\partial\bar{\xi}^{\,\prime}_{s}\hspace{0.03cm}}\biggr].
\]
The function $\Omega$ depends quadratically on the para-Grassmann variable $\bar{\xi}_{l}^{\prime}$, because of this, the expansion is exactly terminated on the third term. The first two terms coincides with the corresponding terms in (\ref{eq:2i}), and so there is a need to analyse only the last term. Let us consider the contribution with $s = 1$ and use an explicit form of the derivative $\partial\hspace{0.02cm}\Omega/\partial\hspace{0.01cm}\bar{\xi}_{1}^{\prime}$, Eq.\,(I.9.14), then
\[
\biggl[\hspace{0.03cm}\bar{\mu}_{1}\hspace{0.03cm},\frac{\partial\hspace{0.03cm}\Omega}
{\partial\bar{\xi}^{\,\prime}_{1}\hspace{0.03cm}}\biggr]
=
-\frac{\!1}{2^{\hspace{0.02cm}2}}\hspace{0.03cm}\Bigl\{\hspace{0.03cm}
[\hspace{0.03cm}\bar{\mu}^{\phantom{\prime}}_{1}, \bar{\xi}^{\,\prime}_{2}\hspace{0.03cm}]\hspace{0.02cm}
[\hspace{0.03cm}\xi^{\phantom{\prime}}_{1}, \xi^{\phantom{\prime}}_{2}\hspace{0.03cm}]
+
[\hspace{0.03cm}\bar{\mu}^{\phantom{\prime}}_{1}, \xi^{\phantom{\prime}}_{2}\hspace{0.03cm}]\hspace{0.02cm}
[\hspace{0.03cm}\bar{\xi}^{\,\prime}_{2}, \xi^{\phantom{\prime}}_{1}\hspace{0.03cm}]
-
[\hspace{0.03cm}\bar{\mu}^{\phantom{\prime}}_{1}, \xi^{\phantom{\prime}}_{1}\hspace{0.03cm}]\hspace{0.02cm}
[\hspace{0.03cm}\bar{\xi}^{\,\prime}_{2}, \xi^{\phantom{\prime}}_{2}\hspace{0.03cm}]\Bigr\}
-
[\hspace{0.03cm}\bar{\mu}^{\phantom{\prime}}_{1}, \xi^{\phantom{\prime}}_{1}\hspace{0.03cm}].
\]
Under the action of the operator $\bigl[\hspace{0.03cm}\bar{\mu}_{l}\hspace{0.03cm},\partial/
\partial\hspace{0.02cm}\bar{\xi}^{\,\prime}_{l}\hspace{0.03cm}\bigr]$ on the last expression only the term with $l = 2$ give a nontrivial contribution. By employing the differentiation rules (I.C.9)\,--\,(I.C.12), further we find
\[
\biggl[\hspace{0.03cm}\bar{\mu}_{2}\hspace{0.03cm},\frac{\partial}
{\partial\bar{\xi}^{\,\prime}_{2}\hspace{0.03cm}}\biggr]\!
\biggl[\hspace{0.03cm}\bar{\mu}_{1}\hspace{0.03cm},\frac{\partial\hspace{0.03cm}\Omega}
{\partial\bar{\xi}^{\,\prime}_{1}\hspace{0.03cm}}\biggr]
=
-\frac{1}{2}\hspace{0.03cm}\Bigl\{\hspace{0.03cm}[\hspace{0.03cm}\bar{\mu}_{1}, \bar{\mu}_{2}\hspace{0.03cm}]\hspace{0.02cm}
[\hspace{0.03cm}\xi_{1}, \xi_{2}\hspace{0.03cm}]
+
[\hspace{0.03cm}\bar{\mu}_{1}, \xi_{2}\hspace{0.03cm}]\hspace{0.02cm}
[\hspace{0.03cm}\bar{\mu}_{2}, \xi_{1}\hspace{0.03cm}]
-
[\hspace{0.03cm}\bar{\mu}_{1}, \xi_{1}\hspace{0.03cm}]\hspace{0.02cm}
[\hspace{0.03cm}\bar{\mu}_{2}, \xi_{2}\hspace{0.03cm}]\Bigr\}.
\]
The second nontrivial contribution follows from the expression $\bigl[\hspace{0.03cm}\bar{\mu}_1,\partial/\partial\hspace{0.02cm}\bar{\xi}_{1}^{\prime} \hspace{0.03cm}\bigr]\bigl[\hspace{0.03cm}\bar{\mu}_2, \partial\hspace{0.03cm}\Omega/ \partial\hspace{0.02cm} \bar{\xi}_{2}^{\prime}\hspace{0.03cm}\bigr]$. By doing so we exactly reproduce the expression (\ref{eq:2i}).\\
\indent The second expression in (\ref{eq:4q}) is analysed similarly, but only here  the rule of action of the right derivative on the function $\Omega$ must be taken into consideration:
\[
\Omega\,\frac{\overleftarrow{\partial}}{\partial\hspace{0.03cm}\xi_{l}}
=
-\frac{\overrightarrow{\partial}\Omega}{\partial\hspace{0.03cm}\xi_{l}}\,.
\]
\indent Based on the representation (\ref{eq:4q}) we write the convolution (\ref{eq:2y}) in the following form:
\begin{equation}
\widetilde{\Omega}\hspace{0.03cm}(\bar{\xi}^{\,\prime},\xi)
=
\Omega\hspace{0.03cm}(\bar{\xi}^{\,\prime},\xi)
\biggl[\iint{\rm e}^{\textstyle-\frac{1}{2}\,[\hspace{0.03cm}\mu,\!\overleftarrow{\partial}\!
/\partial\hspace{0.03cm}\xi\hspace{0.03cm}]}
\,
{\rm e}^{\textstyle-\frac{1}{2}\,[\hspace{0.03cm}\bar{\mu},\mu\hspace{0.03cm}]}
\,
{\rm e}^{\textstyle\frac{1}{2}\,[\hspace{0.03cm}\bar{\mu},\!\overrightarrow{\partial}\!
/\partial\hspace{0.03cm}\bar{\xi}^{\,\prime}\hspace{0.03cm}]}
\hspace{0.03cm}(d\mu)_{2}\hspace{0.03cm}(d\bar{\mu})_{2}
\biggr]\Omega\hspace{0.03cm}(\bar{\xi}^{\,\prime},\xi)
\label{eq:4w}
\end{equation}
or somewhat differently
\[
\widetilde{\Omega}\hspace{0.03cm}(\bar{\xi}^{\,\prime},\xi)
=
\Omega\hspace{0.03cm}(\bar{\xi}^{\,\prime},\xi)
\biggl[\iint{\rm e}^{\textstyle-\frac{1}{2}\,[\hspace{0.03cm}\bar{\mu} - \!\overleftarrow{\partial}\!
/\partial\hspace{0.03cm}\xi,\mu - \!\overrightarrow{\partial}\!
/\partial\hspace{0.03cm}\bar{\xi}^{\,\prime}\hspace{0.03cm}]}
\hspace{0.03cm}(d\mu)_{2}\hspace{0.03cm}(d\bar{\mu})_{2}
\biggr]
{\rm e}^{\textstyle\frac{1}{2}\,[\hspace{0.03cm}\overleftarrow{\partial}\!
/\partial\hspace{0.03cm}\xi,\!\overrightarrow{\partial}\!
/\partial\hspace{0.03cm}\bar{\xi}^{\,\prime}\hspace{0.03cm}]}\,
\Omega\hspace{0.03cm}(\bar{\xi}^{\,\prime},\xi).
\]
If we will follow the ideology of the paper by Daoud \cite{daoud_2003}, then the last step here is a formal integration of the expression in the square brackets, which results in the definition of a star product in the ``standard'' form
\begin{equation}
* = \exp\biggl(\frac{1}{2}\,\sum\limits^{2}_{l\hspace{0.03cm}=\hspace{0.03cm}1}\,
\biggl[\hspace{0.03cm}
\frac{\overleftarrow{\partial}}{\partial\hspace{0.03cm}\xi^{\phantom{\prime}}_{l}}
\hspace{0.03cm},
\frac{\overrightarrow{\partial}}{\partial\hspace{0.03cm}\bar{\xi}^{\,\prime}_{l}}
\hspace{0.03cm}\biggr]\biggr),
\label{eq:4e}
\end{equation}
where the commutator is
\[
\biggl[\hspace{0.03cm}
\frac{\overleftarrow{\partial}}{\partial\hspace{0.03cm}\xi^{\phantom{\prime}}_{l}}
\hspace{0.03cm},
\frac{\overrightarrow{\partial}}{\partial\hspace{0.03cm}\bar{\xi}^{\,\prime}_{l}}
\hspace{0.03cm}\biggr]
=
\frac{\overleftarrow{\partial}}{\partial\hspace{0.03cm}\xi^{\phantom{\prime}}_{l}}
\cdot
\frac{\overrightarrow{\partial}}{\partial\hspace{0.03cm}\bar{\xi}^{\,\prime}_{l}}
\,-\,
\frac{\overrightarrow{\partial}}{\partial\hspace{0.03cm}\bar{\xi}^{\,\prime}_{l}}
\cdot
\frac{\overleftarrow{\partial}}{\partial\hspace{0.03cm}\xi^{\phantom{\prime}}_{l}}\,.
\]
However, the last term in the commutator involves difficulty. For the usual Grassmann variables (i.e. for $p = 1$) we consider that
\begin{equation}
\frac{\overrightarrow{\partial}}{\partial\hspace{0.03cm}\bar{\xi}^{\,\prime}_{l}}
\cdot
\frac{\overleftarrow{\partial}}{\partial\hspace{0.03cm}\xi^{\phantom{\prime}}_{l}}
=
-\frac{\overleftarrow{\partial}}{\partial\hspace{0.03cm}\xi^{\phantom{\prime}}_{l}}
\cdot
\frac{\overrightarrow{\partial}}{\partial\hspace{0.03cm}\bar{\xi}^{\,\prime}_{l}}
\label{eq:4r}
\end{equation}
and thus the definition (\ref{eq:4e}) turns to the expression
\[
* = \exp\biggl(\sum\limits^{2}_{l\hspace{0.03cm}=\hspace{0.03cm}1}\,
\frac{\overleftarrow{\partial}}{\partial\hspace{0.03cm}\xi^{\phantom{\prime}}_{l}}
\cdot
\frac{\overrightarrow{\partial}}{\partial\hspace{0.03cm}\bar{\xi}^{\,\prime}_{l}}
\biggr),
\]
as it was written in \cite{daoud_2003}. However, for the higher para-Grassmann algebras $(p \geq 2$) the procedure (\ref{eq:4r}) is no longer applicable and thus the fundamental problem in the interpretation of the definition (\ref{eq:4e}) arises. For this reason further based on the initial expression (\ref{eq:4w}) we take the following definition of the para-Grassmann star product:
\begin{equation}
* =
\iint{\rm e}^{\textstyle-\frac{1}{2}\,[\hspace{0.03cm}\mu,\!\overleftarrow{\partial}\!
/\partial\hspace{0.03cm}\xi\hspace{0.03cm}]}
\,
{\rm e}^{\textstyle-\frac{1}{2}\,[\hspace{0.03cm}\bar{\mu},\mu\hspace{0.03cm}]}
\,
{\rm e}^{\textstyle\frac{1}{2}\,[\hspace{0.03cm}\bar{\mu},\!\overrightarrow{\partial}\!
/\partial\hspace{0.03cm}\bar{\xi}^{\,\prime}\hspace{0.03cm}]}
\hspace{0.03cm}(d\mu)_{2}\hspace{0.03cm}(d\bar{\mu})_{2}.
\label{eq:4t}
\end{equation}
\indent Let us consider the star product between two para-Grassmann variables $\xi_{s}$ and $\bar{\xi}_{k}^{\prime}$:
\begin{equation}
\xi^{\phantom{\prime}}_{s}\! *\hspace{0.04cm} \bar{\xi}^{\,\prime}_{k} = \xi^{\phantom{\prime}}_{s}\hspace{0.03cm}\bar{\xi}^{\,\prime}_{k}
-
\frac{1}{4}\iint\!\xi^{\phantom{\prime}}_{s}\hspace{0.01cm}
\biggl[\hspace{0.03cm}\mu,\frac{\overleftarrow{\partial}}
{\partial\hspace{0.03cm}\xi}\hspace{0.03cm}\biggr]
\hspace{0.03cm}
{\rm e}^{\textstyle-\frac{1}{2}\,[\hspace{0.03cm}\bar{\mu},\mu\hspace{0.03cm}]}
\biggl[\hspace{0.03cm}\bar{\mu},\frac{\overrightarrow{\partial}}
{\partial\hspace{0.03cm}\bar{\xi}^{\,\prime}}\hspace{0.03cm}\biggr] \bar{\xi}^{\,\prime}_{k}
\,(d\mu)_{2}\hspace{0.03cm}(d\bar{\mu})_{2}\hspace{0.03cm}.
\label{eq:4y}
\end{equation}
By using the differentiation formulae \eqref{ap:B2} and \eqref{ap:B3} from the Appendix B it is not difficult to see that
\begin{equation}
\xi^{\phantom{\prime}}_{s}\hspace{0.01cm}
\biggl[\hspace{0.03cm}\mu,\frac{\overleftarrow{\partial}}
{\partial\hspace{0.03cm}\xi}\hspace{0.03cm}\biggr]
=
-2\hspace{0.03cm}\mu_{s},
\quad
\biggl[\hspace{0.03cm}\bar{\mu},\frac{\overrightarrow{\partial}}
{\partial\hspace{0.03cm}\bar{\xi}^{\,\prime}}\hspace{0.03cm}\biggr] \bar{\xi}^{\,\prime}_{k}
=
2\hspace{0.03cm}\bar{\mu}_{k},
\label{eq:4u}
\end{equation}
and as a consequence, the integral in (\ref{eq:4y}) takes the form:
\[
\iint\!\mu_{s}\hspace{0.03cm}\bar{\mu}_{k}
\,
{\rm e}^{\textstyle-\frac{1}{2}\,[\hspace{0.03cm}\bar{\mu},\mu\hspace{0.03cm}]}
\,(d\mu)_{2}\hspace{0.03cm}(d\bar{\mu})_{2}
\equiv
\delta_{sk}\!\iint\!\mu_{1}\hspace{0.03cm}\bar{\mu}_{1}\,
{\rm e}^{\textstyle-\frac{1}{2}\,[\hspace{0.03cm}\bar{\mu},\mu\hspace{0.03cm}]}
\,(d\mu)_{2}\hspace{0.03cm}(d\bar{\mu})_{2}.
\]
Here, on the right-hand side we have taken into account that a nontrivial contribution follows only from the integrand, which is quadratic in para-Grassmann variables. In the expansion of the exponential in (\ref{eq:3o}) we need only one term, namely
\begin{equation}
-\,\frac{\!1}{2^{\hspace{0.02cm}4}}\,
[\hspace{0.02cm}\bar{\mu}_{1}\hspace{0.02cm},
\mu_{1}\hspace{0.02cm}\hspace{0.02cm}]\hspace{0.03cm}
[\hspace{0.02cm}\bar{\mu}_{2}\hspace{0.02cm},
\mu_{2}\hspace{0.02cm}\hspace{0.02cm}]^{\hspace{0.04cm}2}_{\,\bf ,}
\label{eq:4i}
\end{equation}
then with the help of the integration formulae \eqref{ap:A3}, \eqref{ap:A2} and the definition of $\delta$-function \eqref{ap:A5} we derive
\[
\iint\!\mu_{1}\hspace{0.03cm}\bar{\mu}_{1}
\,
{\rm e}^{\textstyle-\frac{1}{2}\,[\hspace{0.03cm}\bar{\mu},\mu\hspace{0.03cm}]}
\,(d\mu)_{2}\hspace{0.03cm}(d\bar{\mu})_{2}
=
-\,\frac{\!1}{2^{\hspace{0.02cm}4}}\!\iint\!\mu_{1}\hspace{0.03cm}\bar{\mu}_{1}
\,
[\hspace{0.02cm}\bar{\mu}_{1}\hspace{0.02cm},
\mu_{1}\hspace{0.02cm}\hspace{0.02cm}]\hspace{0.03cm}
[\hspace{0.02cm}\bar{\mu}_{2}\hspace{0.02cm},
\mu_{2}\hspace{0.02cm}\hspace{0.02cm}]^{\hspace{0.04cm}2}
\,d^{\hspace{0.03cm}2}\mu_{2}\hspace{0.03cm}
d^{\hspace{0.03cm}2}\mu_{1}\hspace{0.03cm}(d\bar{\mu})_{2}
=
\]
\[
=
-\,\frac{1}{2}\iint\!\delta(\bar{\mu}_{2})\hspace{0.03cm}\bar{\mu}_{1}
\,
[\hspace{0.02cm}\bar{\mu}_{1}\hspace{0.02cm},
\mu_{1}\hspace{0.02cm}\hspace{0.02cm}]\hspace{0.03cm}\mu_{1}
\,d^{\hspace{0.03cm}2}\mu_{1}\hspace{0.03cm}(d\bar{\mu})_{2}
\,+\,
\frac{1}{2}\iint\!\delta(\bar{\mu}_{2})\hspace{0.03cm}
[\hspace{0.02cm}\bar{\mu}_{1}\hspace{0.02cm},
\mu_{1}\hspace{0.02cm}\hspace{0.02cm}]^{2}
\,d^{\hspace{0.03cm}2}\mu_{1}\hspace{0.03cm}(d\bar{\mu})_{2}
=
\]
\[
=
-\,2\!\iint\!\delta(\bar{\mu}_{2})\hspace{0.03cm}\delta(\bar{\mu}_{1})
\hspace{0.03cm}(d\bar{\mu})_{2}
\,+\,
4\!\iint\!\delta(\bar{\mu}_{2})\hspace{0.03cm}\delta(\bar{\mu}_{1})
\hspace{0.03cm}(d\bar{\mu})_{2} = 2.
\]
Thus, we get
\begin{equation}
\iint\!\mu_{s}\hspace{0.03cm}\bar{\mu}_{k}
\,
{\rm e}^{\textstyle-\frac{1}{2}\,[\hspace{0.03cm}\bar{\mu},\mu\hspace{0.03cm}]}
\,(d\mu)_{2}\hspace{0.03cm}(d\bar{\mu})_{2}
= 2\hspace{0.03cm}\delta_{sk}
\label{eq:4o}
\end{equation}
and from (\ref{eq:4y}) finally follows
\[
\xi^{\phantom{\prime}}_{s}\! *\hspace{0.02cm} \bar{\xi}^{\,\prime}_{k} = \xi^{\phantom{\prime}}_{s}\hspace{0.03cm}\bar{\xi}^{\,\prime}_{k}
+ 2\hspace{0.03cm}\delta^{\phantom{\prime}}_{sk}.
\]
In addition, we have
\[
\bar{\xi}^{\,\prime}_{k}\! *\hspace{0.02cm}\xi^{\phantom{\prime}}_{s} =
\bar{\xi}^{\,\prime}_{k}\hspace{0.04cm}\xi^{\phantom{\prime}}_{s}.
\]
\indent One can state a general question of correspondence between the algebra of the creation and annihilation operators $a_{k}^{\pm}$ subject to parafermi statistics of order 2, Eqs.\,(I.3.5)\,--\,(I.3.7) and the algebra of para-Grassmann variables of the same order, in which a product of the elements is defined by the star product (\ref{eq:4t}). Let us present the operator algebra (I.3.5) \,--\,(I.3.7) in a more compact form
\begin{equation}
\check{a}_{i}\hspace{0.03cm}\check{a}_{j}\hspace{0.03cm}\check{a}_{k} + \check{a}_{k}\hspace{0.03cm}\check{a}_{j}\hspace{0.03cm}\check{a}_{i}
=
2\hspace{0.03cm}\check{\delta}_{ij}\hspace{0.04cm}\check{a}_{k} + 2\hspace{0.03cm}\check{\delta}_{kj}\hspace{0.04cm}\check{a}_{i}.
\label{eq:4p}
\end{equation}
Here, the operator $\check{a}_i$ denotes $a_{i}^{+}$ or $a_{i}^{-}$ and $\check{\delta}_{ij} = \delta_{ij}$ when $\check{a}_{i} = a_{i}^{-}(a_{i}^{+})$ and $\check{a}_{j} = a_{j}^{+}(a_{j}^{-})$; otherwise $\check{\delta}_{ij} = 0$. The algebra for the para-Grassmann variables similar in structure to (\ref{eq:4p}) should have the following form:
\begin{equation}
\check{\xi}_{i}\! *\hspace{0.04cm}\check{\xi}_{j}\! *\hspace{0.04cm}\check{\xi}_{k} + \check{\xi}_{k}\! *\hspace{0.04cm}\check{\xi}_{j}\! *\hspace{0.04cm}\check{\xi}_{i}
=
2\hspace{0.03cm}\check{\delta}_{ij}\hspace{0.04cm}\check{\xi}_{k} + 2\hspace{0.03cm}\check{\delta}_{kj}\hspace{0.04cm}\check{\xi}_{i},
\label{eq:4a}
\end{equation}
where the para-Grassmann numbers $\check{\xi}_{i}$ denote $\xi_{i}$ or $\bar{\xi}_{i}^{\prime}$. Here, on the left-hand side we have a triple star product and consequently it is of importance to ascertain that the star product (\ref{eq:4t}) is associative. Let us write out ones again the initial bilinear relations
\begin{equation}
\begin{array}{llll}
&\xi_{i}\! *\hspace{0.04cm}\xi_{j} = \xi_{i}\hspace{0.04cm}\xi_{j},
\qquad
&\xi^{\phantom{\prime}}_{i}\! *\hspace{0.04cm} \bar{\xi}^{\,\prime}_{j} = \xi^{\phantom{\prime}}_{i}\hspace{0.03cm}\bar{\xi}^{\,\prime}_{j} + 2\hspace{0.03cm}\delta_{ij},
\\[1.5ex]
&\bar{\xi}^{\,\prime}_{i}\! *\hspace{0.04cm} \bar{\xi}^{\,\prime}_{j} = \bar{\xi}^{\,\prime}_{i}\hspace{0.03cm}\bar{\xi}^{\,\prime}_{j},
\qquad
&\bar{\xi}^{\,\prime}_{i}\! *\hspace{0.04cm} \xi_{j} = \bar{\xi}^{\,\prime}_{i}\hspace{0.03cm}\xi_{j}. \end{array}
\label{eq:4s}
\end{equation}
In addition, we need the usual trilinear relations for the para-Grassmann numbers of order $p = 2$:
\begin{equation}
\check{\xi}_{i}\hspace{0.04cm}\check{\xi}_{j}\hspace{0.04cm}\check{\xi}_{k} + \check{\xi}_{k}\hspace{0.04cm}\check{\xi}_{j}\hspace{0.04cm}\check{\xi}_{i}
= 0.
\label{eq:4d}
\end{equation}
\indent From (\ref{eq:4s}) and (\ref{eq:4d}) trivially follows a validity of the relations:
\[
\begin{split}
&\xi_{i}\! *\hspace{0.04cm}\xi_{j}\! *\hspace{0.04cm}\xi_{k} + \xi_{k}\! *\hspace{0.04cm}\xi_{j}\! *\hspace{0.04cm}\xi_{i} = 0, \\[1ex]
&\bar{\xi}^{\,\prime}_{i}\! *\hspace{0.04cm}\bar{\xi}^{\,\prime}_{j}\! *\hspace{0.04cm}\bar{\xi}^{\,\prime}_{k} + \bar{\xi}^{\,\prime}_{k}\! *\hspace{0.04cm}\bar{\xi}^{\,\prime}_{j}\! *\hspace{0.04cm}\bar{\xi}^{\,\prime}_{i} = 0.
\end{split}
\]
Here, the associativity of the triple star product is also evident. These formulae are a classical analogue of the operator relation
\[
a^{\pm}_{i}\hspace{0.03cm}a^{\pm}_{j}\hspace{0.03cm}a^{\pm}_{k} + a^{\pm}_{k}\hspace{0.03cm}a^{\pm}_{j}\hspace{0.03cm}a^{\pm}_{i}
= 0.
\]
\indent Now we consider a more complicated expression
\begin{equation}
\bar{\xi}^{\,\prime}_{i}\! *\hspace{0.04cm}\xi^{\phantom{\prime}}_{j}\! *\hspace{0.04cm}\xi^{\phantom{\prime}}_{k}
+
\xi^{\phantom{\prime}}_{k}\! *\hspace{0.04cm}\xi^{\phantom{\prime}}_{j}\! *\hspace{0.04cm}\bar{\xi}^{\,\prime}_{i}.
\label{eq:4f}
\end{equation}
Let us analyse the first term. We put parentheses on the first pair of multipliers, then by virtue of (\ref{eq:4s}) and the definition (\ref{eq:4t}) we have
\[
(\bar{\xi}^{\,\prime}_{i}\! *\hspace{0.04cm}\xi^{\phantom{\prime}}_{j})\! *\hspace{0.04cm}\xi^{\phantom{\prime}}_{k}
=
(\bar{\xi}^{\,\prime}_{i}\hspace{0.04cm}\xi^{\phantom{\prime}}_{j})\! *\hspace{0.04cm}\xi^{\phantom{\prime}}_{k}
=
\bar{\xi}^{\,\prime}_{i}\hspace{0.04cm}\xi^{\phantom{\prime}}_{j}
\hspace{0.04cm}\xi^{\phantom{\prime}}_{k}.
\]
Further, we put parentheses on the second pair of multipliers
\[
\bar{\xi}^{\,\prime}_{i}\! *\hspace{0.04cm}(\xi^{\phantom{\prime}}_{j}\! *\hspace{0.04cm}\xi^{\phantom{\prime}}_{k})
=
\bar{\xi}^{\,\prime}_{i}\! *\hspace{0.04cm}(\xi^{\phantom{\prime}}_{j}\hspace{0.04cm}\xi^{\phantom{\prime}}_{k})
=
\bar{\xi}^{\,\prime}_{i}\hspace{0.04cm}\xi^{\phantom{\prime}}_{j}
\hspace{0.04cm}\xi^{\phantom{\prime}}_{k}.
\]
We obtain the same expression. Let us analyse the second term in (\ref{eq:4f}). The first arrangement of parentheses gives
\begin{equation}
(\xi^{\phantom{\prime}}_{k}\! *\hspace{0.04cm}\xi^{\phantom{\prime}}_{j})\! *\hspace{0.04cm}\bar{\xi}^{\,\prime}_{i}
=
(\xi^{\phantom{\prime}}_{k}\hspace{0.04cm}\xi^{\phantom{\prime}}_{j})\! *\hspace{0.04cm}\bar{\xi}^{\,\prime}_{i}
=
\xi^{\phantom{\prime}}_{k}\hspace{0.04cm}\xi^{\phantom{\prime}}_{j}
\hspace{0.04cm}\bar{\xi}^{\,\prime}_{i}
-
\frac{1}{4}\!\iint(\xi^{\phantom{\prime}}_{k}
\hspace{0.04cm}\xi^{\phantom{\prime}}_{j})\hspace{0.01cm}
\biggl[\hspace{0.03cm}\mu,\frac{\overleftarrow{\partial}}
{\partial\hspace{0.03cm}\xi}\hspace{0.03cm}\biggr]
\hspace{0.03cm}
{\rm e}^{\textstyle-\frac{1}{2}\,[\hspace{0.03cm}\bar{\mu},\mu\hspace{0.03cm}]}
\biggl[\hspace{0.03cm}\bar{\mu},\frac{\overrightarrow{\partial}}
{\partial\hspace{0.03cm}\bar{\xi}^{\,\prime}}\hspace{0.03cm}\biggr] \bar{\xi}^{\,\prime}_{i}
\,(d\mu)_{2}\hspace{0.03cm}(d\bar{\mu})_{2}\hspace{0.03cm}.
\label{eq:4g}
\end{equation}
Let us calculate the derivative of the product $(\xi_{k}\hspace{0.02cm}\xi_{j})$. Using the definition
of para-Grassmann numbers, Eq.\,(\ref{eq:4d}), and the differentiation formula \eqref{ap:B4}, we find
\[
(\xi_{k}\hspace{0.04cm}\xi_{j})
\hspace{0.01cm}
\biggl[\hspace{0.03cm}\mu,\frac{\overleftarrow{\partial}}
{\partial\hspace{0.03cm}\xi}\hspace{0.03cm}\biggr]
=
\sum\limits_{l}\,\biggl\{(\xi_{k}\hspace{0.04cm}\xi_{j}\hspace{0.04cm}\mu_{l})\,
\frac{\overleftarrow{\partial}}{\partial\hspace{0.03cm}\xi_{l}}
-
(\xi_{k}\hspace{0.04cm}\xi_{j})\,\frac{\overleftarrow{\partial}}{\partial\hspace{0.03cm}\xi_{l}}
\,\mu_{l}\biggr\}
=
\]
\[
=
\sum\limits_{l}\,\biggl\{-\mu_{l}\biggl(\xi_{j}\hspace{0.04cm}\xi_{k}\,
\frac{\overleftarrow{\partial}}{\partial\hspace{0.03cm}\xi_{l}}\biggr)
-
\biggl(\xi_{k}\hspace{0.04cm}\xi_{j}\,\frac{\overleftarrow{\partial}}
{\partial\hspace{0.03cm}\xi_{l}}\biggr)\mu_{l}\biggr\}
=
-2\hspace{0.04cm}(\mu_{k}\hspace{0.04cm}\xi_{j} + \xi_{k}\hspace{0.03cm}\mu_{j}).
\]
We use the second formula in (\ref{eq:4u}) for the second derivative in (\ref{eq:4g}). As a result, the integral (\ref{eq:4g}) takes the form
\begin{equation}
\iint\!\mu_{k}\hspace{0.04cm}\xi_{j}\;
{\rm e}^{\textstyle-\frac{1}{2}\,[\hspace{0.03cm}\bar{\mu},\mu\hspace{0.03cm}]}
\bar{\mu}_{i}\,(d\mu)_{2}\hspace{0.03cm}(d\bar{\mu})_{2}
\,+\,
\xi_{k}\!\!\iint\!\mu_{j}\hspace{0.04cm}\bar{\mu}_{i}\;
{\rm e}^{\textstyle-\frac{1}{2}\,[\hspace{0.03cm}\bar{\mu},\mu\hspace{0.03cm}]}
\,(d\mu)_{2}\hspace{0.03cm}(d\bar{\mu})_{2}\,.
\hspace{0.4cm}
\label{eq:4h}
\end{equation}
The second integral here, by virtue of (\ref{eq:4o}) equals $2\hspace{0.02cm}\xi_{k}\hspace{0.02cm}\delta_{ij}$. We present the first integral as
\begin{equation}
\xi_{j}\!\!\iint\!\mu_{k}\hspace{0.04cm}\bar{\mu}_{i}\,
{\rm e}^{\textstyle-\frac{1}{2}\,[\hspace{0.03cm}\bar{\mu},\mu\hspace{0.03cm}]}
\,(d\mu)_{2}\hspace{0.03cm}(d\bar{\mu})_{2}
\,+\,
\iint[\hspace{0.03cm}\mu_{k},\xi_{j}\hspace{0.03cm}]
\,\bar{\mu}_{i}\,
{\rm e}^{\textstyle-\frac{1}{2}\,[\hspace{0.03cm}\bar{\mu},\mu\hspace{0.03cm}]}
\,(d\mu)_{2}\hspace{0.03cm}(d\bar{\mu})_{2}\,.
\label{eq:4j}
\end{equation}
The first integral in (\ref{eq:4j}) by the same formula (\ref{eq:4o}) equals $2\hspace{0.02cm}\xi_{j}\hspace{0.03cm}\delta_{ki}$, and the second one is calculated as follows:
\[
\iint[\hspace{0.03cm}\mu_{k},\xi_{j}\hspace{0.03cm}]
\,\bar{\mu}_{i}\,
{\rm e}^{\textstyle-\frac{1}{2}\,[\hspace{0.03cm}\bar{\mu},\mu\hspace{0.03cm}]}
\,(d\mu)_{2}\hspace{0.03cm}(d\bar{\mu})_{2}
=
\delta_{ki}\!\iint[\hspace{0.03cm}\mu_{1},\xi_{j}\hspace{0.03cm}]
\,\bar{\mu}_{1}\,
{\rm e}^{\textstyle-\frac{1}{2}\,[\hspace{0.03cm}\bar{\mu},\mu\hspace{0.03cm}]}
\,(d\mu)_{2}\hspace{0.03cm}(d\bar{\mu})_{2}
=
\]
\[
\begin{split}
&=
-\frac{\!1}{2^{4}}\,\delta_{ki}\!\iint[\hspace{0.03cm}\mu_{1},\xi_{j}\hspace{0.03cm}]
\,\bar{\mu}_{1}\,
[\hspace{0.02cm}\bar{\mu}_{1}\hspace{0.02cm},
\mu_{1}\hspace{0.02cm}\hspace{0.02cm}]\hspace{0.03cm}
[\hspace{0.02cm}\bar{\mu}_{2}\hspace{0.02cm},
\mu_{2}\hspace{0.02cm}\hspace{0.02cm}]^{\hspace{0.04cm}2}
\hspace{0.03cm}(d\bar{\mu})_{2}
= \\[1ex]
&=
-\frac{1}{2}\,\delta_{ki}\!
\iint\!\delta(\bar{\mu}_{2})[\hspace{0.03cm}\mu_{1},\xi_{j}\hspace{0.03cm}]
\,\bar{\mu}_{1}\,
[\hspace{0.02cm}\bar{\mu}_{1}\hspace{0.02cm},
\mu_{1}\hspace{0.02cm}\hspace{0.02cm}]\hspace{0.03cm}
\,d^{\hspace{0.03cm}2}\mu_{1}\hspace{0.03cm}(d\bar{\mu})_{2}
= \\[1ex]
&=
-\hspace{0.01cm}\delta_{ki}\!\iint\!\delta(\bar{\mu}_{2})\,\bar{\mu}_{1}\,
\{\hspace{0.02cm}\bar{\mu}_{1}\hspace{0.02cm},\xi_{j}\hspace{0.02cm}\hspace{0.02cm}\}
\,d^{\hspace{0.03cm}2}\bar{\mu}_{1}\hspace{0.04cm}
d^{\hspace{0.03cm}2}\bar{\mu}_{2}
= -2\hspace{0.04cm}\delta_{ki}\hspace{0.04cm}\xi_{j}.
\end{split}
\]
Here, we have used the integration formulae in Appendix A, and in the expansion of exponential (\ref{eq:3o}) we have kept only the term (\ref{eq:4i}). Therefore, for (\ref{eq:4j}) we have
\[
2\hspace{0.04cm}\xi_{j}\hspace{0.04cm}\delta_{ki}
-
2\hspace{0.04cm}\delta_{ki}\hspace{0.04cm}\xi_{j} = 0.
\]
Curiously, the first integral in (\ref{eq:4h}) vanishes only for $p = 2$, as it takes place for the derivatives \eqref{ap:B2}.
Considering the aforementioned, the expression (\ref{eq:4g}) takes the from
\begin{equation}
(\xi^{\phantom{\prime}}_{k}\! *\hspace{0.02cm}\xi^{\phantom{\prime}}_{j})\! *\hspace{0.02cm}\bar{\xi}^{\,\prime}_{i}
=
\xi^{\phantom{\prime}}_{k}\hspace{0.04cm}\xi^{\phantom{\prime}}_{j}
\hspace{0.04cm}\bar{\xi}^{\,\prime}_{i} + 2\hspace{0.04cm}\delta^{\phantom{\prime}}_{ji}\hspace{0.04cm}\xi^{\phantom{\prime}}_{k}.
\label{eq:4k}
\end{equation}
\indent We put parenthesis in a different way. Based on (\ref{eq:4s}) we have
\begin{equation}
\xi^{\phantom{\prime}}_{k}\! *\hspace{0.02cm}(\xi^{\phantom{\prime}}_{j}\! *\hspace{0.04cm}\bar{\xi}^{\,\prime}_{i})
=
\xi^{\phantom{\prime}}_{k}\! *\hspace{0.02cm}(\xi^{\phantom{\prime}}_{j}\hspace{0.04cm}\bar{\xi}^{\,\prime}_{i})
+
2\hspace{0.04cm}\delta^{\phantom{\prime}}_{ji}\hspace{0.04cm}\xi^{\phantom{\prime}}_{k}.
\label{eq:4l}
\end{equation}
We write the first term on the right-hand side:
\begin{equation}
\xi^{\phantom{\prime}}_{k}\! *\hspace{0.02cm}(\xi^{\phantom{\prime}}_{j}\hspace{0.04cm}\bar{\xi}^{\,\prime}_{i})
=
\xi^{\phantom{\prime}}_{k}\hspace{0.04cm}\xi^{\phantom{\prime}}_{j}
\hspace{0.04cm}\bar{\xi}^{\,\prime}_{i}
\,-\,
\frac{1}{4}\!\iint\!\xi^{\phantom{\prime}}_{k}
\hspace{0.01cm}
\biggl[\hspace{0.03cm}\mu,\frac{\overleftarrow{\partial}}
{\partial\hspace{0.03cm}\xi}\hspace{0.03cm}\biggr]
\,
{\rm e}^{\textstyle-\frac{1}{2}\,[\hspace{0.03cm}\bar{\mu},\mu\hspace{0.03cm}]}
\biggl[\hspace{0.03cm}\bar{\mu},\frac{\overrightarrow{\partial}}
{\partial\hspace{0.03cm}\bar{\xi}^{\,\prime}}\hspace{0.03cm}\biggr]
(\xi^{\phantom{\prime}}_{j}\hspace{0.03cm}\bar{\xi}^{\,\prime}_{i})
\,(d\mu)_{2}\hspace{0.03cm}(d\bar{\mu})_{2}\hspace{0.03cm}.
\label{eq:4z}
\end{equation}
Here we are faced with the need to calculate the following derivative:
\[
\biggl[\hspace{0.03cm}\bar{\mu},\frac{\overrightarrow{\partial}}
{\partial\hspace{0.03cm}\bar{\xi}^{\,\prime}}\hspace{0.03cm}\biggr]
(\xi^{\phantom{\prime}}_{j}\hspace{0.03cm}\bar{\xi}^{\,\prime}_{i})
=
\sum\limits_{l}\hspace{0.03cm}\biggl\{
\bar{\mu}^{\phantom{\prime}}_{l}\,\frac{\overrightarrow{\partial}}
{\partial\hspace{0.03cm}\bar{\xi}^{\,\prime}_{l}}\,
(\xi^{\phantom{\prime}}_{j}\hspace{0.03cm}\bar{\xi}^{\,\prime}_{i})
-
\frac{\overrightarrow{\partial}}
{\partial\hspace{0.03cm}\bar{\xi}^{\,\prime}_{l}}\,
(\bar{\mu}^{\phantom{\prime}}_{l}\hspace{0.03cm}\xi^{\phantom{\prime}}_{j}
\hspace{0.03cm}\bar{\xi}^{\,\prime}_{i})\biggr\}
=
\sum\limits_{l}\biggl(
\frac{\overrightarrow{\partial}}
{\partial\hspace{0.03cm}\bar{\xi}^{\,\prime}_{l}}\,\bar{\xi}^{\,\prime}_{i}\biggr)
\hspace{0.03cm}\xi^{\phantom{\prime}}_{j}\hspace{0.03cm}\bar{\mu}^{\phantom{\prime}}_{l}
=
2\hspace{0.03cm}\xi^{\phantom{\prime}}_{j}\hspace{0.03cm}\bar{\mu}^{\phantom{\prime}}_{i}.
\]
Taking into account the preceding expression and the first formula in (\ref{eq:4u}), instead of (\ref{eq:4z}) we find
\[
\xi^{\phantom{\prime}}_{k}\! *\hspace{0.02cm}(\xi^{\phantom{\prime}}_{j}\hspace{0.04cm}\bar{\xi}^{\,\prime}_{i})
=
\xi^{\phantom{\prime}}_{k}\hspace{0.04cm}\xi^{\phantom{\prime}}_{j}
\hspace{0.04cm}\bar{\xi}^{\,\prime}_{i}
\,+
\iint\!\mu_{k}\,{\rm e}^{\textstyle-\frac{1}{2}\,[\hspace{0.03cm}\bar{\mu},\mu\hspace{0.03cm}]}\,
\xi_{j}\hspace{0.03cm}\bar{\mu}_{i}
\hspace{0.03cm}(d\mu)_{2}\hspace{0.03cm}(d\bar{\mu})_{2}\hspace{0.03cm}.
\]
Here, the integral is equal to the first integral in (\ref{eq:4h}) and therefore, by virtue of the
previous considerations, it vanishes. Thus, (\ref{eq:4l}) coincides with (\ref{eq:4k}) and instead
of (\ref{eq:4f}) we can write the following expression:
\[
\bar{\xi}^{\,\prime}_{i}\! *\hspace{0.04cm}\xi^{\phantom{\prime}}_{j}\! *\hspace{0.04cm}\xi^{\phantom{\prime}}_{k}
+
\xi^{\phantom{\prime}}_{k}\! *\hspace{0.04cm}\xi^{\phantom{\prime}}_{j}\! *\hspace{0.04cm}\bar{\xi}^{\,\prime}_{i}
=
2\hspace{0.04cm}\delta^{\phantom{\prime}}_{ij}\hspace{0.04cm}\xi^{\phantom{\prime}}_{k}.
\]
It is a classical analogue of the operator relation:
\[
a^{+}_{i}\hspace{0.03cm}a^{-}_{j}\hspace{0.03cm}a^{-}_{k}
+
a^{-}_{k}\hspace{0.03cm}a^{-}_{j}\hspace{0.03cm}a^{+}_{i}
= 2\hspace{0.04cm}\delta^{\phantom{-}}_{ij}\hspace{0.04cm}a^{-}_{k}.
\]
\indent An analysis of the remaining relations is performed by a similar way. Thus, for example, the operator relation
\[
a^{+}_{i}\hspace{0.03cm}a^{-}_{j}\hspace{0.03cm}a^{+}_{k}
+
a^{+}_{k}\hspace{0.03cm}a^{-}_{j}\hspace{0.03cm}a^{+}_{i}
= 2\hspace{0.04cm}\delta^{\phantom{+}}_{ij}\hspace{0.04cm}a^{+}_{k}
+
2\hspace{0.04cm}\delta^{\phantom{+}}_{kj}\hspace{0.04cm}a^{+}_{i}
\]
will correspond to the expression
\[
\bar{\xi}^{\,\prime}_{i}\! *\hspace{0.04cm}\xi^{\phantom{\prime}}_{j}\! *\hspace{0.04cm}\bar{\xi}^{\,\prime}_{k}
+
\bar{\xi}^{\,\prime}_{k}\! *\hspace{0.04cm}\xi^{\phantom{\prime}}_{j}\! *\hspace{0.04cm}\bar{\xi}^{\,\prime}_{i}
=
2\hspace{0.04cm}\delta^{\phantom{\prime}}_{ij}\hspace{0.04cm}\bar{\xi}^{\,\prime}_{k}
+
2\hspace{0.04cm}\delta^{\phantom{\prime}}_{kj}\hspace{0.04cm}\bar{\xi}^{\,\prime}_{i},
\]
etc. By doing so we observe complete coincidence between the para-Fermi operator algebra (\ref{eq:4p}) and the algebra for para-Grassmann variables (\ref{eq:4a}) with the star product.

%%%%%%%%%%%%%%%%%%%%%%%% section 5 %%%%%%%%%%%%%%%%%%%%%%%%%%%%

\section{Triple star product $\Omega*\Omega*\Omega$}
\setcounter{equation}{0}
\label{section_5}

In terms of the definition of the star product (\ref{eq:4t}) the integral convolution (\ref{eq:2y}) is written in the form
\begin{equation}
\widetilde{\Omega}\hspace{0.03cm}(\bar{\xi}^{\,\prime},\xi)
=
\Omega\hspace{0.03cm}(\bar{\xi}^{\,\prime},\xi)\! *\hspace{0.02cm}
\Omega\hspace{0.03cm}(\bar{\xi}^{\,\prime},\xi).
\label{eq:5q}
\end{equation}
From the other hand, in sections 7 and 12 of Part I by a direct calculation we prove a validity of the expansion
\begin{equation}
\Omega\hspace{0.03cm}(\bar{\xi}^{\,\prime},\xi)
=
\Omega\hspace{0.03cm}(\bar{\xi}^{\,\prime},\xi)\! *\hspace{0.02cm}
\widetilde{\Omega}\hspace{0.03cm}(\bar{\xi}^{\,\prime},\xi).
\label{eq:5w}
\end{equation}
Thus, if the relation (\ref{eq:5q}) was indeed valid, then by virtue of (\ref{eq:5w}) the following equality with triple star product would take place:
\begin{equation}
\Omega\hspace{0.03cm}(\bar{\xi}^{\,\prime},\xi)
=
\Omega\hspace{0.03cm}(\bar{\xi}^{\,\prime},\xi)\hspace{0.01cm}\! *\hspace{0.02cm}
\Omega\hspace{0.03cm}(\bar{\xi}^{\,\prime},\xi)\hspace{0.01cm}\! *\hspace{0.02cm}
\Omega\hspace{0.03cm}(\bar{\xi}^{\,\prime},\xi).
\label{eq:5e}
\end{equation}
This expression is a classical analog of the operator equality
\[
a^{3}_{0} = a^{\phantom{3}}_{0}.
\]
The consequence of this operator equality is a possibility to present the operator $\hat{A}$, Eq.\,(I.10.7), in the exponential form (see Eq.\,(I.2.14))
\begin{equation}
\hat{A} = \alpha\Bigl[\hspace{0.03cm}\hat{I} - \Bigl(\hspace{0.02cm}\frac{i\sqrt{3}}{2}\hspace{0.02cm}\Bigr)
\hspace{0.02cm} a_{0} - \frac{3}{2}\,(a_{0})^{2}\hspace{0.02cm}\Bigr]
\equiv
\alpha\hspace{0.03cm}{\rm e}^{\textstyle-i\hspace{0.02cm}\frac{2\pi}{3}\,a_{0}}.
\label{eq:5r}
\end{equation}
The matrix element of this operator as it was defined by the expression (I.9.23) in view of (\ref{eq:5q}) and (\ref{eq:5e}) also can be presented in the compact form:
\[
\langle\hspace{0.02cm}\bar{\xi}^{\,\prime}\hspace{0.02cm}|\,
\hat{A}|\,\xi\hspace{0.02cm}\rangle = {\cal A}\hspace{0.03cm}(\bar{\xi}^{\,\prime},\xi)\,
\langle\hspace{0.02cm}\bar{\xi}^{\,\prime}\hspace{0.02cm}|\,\xi\hspace{0.02cm}\rangle,
\]
where
\[
{\cal A}\hspace{0.03cm}(\bar{\xi}^{\,\prime},\xi)
=
\alpha\Bigl[\hspace{0.03cm}I - \Bigl(\hspace{0.02cm}\frac{i\sqrt{3}}{2}\hspace{0.02cm}\Bigr)
\hspace{0.02cm} \Omega - \frac{3}{2}\;\Omega\hspace{0.04cm}\! *\hspace{0.01cm}
\Omega\hspace{0.02cm}\Bigr]
\equiv
\alpha\hspace{0.03cm}
{\rm e}^{\textstyle-i\hspace{0.02cm}\frac{2\pi}{3}\,
\Omega\hspace{0.03cm}(\bar{\xi}^{\,\prime},\xi)}_{\hspace{0.04cm}*}.
\]
Here,
\[
{\rm e}^{\textstyle x}_{\hspace{0.04cm}*} = \sum\limits^{\infty}_{k\hspace{0.02cm}=\hspace{0.02cm}0}
\frac{\!1}{k!}\,(x)^{k}_{*}, \quad (x)^{k}_{*} \equiv \underbrace{(x*x*\ldots*x)}_{k\, {\rm times}}
\]
is the star exponential \cite{daoud_2003}. Thus, we only need to verify a validity of (\ref{eq:5q}). In our reasoning we will follow consideration of section 12 of the Part I.\\
\indent Let us introduce the notations
\begin{equation}
x \equiv \displaystyle\frac{1}{2}\;[\hspace{0.03cm}\bar{\xi}^{\,\prime}_{1}, \xi^{\phantom{\prime}}_{1}\hspace{0.02cm}],
\quad
y\equiv \displaystyle\frac{1}{2}\;[\hspace{0.03cm}\bar{\xi}^{\,\prime}_{2}, \xi^{\phantom{\prime}}_{2}\hspace{0.02cm}]
\label{eq:5t}
\end{equation}
and present the function $\Omega$ in the form
\[
\Omega\hspace{0.03cm}(\bar{\xi}^{\,\prime},\xi)
=
\Delta\Omega\hspace{0.03cm}(\bar{\xi}^{\,\prime},\xi) - (x + y -1),
\]
where
\[
\Delta\Omega\hspace{0.03cm}(\bar{\xi}^{\,\prime},\xi)
=
-\frac{\!1}{2^{\hspace{0.03cm}3}}\,[\hspace{0.03cm}\bar{\xi}^{\,\prime}_{1}, \bar{\xi}^{\,\prime}_{2}\hspace{0.03cm}]\hspace{0.03cm}
[\hspace{0.03cm}\xi^{\phantom{\prime}}_{1}, \xi^{\phantom{\prime}}_{2}\hspace{0.03cm}]
-
\frac{\!1}{2^{\hspace{0.03cm}3}}\,[\hspace{0.03cm}\bar{\xi}^{\,\prime}_{1}, \xi^{\phantom{\prime}}_{2}\hspace{0.03cm}]\hspace{0.03cm}
[\hspace{0.03cm}\bar{\xi}^{\,\prime}_{2}, \xi^{\phantom{\prime}}_{1}\hspace{0.03cm}]
+ \frac{1}{2}\,x\hspace{0.02cm}y.
\]
From here, in particular it follows
\begin{equation}
\bigl[\hspace{0.03cm}\Omega\hspace{0.03cm}(\bar{\xi}^{\,\prime},\xi)\bigr]^{2}
=
\bigl(\Delta\Omega\hspace{0.03cm}(\bar{\xi}^{\,\prime},\xi)\bigr)^{2}
- 2\hspace{0.03cm}(x + y -1)\Delta\Omega\hspace{0.03cm}(\bar{\xi}^{\,\prime},\xi)
+ (x + y -1)^{2}.
\label{eq:5y}
\end{equation}
Our purpose is to rewrite the expression obtained in sections 2 and 3 for the star product $\Omega\hspace{0.04cm}\! *\hspace{0.01cm}\Omega$, Eq.\,(\ref{eq:3i}), in terms of the variables $x$ and $y$ with subsequent comparison with (\ref{eq:2r}). To do so, we use the formulae obtained early (I.12.7)\,--\,(I.12.10). On the strength of these formulae for the term in (\ref{eq:5y}) linear in $\Delta\Omega$ we have
\[
(x + y -1)\Delta\Omega\hspace{0.03cm}(\bar{\xi}^{\,\prime},\xi)
=
\frac{\!1}{2^{\hspace{0.03cm}3}}\,\bigl(\hspace{0.02cm}[\hspace{0.03cm}\bar{\xi}^{\,\prime}_{1}, \bar{\xi}^{\,\prime}_{2}\hspace{0.03cm}]\hspace{0.03cm}
[\hspace{0.03cm}\xi^{\phantom{\prime}}_{1}, \xi^{\phantom{\prime}}_{2}\hspace{0.03cm}]
+
[\hspace{0.03cm}\bar{\xi}^{\,\prime}_{1}, \xi^{\phantom{\prime}}_{2}\hspace{0.03cm}]\hspace{0.03cm}
[\hspace{0.03cm}\bar{\xi}^{\,\prime}_{2}, \xi^{\phantom{\prime}}_{1}\hspace{0.03cm}]\hspace{0.02cm}\bigr)
- \frac{1}{2}\,x\hspace{0.02cm}y + (x + y)\hspace{0.02cm}x\hspace{0.01cm}y.
\]
For the term in (\ref{eq:5y}) quadratic in $\Delta\Omega$ the same formulae lead to the following expression:
\begin{equation}
\bigl(\Delta\Omega\hspace{0.03cm}(\bar{\xi}^{\,\prime},\xi)\bigr)^{2}
=
\frac{\!\!1}{2^{\hspace{0.03cm}6}}\,[\hspace{0.03cm}\bar{\xi}^{\,\prime}_{1}, \bar{\xi}^{\,\prime}_{2}\hspace{0.03cm}]^{\hspace{0.03cm}2}\hspace{0.03cm}
[\hspace{0.03cm}\xi^{\phantom{\prime}}_{1}, \xi^{\phantom{\prime}}_{2}\hspace{0.03cm}]^{\hspace{0.02cm}2}
+
\frac{\!1}{2^{\hspace{0.03cm}6}}\,[\hspace{0.03cm}\bar{\xi}^{\,\prime}_{1}, \xi^{\phantom{\prime}}_{2}\hspace{0.03cm}]^{\hspace{0.02cm}2}\hspace{0.03cm}
[\hspace{0.03cm}\bar{\xi}^{\,\prime}_{2}, \xi^{\phantom{\prime}}_{1}\hspace{0.03cm}]^{\hspace{0.03cm}2}
+
\frac{\!1}{2^{\hspace{0.02cm}2}}\,x^{\hspace{0.02cm}2}\hspace{0.02cm}y^{\hspace{0.02cm}2}
\,+
\label{eq:5u}
\end{equation}
\[
+
\,\frac{\!\!1}{2^{\hspace{0.03cm}5}}\,[\hspace{0.03cm}\bar{\xi}^{\,\prime}_{1}, \bar{\xi}^{\,\prime}_{2}\hspace{0.03cm}]\hspace{0.03cm}
[\hspace{0.03cm}\xi^{\phantom{\prime}}_{1}, \xi^{\phantom{\prime}}_{2}\hspace{0.03cm}]
\hspace{0.03cm}
[\hspace{0.03cm}\bar{\xi}^{\,\prime}_{1}, \xi^{\phantom{\prime}}_{2}\hspace{0.03cm}]\hspace{0.03cm}
[\hspace{0.03cm}\bar{\xi}^{\,\prime}_{2}, \xi^{\phantom{\prime}}_{1}\hspace{0.03cm}]
-
\frac{\!\!1}{2^{\hspace{0.03cm}3}}\,[\hspace{0.03cm}\bar{\xi}^{\,\prime}_{1}, \bar{\xi}^{\,\prime}_{2}\hspace{0.03cm}]\hspace{0.03cm}
[\hspace{0.03cm}\xi^{\phantom{\prime}}_{1}, \xi^{\phantom{\prime}}_{2}\hspace{0.03cm}]
\hspace{0.03cm}x\hspace{0.02cm}y
-
\frac{\!1}{2^{\hspace{0.03cm}3}}\,[\hspace{0.03cm}\bar{\xi}^{\,\prime}_{1}, \xi^{\phantom{\prime}}_{2}\hspace{0.03cm}]\hspace{0.03cm}
[\hspace{0.03cm}\bar{\xi}^{\,\prime}_{2}, \xi^{\phantom{\prime}}_{1}\hspace{0.03cm}]
\hspace{0.03cm}x\hspace{0.02cm}y.
\]
In view of (I.12.10) the last two terms here equal $x^{2} y^{2}/4$. Let us consider the first term. By algebra of para-Grassmann numbers of order $p = 2$, Eq.\,(\ref{eq:4d}), this term can be presented as follows:
\[
\begin{split}
&[\hspace{0.03cm}\bar{\xi}^{\,\prime}_{1}, \bar{\xi}^{\,\prime}_{2}\hspace{0.03cm}]^{\hspace{0.03cm}2}\hspace{0.03cm}
[\hspace{0.03cm}\xi^{\phantom{\prime}}_{1}, \xi^{\phantom{\prime}}_{2}\hspace{0.03cm}]^{\hspace{0.02cm}2}
=
(\hspace{0.03cm}\bar{\xi}^{\,\prime}_{1}\hspace{0.03cm}\bar{\xi}^{\,\prime}_{2}\hspace{0.03cm}
\bar{\xi}^{\,\prime}_{2}\hspace{0.03cm}\bar{\xi}^{\,\prime}_{1}
+
\bar{\xi}^{\,\prime}_{2}\hspace{0.03cm}\bar{\xi}^{\,\prime}_{1}\hspace{0.03cm}
\bar{\xi}^{\,\prime}_{1}\hspace{0.03cm}\bar{\xi}^{\,\prime}_{2}\hspace{0.03cm})
\hspace{0.03cm}
(\hspace{0.03cm}\xi^{\phantom{\prime}}_{1}\hspace{0.03cm}\xi^{\phantom{\prime}}_{2}
\hspace{0.03cm}\xi^{\phantom{\prime}}_{2}\hspace{0.03cm}\xi^{\phantom{\prime}}_{1}
+
\hspace{0.03cm}\xi^{\phantom{\prime}}_{2}\hspace{0.03cm}\xi^{\phantom{\prime}}_{1}
\hspace{0.03cm}\xi^{\phantom{\prime}}_{1}\hspace{0.03cm}\xi^{\phantom{\prime}}_{2})
= \\[1.2ex]
= 2^{\hspace{0.02cm}2}\hspace{0.02cm}(\bar{\xi}^{\,\prime}_{1}&)^{\hspace{0.02cm}2}
\hspace{0.02cm}(\bar{\xi}^{\,\prime}_{2})^{\hspace{0.02cm}2}
\hspace{0.02cm}(\xi^{\phantom{\prime}}_{1})^{\hspace{0.02cm}2}
\hspace{0.02cm}(\xi^{\phantom{\prime}}_{2})^{\hspace{0.02cm}2}
=
2^{\hspace{0.02cm}2}\hspace{0.02cm}(\bar{\xi}^{\,\prime}_{1})^{\hspace{0.02cm}2}
\hspace{0.02cm}(\xi^{\phantom{\prime}}_{1})^{\hspace{0.02cm}2}
\hspace{0.02cm}(\bar{\xi}^{\,\prime}_{2})^{\hspace{0.02cm}2}
\hspace{0.02cm}(\xi^{\phantom{\prime}}_{2})^{\hspace{0.02cm}2}
=
[\hspace{0.03cm}\bar{\xi}^{\,\prime}_{1}, \xi^{\phantom{\prime}}_{1}\hspace{0.02cm}]^{\hspace{0.03cm}2}\hspace{0.03cm}
[\hspace{0.03cm}\bar{\xi}^{\,\prime}_{2}, \xi^{\phantom{\prime}}_{2}\hspace{0.02cm}]^{\hspace{0.02cm}2}
=
2^{\hspace{0.02cm}4}\hspace{0.02cm}x^{\hspace{0.02cm}2}\hspace{0.02cm}
y^{\hspace{0.02cm}2}.
\end{split}
\]
Here, at the last step we have used Eqs.\,(I.12.7) and (I.12.8). Similar reasoning for the second term in (\ref{eq:5u}) results in the equality
\[
[\hspace{0.03cm}\bar{\xi}^{\,\prime}_{1}, \xi^{\phantom{\prime}}_{2}\hspace{0.03cm}]^{\hspace{0.02cm}2}\hspace{0.03cm}
[\hspace{0.03cm}\bar{\xi}^{\,\prime}_{2}, \xi^{\phantom{\prime}}_{1}\hspace{0.03cm}]^{\hspace{0.03cm}2}
= 2^{\hspace{0.02cm}4}\hspace{0.02cm}x^{\hspace{0.02cm}2}\hspace{0.01cm}
y^{\hspace{0.02cm}2}.
\]
We need only to consider the mixed contribution to (\ref{eq:5u}), which we present as a product of two multiplies
\[
\bigl(\hspace{0.03cm}[\hspace{0.03cm}\bar{\xi}^{\,\prime}_{1}, \bar{\xi}^{\,\prime}_{2}\hspace{0.03cm}]\hspace{0.03cm}
[\hspace{0.03cm}\bar{\xi}^{\,\prime}_{1}, \xi^{\phantom{\prime}}_{2}\hspace{0.03cm}]\hspace{0.03cm}
\hspace{0.03cm}\bigr)
\bigl(\hspace{0.03cm}
[\hspace{0.03cm}\xi^{\phantom{\prime}}_{1}, \xi^{\phantom{\prime}}_{2}\hspace{0.03cm}]
\hspace{0.03cm}
[\hspace{0.03cm}\bar{\xi}^{\,\prime}_{2}, \xi^{\phantom{\prime}}_{1}\hspace{0.03cm}]
\hspace{0.03cm}\bigr).
\]
For the first factor, by virtue of algebra (\ref{eq:4d}), we have
\[
[\hspace{0.03cm}\bar{\xi}^{\,\prime}_{1}, \bar{\xi}^{\,\prime}_{2}\hspace{0.03cm}]\hspace{0.03cm}
[\hspace{0.03cm}\bar{\xi}^{\,\prime}_{1}, \xi^{\phantom{\prime}}_{2}\hspace{0.03cm}]
=
- \bar{\xi}^{\,\prime}_{1}\hspace{0.03cm}\bar{\xi}^{\,\prime}_{2}\hspace{0.03cm}
\xi^{\phantom{\prime}}_{2}\hspace{0.03cm}\bar{\xi}^{\,\prime}_{1}
-
\bar{\xi}^{\,\prime}_{2}\hspace{0.03cm}\bar{\xi}^{\,\prime}_{1}\hspace{0.03cm}
\bar{\xi}^{\,\prime}_{1}\hspace{0.03cm}\xi^{\phantom{\prime}}_{2}
=
(\bar{\xi}^{\,\prime}_{1})^{\hspace{0.02cm}2}
\bigl(\xi^{\phantom{\prime}}_{2}\hspace{0.03cm}\bar{\xi}^{\,\prime}_{2}
+
\bar{\xi}^{\,\prime}_{2}\hspace{0.03cm}\xi^{\phantom{\prime}}_{2}\bigr)
\]
and for the second one, correspondingly, we get
\[
\hspace{0.15cm}
[\hspace{0.03cm}\xi^{\phantom{\prime}}_{1}, \xi^{\phantom{\prime}}_{2}\hspace{0.03cm}]
\hspace{0.03cm}
[\hspace{0.03cm}\bar{\xi}^{\,\prime}_{2}, \xi^{\phantom{\prime}}_{1}\hspace{0.03cm}]
=
\xi^{\phantom{\prime}}_{1}\hspace{0.03cm}\xi^{\phantom{\prime}}_{2}
\hspace{0.03cm}\bar{\xi}^{\,\prime}_{2}\hspace{0.03cm}\xi^{\phantom{\prime}}_{1}
+
\xi^{\phantom{\prime}}_{2}\hspace{0.03cm}\xi^{\phantom{\prime}}_{1}
\hspace{0.03cm}\xi^{\phantom{\prime}}_{1}\hspace{0.03cm}\bar{\xi}^{\,\prime}_{2}
=
- (\xi^{\phantom{\prime}}_{1})^{\hspace{0.02cm}2}
\bigl(\xi^{\phantom{\prime}}_{2}\hspace{0.03cm}\bar{\xi}^{\,\prime}_{2}
+
\bar{\xi}^{\,\prime}_{2}\hspace{0.03cm}\xi^{\phantom{\prime}}_{2}\bigr).
\]
Their multiplication gives us the required expression
\[
[\hspace{0.03cm}\bar{\xi}^{\,\prime}_{1}, \bar{\xi}^{\,\prime}_{2}\hspace{0.03cm}]\hspace{0.03cm}
[\hspace{0.03cm}\xi^{\phantom{\prime}}_{1}, \xi^{\phantom{\prime}}_{2}\hspace{0.03cm}]
\hspace{0.03cm}
[\hspace{0.03cm}\bar{\xi}^{\,\prime}_{1}, \xi^{\phantom{\prime}}_{2}\hspace{0.03cm}]\hspace{0.03cm}
[\hspace{0.03cm}\bar{\xi}^{\,\prime}_{2}, \xi^{\phantom{\prime}}_{1}\hspace{0.03cm}]
=
2\hspace{0.03cm}(\bar{\xi}^{\,\prime}_{1})^{\hspace{0.02cm}2}
\hspace{0.02cm}(\xi^{\phantom{\prime}}_{1})^{\hspace{0.02cm}2}
\hspace{0.02cm}(\bar{\xi}^{\,\prime}_{2})^{\hspace{0.02cm}2}
\hspace{0.02cm}(\xi^{\phantom{\prime}}_{2})^{\hspace{0.02cm}2}
=
\frac{1}{2}\,[\hspace{0.03cm}\bar{\xi}^{\,\prime}_{1}, \xi^{\phantom{\prime}}_{1}\hspace{0.02cm}]^{\hspace{0.03cm}2}\hspace{0.03cm}
[\hspace{0.03cm}\bar{\xi}^{\,\prime}_{2}, \xi^{\phantom{\prime}}_{2}\hspace{0.02cm}]^{\hspace{0.02cm}2}
=
2^{\hspace{0.02cm}3}\hspace{0.02cm}x^{\hspace{0.02cm}2}\hspace{0.02cm}
y^{\hspace{0.02cm}2}.
\]
Taking into consideration the aforementioned, we can write the expression (\ref{eq:5u}) in a very simple form:
\[
\bigl(\Delta\Omega\hspace{0.03cm}(\bar{\xi}^{\,\prime},\xi)\bigr)^{2}
=
\frac{3}{2}\,x^{\hspace{0.02cm}2}\hspace{0.01cm}y^{\hspace{0.02cm}2}.
\]
Therefore, the square of the function $\Omega$, Eq.\,(\ref{eq:5y}), can be written in the following form
\begin{equation}
\begin{split}
\bigl[\hspace{0.03cm}\Omega\hspace{0.03cm}(\bar{\xi}^{\,\prime},\xi)\bigr]^{2}
&=
\frac{3}{2}\,x^{\hspace{0.02cm}2}\hspace{0.01cm}y^{\hspace{0.02cm}2}
-
2\hspace{0.03cm}(x^{\hspace{0.02cm}2}\hspace{0.01cm}y + x\hspace{0.01cm}y^{\hspace{0.02cm}2})
+
3\hspace{0.03cm}x\hspace{0.01cm}y + (x^{\hspace{0.02cm}2} + y^{\hspace{0.02cm}2})
- 2\hspace{0.03cm}(x + y) - 2\,- \\[1ex]
&-\,\frac{\!1}{2^{\hspace{0.02cm}2}}\,\bigl(\hspace{0.02cm}[\hspace{0.03cm}\bar{\xi}^{\,\prime}_{1}, \bar{\xi}^{\,\prime}_{2}\hspace{0.03cm}]\hspace{0.03cm}
[\hspace{0.03cm}\xi^{\phantom{\prime}}_{1}, \xi^{\phantom{\prime}}_{2}\hspace{0.03cm}]
+
[\hspace{0.03cm}\bar{\xi}^{\,\prime}_{1}, \xi^{\phantom{\prime}}_{2}\hspace{0.03cm}]\hspace{0.03cm}
[\hspace{0.03cm}\bar{\xi}^{\,\prime}_{2}, \xi^{\phantom{\prime}}_{1}\hspace{0.03cm}]\hspace{0.02cm}\bigr).
\end{split}
\label{eq:5i}
\end{equation}
\indent Now we turn to the consideration of the term with the derivatives in (\ref{eq:3i}). To be specific, let us consider a contribution of the form
\begin{equation}
-\frac{1}{2}\,\biggl[\,\frac{\partial\hspace{0.03cm}
\Omega\hspace{0.03cm}(\bar{\xi}^{\,\prime},\xi)}
{\partial\hspace{0.03cm}\xi^{\phantom{\prime}}_{1}}\hspace{0.04cm}, \frac{\partial\hspace{0.03cm}\Omega\hspace{0.03cm}(\bar{\xi}^{\,\prime},\xi)}
{\partial\hspace{0.03cm}\bar{\xi}^{\,\prime}_{1}}\hspace{0.03cm}\,\biggr].
\label{eq:5o}
\end{equation}
Here, the derivations are defined by the general expressions (I.9.11) and (I.9.14) and equal
\[
\begin{split}
&\frac{\partial\hspace{0.03cm}\Omega\hspace{0.03cm}(\bar{\xi}^{\,\prime},\xi)}
{\partial\hspace{0.03cm}\xi^{\phantom{\prime}}_{1}}
=
-\frac{\!1}{2^{\hspace{0.02cm}2}}\,\bigl(\hspace{0.03cm}[\hspace{0.03cm}\bar{\xi}^{\,\prime}_{1}, \bar{\xi}^{\,\prime}_{2}\hspace{0.03cm}]\hspace{0.04cm}
\xi^{\phantom{\prime}}_{2}
-
\bar{\xi}^{\,\prime}_{2}\hspace{0.04cm}[\hspace{0.03cm}\bar{\xi}^{\,\prime}_{1}, \xi^{\phantom{\prime}}_{2}\hspace{0.03cm}]\hspace{0.02cm}
+
\bar{\xi}^{\,\prime}_{1}\hspace{0.04cm}[\hspace{0.03cm}\bar{\xi}^{\,\prime}_{2}, \xi^{\phantom{\prime}}_{2}\hspace{0.03cm}]\hspace{0.03cm}\bigr)
+ \bar{\xi}^{\,\prime}_{1}, \\[1ex]
&\frac{\partial\hspace{0.03cm}\Omega\hspace{0.03cm}(\bar{\xi}^{\,\prime},\xi)}
{\partial\hspace{0.03cm}\bar{\xi}^{\,\prime}_{1}}
=
-\frac{\!1}{2^{\hspace{0.02cm}2}}\,\bigl(\hspace{0.03cm} \bar{\xi}^{\,\prime}_{2}\hspace{0.04cm}
[\hspace{0.03cm}\xi^{\phantom{\prime}}_{1}, \xi^{\phantom{\prime}}_{2}\hspace{0.03cm}]
+
[\hspace{0.03cm}\bar{\xi}^{\,\prime}_{2}, \xi^{\phantom{\prime}}_{1}\hspace{0.03cm}]
\hspace{0.04cm}\xi^{\phantom{\prime}}_{2}
-
[\hspace{0.03cm}\bar{\xi}^{\,\prime}_{2}, \xi^{\phantom{\prime}}_{2}\hspace{0.03cm}]
\hspace{0.04cm}\xi^{\phantom{\prime}}_{1}\hspace{0.03cm}\bigr)
- \xi^{\phantom{\prime}}_{1}.
\end{split}
\]
The substitution of these derivatives into the commutator (\ref{eq:5o}) after transformations similar previous ones, results in the following expression:
\[
-\frac{1}{2}\,\biggl[\,\frac{\partial\hspace{0.03cm}
\Omega\hspace{0.03cm}(\bar{\xi}^{\,\prime},\xi)}
{\partial\hspace{0.03cm}\xi^{\phantom{\prime}}_{1}}\hspace{0.04cm}, \frac{\partial\hspace{0.03cm}\Omega\hspace{0.03cm}(\bar{\xi}^{\,\prime},\xi)}
{\partial\hspace{0.03cm}\bar{\xi}^{\,\prime}_{1}}\hspace{0.03cm}\,\biggr]
=
-\frac{1}{2}\,x\hspace{0.01cm}y^{\hspace{0.02cm}2}
+ x\hspace{0.01cm}y - x
-
\frac{\!1}{2^{\hspace{0.02cm}2}}\,\bigl(\hspace{0.02cm}[\hspace{0.03cm}\bar{\xi}^{\,\prime}_{1}, \bar{\xi}^{\,\prime}_{2}\hspace{0.03cm}]\hspace{0.03cm}
[\hspace{0.03cm}\xi^{\phantom{\prime}}_{1}, \xi^{\phantom{\prime}}_{2}\hspace{0.03cm}]
+
[\hspace{0.03cm}\bar{\xi}^{\,\prime}_{1}, \xi^{\phantom{\prime}}_{2}\hspace{0.03cm}]\hspace{0.03cm}
[\hspace{0.03cm}\bar{\xi}^{\,\prime}_{2}, \xi^{\phantom{\prime}}_{1}\hspace{0.03cm}]\hspace{0.02cm}\bigr).
\hspace{2cm}
\]
The second commutator additional to (\ref{eq:5o}) with the derivatives with respect to $\xi_{2}$ and $\bar{\xi}_{2}^{\prime}$ is obtained from this expression by the replacement $x \rightleftarrows y$ and then the term with derivatives in (\ref{eq:3i}) takes its final form:
\begin{equation}
-\frac{1}{2}\,\biggl[\,\frac{\partial\hspace{0.03cm}
\Omega\hspace{0.03cm}(\bar{\xi}^{\,\prime},\xi)}
{\partial\hspace{0.03cm}\xi}\hspace{0.03cm}, \frac{\partial\hspace{0.03cm}\Omega\hspace{0.03cm}(\bar{\xi}^{\,\prime},\xi)}
{\partial\hspace{0.03cm}\bar{\xi}^{\,\prime}}\hspace{0.03cm}\,\biggr]
\!=\!
\frac{1}{2}\,(x^{\hspace{0.02cm}2}\hspace{0.01cm}y + x\hspace{0.01cm}y^{\hspace{0.02cm}2})
-
2\hspace{0.03cm}x\hspace{0.01cm}y + (x + y)
+\frac{1}{2}\,\bigl(\hspace{0.02cm}[\hspace{0.03cm}\bar{\xi}^{\,\prime}_{1}, \bar{\xi}^{\,\prime}_{2}\hspace{0.03cm}]\hspace{0.03cm}
[\hspace{0.03cm}\xi^{\phantom{\prime}}_{1}, \xi^{\phantom{\prime}}_{2}\hspace{0.03cm}]
+
[\hspace{0.03cm}\bar{\xi}^{\,\prime}_{1}, \xi^{\phantom{\prime}}_{2}\hspace{0.03cm}]\hspace{0.03cm}
[\hspace{0.03cm}\bar{\xi}^{\,\prime}_{2}, \xi^{\phantom{\prime}}_{1}\hspace{0.03cm}]\hspace{0.03cm}\bigr).
\label{eq:5p}
\end{equation}
Substituting the obtained expressions (\ref{eq:5i}) and (\ref{eq:5p}) into (\ref{eq:3i}) and collecting similar terms, we derive an explicit form of the convolution $\Omega\hspace{0.02cm}\!*\hspace{0.01cm}\Omega$:
\begin{align}
\Omega\hspace{0.03cm}(\bar{\xi}^{\,\prime},\xi)\! *\hspace{0.02cm}
\Omega\hspace{0.03cm}(\bar{\xi}^{\,\prime},\xi)
&=
\frac{3}{2}\,x^{\hspace{0.02cm}2}\hspace{0.01cm}y^{\hspace{0.02cm}2}
-
\frac{3}{2}\,(x^{\hspace{0.02cm}2}\hspace{0.01cm}y + x\hspace{0.01cm}y^{\hspace{0.02cm}2})
+
\frac{3}{2}\,x\hspace{0.01cm}y + (x^{\hspace{0.02cm}2} + y^{\hspace{0.02cm}2})
- (x + y) + 1\,- \label{eq:5a} \\[1ex]
&-\frac{\!1}{2^{\hspace{0.02cm}3}}\,\bigl(\hspace{0.02cm}[\hspace{0.03cm}\bar{\xi}^{\,\prime}_{1}, \bar{\xi}^{\,\prime}_{2}\hspace{0.03cm}]\hspace{0.03cm}
[\hspace{0.03cm}\xi^{\phantom{\prime}}_{1}, \xi^{\phantom{\prime}}_{2}\hspace{0.03cm}]
-
[\hspace{0.03cm}\bar{\xi}^{\,\prime}_{1}, \xi^{\phantom{\prime}}_{2}\hspace{0.03cm}]\hspace{0.03cm}
[\hspace{0.03cm}\bar{\xi}^{\,\prime}_{2}, \xi^{\phantom{\prime}}_{1}\hspace{0.03cm}]\hspace{0.02cm}\bigr).\notag
\end{align}
This expression should be compared with the function $\widetilde{\Omega}(\bar{\xi}^{\prime}, \xi)$ as it is defined by Eq.\,(\ref{eq:2r}). In terms of the notations (\ref{eq:5t}) this function has the following form:
\begin{equation}
\widetilde{\Omega}\hspace{0.03cm}(\bar{\xi}^{\,\prime},\xi)
=
2\hspace{0.03cm}x^{\hspace{0.02cm}2}\hspace{0.01cm}y^{\hspace{0.02cm}2}
-
2\hspace{0.03cm}(x^{\hspace{0.02cm}2}\hspace{0.01cm}y + x\hspace{0.01cm}y^{\hspace{0.02cm}2})
+
2\hspace{0.03cm}x\hspace{0.01cm}y + (x^{\hspace{0.02cm}2} + y^{\hspace{0.02cm}2})
- (x + y) + 1.
\label{eq:5s}
\end{equation}
In spite of very close similarity of these two expressions, there is no coincidence in a literal sense, and thus the equality (\ref{eq:5q}) does not take place. This is an additional argument to conclusions of section 6 of Part I that the Geyer operator $a_{0}^{2}$ as it was defined by Eq.\,(I.B.17) should be considered simply as a symbol. It is not the result of the multiplication of two operators $a_{0}\cdot a_{0}$.
In the next sections we will analyse this problem in more detail.

%%%%%%%%%%%%%%%%%%%%%%%% section 6 %%%%%%%%%%%%%%%%%%%%%%%%%%%%

\section{Connection between the operators $\hat{\omega}^{\,2}$ and $a^{2}_{0}$ revised}
\setcounter{equation}{0}
\label{section_6}

In section 8 of Part I we have defined a connection between the operators $\hat{\omega}^{2}$ and $a_0^2$. For ease of reading, hereafter the operator $\hat{\omega}^{2}$ will be named the Harish-Chandra operator and  the operator $a_0^2$ will the Geyer operator. In establishing the connection between these two operators, we have used the matrix relation derived by Harish-Chandra \cite{harish-chandra_1947}, namely (see Appendix A in Part I)
\begin{equation}
B = 3 - \omega^{2},
\label{eq:6q}
\end{equation}
which enables one to put the operator $\hat{\omega}^{2}$ in the following form:
\begin{equation}
\hat{\omega}^{\,2} = \frac{1}{2}\,(1 + \widehat{\boldsymbol{\eta}}_{5}).
\label{eq:6w}
\end{equation}
(see the notations in section 8 of Part I). A particular consequence of this representation was the connection
\begin{equation}
\hat{\omega}^{\,2} = a^{\hspace{0.02cm}2}_{0}.
\label{eq:6e}
\end{equation}
However, as it was shown in the previous section this relation results in a contradiction. It should be particularly emphasized that on the left-hand side of (\ref{eq:6e}) is the {\it square} of the operator $\hat{\omega}$ as it was defined by Eq.\,(I.6.19), and on the right-hand side does the operator $a_{0}^{2}$, which was introduced into consideration by Geyer \cite{geyer_1968}. The latter as it will be clear from the subsequent discussion, should be considered simply as a symbol, but not the square of a certain operator.\\ 
\indent Let us give up the relation (\ref{eq:6q}) and consider what would be the consequences of this. For that purpose, we return to the formula (I.8.4). On the left-hand side of this formula we rewrite the second term as follows:
\begin{equation}
\widehat{\boldsymbol{\eta}}_{\mu_{1}}\widehat{\boldsymbol{\eta}}_{\mu_{2}} +
\widehat{\boldsymbol{\eta}}_{\mu_{1}}\widehat{\boldsymbol{\eta}}_{\mu_{3}} + 
\widehat{\boldsymbol{\eta}}_{\mu_{1}}\widehat{\boldsymbol{\eta}}_{\mu_{4}} +
\widehat{\boldsymbol{\eta}}_{\mu_{2}}\widehat{\boldsymbol{\eta}}_{\mu_{3}} + 
\widehat{\boldsymbol{\eta}}_{\mu_{2}}\widehat{\boldsymbol{\eta}}_{\mu_{4}} +
\widehat{\boldsymbol{\eta}}_{\mu_{3}}\widehat{\boldsymbol{\eta}}_{\mu_{4}}
=
\label{eq:6r}
\end{equation}
\[
=
(\widehat{\boldsymbol{\eta}}_{\mu_{1}}\widehat{\boldsymbol{\eta}}_{\mu_{2}}
+
\widehat{\boldsymbol{\eta}}_{\mu_{3}}\widehat{\boldsymbol{\eta}}_{\mu_{4}})
+
(\widehat{\boldsymbol{\eta}}_{\mu_{1}} + \widehat{\boldsymbol{\eta}}_{\mu_{2}})
(\widehat{\boldsymbol{\eta}}_{\mu_{3}} + \widehat{\boldsymbol{\eta}}_{\mu_{4}}).
\]
To be specific, we fix the values of the indices as follows: $\mu_k\equiv k,\,k=1, 2, 3, 4$. The use of formulae (I.8.9) and (I.8.10) for the first term on the right-hand side in (\ref{eq:6r}) gives us
\begin{equation}
\widehat{\boldsymbol{\eta}}_{1}\widehat{\boldsymbol{\eta}}_{2}
+
\widehat{\boldsymbol{\eta}}_{3}\widehat{\boldsymbol{\eta}}_{4}
=
-2\hspace{0.03cm}(\hspace{0.03cm}N^{\hspace{0.03cm}2}_{1} + N^{\hspace{0.03cm}2}_{2})
+ 2.
\label{eq:6t}
\end{equation}
Further, by using the connection between the creation and annihilation operators $a_{k}^{\pm}$ and the operators $\hat{\beta}_{\mu}$, Eq.\,(I.3.4), we derive
\begin{equation}
\begin{split}
&\hat{\beta}^{\hspace{0.03cm}2}_{1} = \frac{1}{4}\,\bigl(\hspace{0.03cm}(a^{+}_{1})^{\hspace{0.03cm}2} + (a^{-}_{1})^{\hspace{0.03cm}2} + \{a^{+}_{1},a^{-}_{1}\}\bigr), \\[1ex]
&\hat{\beta}^{\hspace{0.03cm}2}_{2} = \frac{1}{4}\,\bigl(\hspace{0.03cm}(a^{+}_{1})^{\hspace{0.03cm}2} + (a^{-}_{1})^{\hspace{0.03cm}2} - \{a^{+}_{1},a^{-}_{1}\}\bigr).
\end{split}
\label{eq:6y}
\end{equation}
Recalling the definition of $\boldsymbol{\eta}$-matrices: $\boldsymbol{\eta}_{\mu} = 2 \beta_{\mu}^{2} - 1$, we get further
\begin{equation}
\widehat{\boldsymbol{\eta}}_{1} + \widehat{\boldsymbol{\eta}}_{2} =
2\hspace{0.03cm}(\hat{\beta}^{\hspace{0.03cm}2}_{1} + \hat{\beta}^{\hspace{0.03cm}2}_{2}) - 2
= \{a^{+}_{1},a^{-}_{1}\} - 2,
\label{eq:6u}
\end{equation}
and similar
\begin{equation}
\widehat{\boldsymbol{\eta}}_{3} + \widehat{\boldsymbol{\eta}}_{4} =
2\hspace{0.03cm}(\hat{\beta}^{\hspace{0.03cm}2}_{3} + \hat{\beta}^{\hspace{0.03cm}2}_{4}) - 2
= \{a^{+}_{2},a^{-}_{2}\} - 2.
\label{eq:6i}
\end{equation}
Substituting (\ref{eq:6t})\,--\,(\ref{eq:6i}) into (\ref{eq:6r}), and then into (I.8.4) and finally into (I.8.3), instead of (\ref{eq:6e}), we derive
\begin{equation}
\hat{\omega}^{\,2} = \frac{3}{4}\,a^{\hspace{0.02cm}2}_{0}
+
\frac{1}{4}\hspace{0.03cm}\Bigl[\bigl(\hspace{0.02cm}N^{\hspace{0.03cm}2}_{1} + N^{\hspace{0.03cm}2}_{2} - 1\bigr)
-
2\Bigl(\hspace{0.03cm}\frac{1}{2}\,\{a^{+}_{1},a^{-}_{1}\} - 1\Bigr)
\Bigl(\hspace{0.03cm}\frac{1}{2}\,\{a^{+}_{2},a^{-}_{2}\} - 1\Bigr)\Bigr].
\label{eq:6o}
\end{equation}
Here, we have taken into account the connection between the operator $a_{0}^{2}$ and the operator 
$\boldsymbol{\hat{\eta}}_{5}$, Eq.\,(I.8.11).\\
\indent Let us define a matrix element of the expression (\ref{eq:6o}) in the basis of parafermion coherent states. The matrix element of the first term on the right-hand side is given by the expressions (\ref{eq:2e}) and (\ref{eq:2r}) (or (\ref{eq:5s})). The matrix elements of the operators $N_{1}^{2}$ and $N_{2}^{2}$ are defined by the formula (I.9.16) and in terms of the variables $x$ and $y$ have the form
\begin{equation}
\langle\hspace{0.02cm}\bar{\xi}^{\,\prime}\hspace{0.02cm}|\,
N^{\hspace{0.02cm}2}_{1}|\,\xi\hspace{0.03cm}\rangle
=
(x^{\hspace{0.02cm}2} - x +1)\hspace{0.02cm}
\langle\hspace{0.02cm}\bar{\xi}^{\,\prime}\hspace{0.02cm}|\,\xi\hspace{0.02cm}\rangle,
\quad
\langle\hspace{0.02cm}\bar{\xi}^{\,\prime}\hspace{0.02cm}|\,
N^{\hspace{0.02cm}2}_{2}|\,\xi\hspace{0.03cm}\rangle
=
(y^{\hspace{0.02cm}2} - y +1)\hspace{0.02cm}
\langle\hspace{0.02cm}\bar{\xi}^{\,\prime}\hspace{0.02cm}|\,\xi\hspace{0.02cm}\rangle.
\label{eq:6p}
\end{equation}
We have only to define the matrix element of the last term in (\ref{eq:6o}). Towards this end, we will need the following formulae:
\begin{equation}
\begin{split}
&a^{-}_{k}a^{+}_{n}|\,\xi\hspace{0.02cm}\rangle =
(\hspace{0.02cm}2\hspace{0.03cm}\delta_{kn} + \xi^{\phantom{+}}_{k}\!a^{+}_{n})|\,\xi\hspace{0.02cm}\rangle, \\[1.2ex]
&\langle\hspace{0.02cm}\bar{\xi}^{\,\prime}\hspace{0.02cm}|\,a^{-}_{n}a^{+}_{k}
=
\langle\hspace{0.02cm}\bar{\xi}^{\,\prime}\hspace{0.02cm}|\hspace{0.03cm}
(\hspace{0.02cm}a^{-}_{n}\hspace{0.03cm}\bar{\xi}^{\,\prime}_{k} + 2\hspace{0.03cm}\delta_{kn}).
\end{split}
\label{eq:6a}
\end{equation}
Then for the anticommutators $\{a_{k}^{+}, a_{k}^{-}\}$ by virtue of these formulae we have
\begin{equation}
\langle\hspace{0.02cm}\bar{\xi}^{\,\prime}\hspace{0.02cm}|\hspace{0.03cm}
\{a^{+}_{k},a^{-}_{k}\}|\hspace{0.04cm}\xi\hspace{0.03cm}\rangle
=
\bigl(\{\bar{\xi}^{\,\prime}_{k},\xi^{\phantom{\prime}}_{k}\} + 2\hspace{0.02cm}\bigr)\hspace{0.02cm}
\langle\hspace{0.02cm}\bar{\xi}^{\,\prime}\hspace{0.02cm}|\,\xi\hspace{0.02cm}\rangle.
\label{eq:6s}
\end{equation}
\indent Let us consider the matrix element of a product of two commutators, which we present as a sum of four terms
\begin{equation}
\langle\hspace{0.02cm}\bar{\xi}^{\,\prime}\hspace{0.02cm}|\hspace{0.03cm}
\{a^{+}_{1},a^{-}_{1}\}\hspace{0.02cm}\{a^{+}_{2},a^{-}_{2}\}
|\hspace{0.04cm}\xi\hspace{0.03cm}\rangle
=
\label{eq:6d}
\end{equation}
\[
=
\langle\hspace{0.02cm}\bar{\xi}^{\,\prime}\hspace{0.02cm}|\,
a^{+}_{1}a^{-}_{1}a^{+}_{2}a^{-}_{2}|\hspace{0.04cm}\xi\hspace{0.03cm}\rangle
+
\langle\hspace{0.02cm}\bar{\xi}^{\,\prime}\hspace{0.02cm}|\,
a^{+}_{1}a^{-}_{1}a^{-}_{2}a^{+}_{2}|\hspace{0.04cm}\xi\hspace{0.03cm}\rangle
+
\langle\hspace{0.02cm}\bar{\xi}^{\,\prime}\hspace{0.02cm}|\,
a^{-}_{1}a^{+}_{1}a^{+}_{2}a^{-}_{2}|\hspace{0.04cm}\xi\hspace{0.03cm}\rangle
+
\langle\hspace{0.02cm}\bar{\xi}^{\,\prime}\hspace{0.02cm}|\,
a^{-}_{1}a^{+}_{1}a^{-}_{2}a^{+}_{2}|\hspace{0.04cm}\xi\hspace{0.03cm}\rangle.
\]
For the first three terms, by using the relations (\ref{eq:6a}) it is not difficult to obtain, correspondingly
\[
\begin{split}
&\langle\hspace{0.02cm}\bar{\xi}^{\,\prime}\hspace{0.02cm}|\,
a^{+}_{1}a^{-}_{1}a^{+}_{2}a^{-}_{2}|\hspace{0.04cm}\xi\hspace{0.03cm}\rangle
=
\bar{\xi}^{\,\prime}_{1}\hspace{0.03cm}\xi^{\phantom{\prime}}_{1}\hspace{0.03cm}
\bar{\xi}^{\,\prime}_{2}\hspace{0.03cm}\xi^{\phantom{\prime}}_{2}\hspace{0.03cm}
\langle\hspace{0.02cm}\bar{\xi}^{\,\prime}\hspace{0.02cm}|\,\xi\hspace{0.02cm}\rangle,
\\[1.3ex]
&\langle\hspace{0.02cm}\bar{\xi}^{\,\prime}\hspace{0.02cm}|\,
a^{+}_{1}a^{-}_{1}a^{-}_{2}a^{+}_{2}|\hspace{0.04cm}\xi\hspace{0.03cm}\rangle
=
(-\hspace{0.03cm}\bar{\xi}^{\,\prime}_{1}\hspace{0.03cm}\bar{\xi}^{\,\prime}_{2}\hspace{0.03cm}
\xi^{\phantom{\prime}}_{2}\hspace{0.03cm}\xi^{\phantom{\prime}}_{1}
+
2\hspace{0.03cm}\bar{\xi}^{\,\prime}_{1}\hspace{0.03cm}\xi^{\phantom{\prime}}_{1} )
\hspace{0.03cm}
\langle\hspace{0.02cm}\bar{\xi}^{\,\prime}\hspace{0.02cm}|\,\xi\hspace{0.02cm}\rangle,
\\[1.3ex]
&\langle\hspace{0.02cm}\bar{\xi}^{\,\prime}\hspace{0.02cm}|\,
a^{-}_{1}a^{+}_{1}a^{+}_{2}a^{-}_{2}|\hspace{0.04cm}\xi\hspace{0.03cm}\rangle
=
(-\hspace{0.03cm}\bar{\xi}^{\,\prime}_{2}\hspace{0.03cm}\bar{\xi}^{\,\prime}_{1}\hspace{0.03cm}
\xi^{\phantom{\prime}}_{1}\hspace{0.03cm}\xi^{\phantom{\prime}}_{2}
+
2\hspace{0.03cm}\bar{\xi}^{\,\prime}_{2}\hspace{0.03cm}\xi^{\phantom{\prime}}_{2})
\hspace{0.03cm}
\langle\hspace{0.02cm}\bar{\xi}^{\,\prime}\hspace{0.02cm}|\,\xi\hspace{0.02cm}\rangle.
\end{split}
\]
The analysis of the last term in (\ref{eq:6a}) is somewhat more complicated. Here, from (\ref{eq:6a}) we have
\[
\begin{split}
&\langle\hspace{0.02cm}\bar{\xi}^{\,\prime}\hspace{0.02cm}|\,
a^{-}_{1}a^{+}_{1}a^{-}_{2}a^{+}_{2}|\hspace{0.04cm}\xi\hspace{0.03cm}\rangle
=
\langle\hspace{0.02cm}\bar{\xi}^{\,\prime}\hspace{0.02cm}|\,
(a^{-}_{1}\hspace{0.03cm}\bar{\xi}^{\,\prime}_{1} + 2)
(\xi^{\phantom{\prime}}_{2}\hspace{0.03cm}a^{+}_{2} + 2)|\hspace{0.04cm}\xi\hspace{0.03cm}\rangle
= \\[1.2ex]
&\langle\hspace{0.02cm}\bar{\xi}^{\,\prime}\hspace{0.02cm}|\,
a^{-}_{1}\bar{\xi}^{\,\prime}_{1}\hspace{0.03cm}\xi^{\phantom{\prime}}_{2}
\hspace{0.03cm}a^{+}_{2}|\hspace{0.04cm}\xi\hspace{0.03cm}\rangle
+
2\hspace{0.03cm}(\xi^{\phantom{\prime}}_{1}\hspace{0.03cm}
\bar{\xi}^{\,\prime}_{1} + \xi^{\phantom{\prime}}_{2}\hspace{0.03cm}\bar{\xi}^{\,\prime}_{2})
\hspace{0.03cm}
\langle\hspace{0.02cm}\bar{\xi}^{\,\prime}\hspace{0.02cm}|\,\xi\hspace{0.02cm}\rangle
+
4\hspace{0.04cm}
\langle\hspace{0.02cm}\bar{\xi}^{\,\prime}\hspace{0.02cm}|\,\xi\hspace{0.02cm}\rangle.
\end{split}
\]
We transform the expression in the first term by using the permutation rules (I.C.8)
\[
a^{-}_{1}\bar{\xi}^{\,\prime}_{1}\hspace{0.03cm}\xi^{\phantom{\prime}}_{2}
\hspace{0.03cm}a^{+}_{2}
=
- a^{-}_{1}a^{+}_{2}\hspace{0.03cm}\xi^{\phantom{\prime}}_{2}\hspace{0.03cm}
\bar{\xi}^{\,\prime}_{1},
\]
then
\[
\langle\hspace{0.02cm}\bar{\xi}^{\,\prime}\hspace{0.02cm}|\,
a^{-}_{1}\bar{\xi}^{\,\prime}_{1}\hspace{0.03cm}\xi^{\phantom{\prime}}_{2}
\hspace{0.03cm}a^{+}_{2}|\hspace{0.04cm}\xi\hspace{0.03cm}\rangle
=
\xi^{\phantom{\prime}}_{2}\hspace{0.04cm}\bar{\xi}^{\,\prime}_{2}\hspace{0.03cm}
\xi^{\phantom{\prime}}_{1}\hspace{0.03cm}\bar{\xi}^{\,\prime}_{1}\hspace{0.04cm}
\langle\hspace{0.02cm}\bar{\xi}^{\,\prime}\hspace{0.02cm}|\,\xi\hspace{0.02cm}\rangle.
\]
Taking into account the obtained expressions (\ref{eq:6s}) and (\ref{eq:6d}), we derive an explicit form of the matrix element of the last term in (\ref{eq:6o}):
\begin{equation}
\langle\hspace{0.02cm}\bar{\xi}^{\,\prime}\hspace{0.02cm}|
\Bigl(\hspace{0.03cm}\frac{1}{2}\,\{a^{+}_{1},a^{-}_{1}\} - 1\Bigr)\!
\Bigl(\hspace{0.03cm}\frac{1}{2}\,\{a^{+}_{2},a^{-}_{2}\} - 1\Bigr)|\,\xi\hspace{0.03cm}\rangle
=
\label{eq:6f}
\end{equation}
\[
=
\Bigl[\,\frac{1}{4}\,\bigl(\hspace{0.03cm}\bar{\xi}^{\,\prime}_{1}
\hspace{0.03cm}\xi^{\phantom{\prime}}_{1}\hspace{0.03cm}
\bar{\xi}^{\,\prime}_{2}\hspace{0.03cm}\xi^{\phantom{\prime}}_{2}
-
\bar{\xi}^{\,\prime}_{1}\hspace{0.03cm}\bar{\xi}^{\,\prime}_{2}\hspace{0.03cm}
\xi^{\phantom{\prime}}_{2}\hspace{0.03cm}\xi^{\phantom{\prime}}_{1}
-
\bar{\xi}^{\,\prime}_{2}\hspace{0.03cm}\bar{\xi}^{\,\prime}_{1}\hspace{0.03cm}
\xi^{\phantom{\prime}}_{1}\hspace{0.03cm}\xi^{\phantom{\prime}}_{2}
+
\xi^{\phantom{\prime}}_{2}\hspace{0.04cm}\bar{\xi}^{\,\prime}_{2}\hspace{0.03cm}
\xi^{\phantom{\prime}}_{1}\hspace{0.03cm}\bar{\xi}^{\,\prime}_{1}
%\,+
%\]
%\[
+\,2\hspace{0.03cm}\{\hspace{0.02cm}\bar{\xi}^{\,\prime}_{1},\xi^{\phantom{\prime}}_{1}\}
+
2\hspace{0.04cm}\{\hspace{0.02cm}\bar{\xi}^{\,\prime}_{2},\xi^{\phantom{\prime}}_{2}\} + 4\hspace{0.02cm}\bigr)
\,-
\]
\[
-\,\frac{1}{2}\,\bigl(\hspace{0.01cm}\{\hspace{0.02cm}
\bar{\xi}^{\,\prime}_{1},\xi^{\phantom{\prime}}_{1}\}
+
\{\hspace{0.02cm}\bar{\xi}^{\,\prime}_{2},\xi^{\phantom{\prime}}_{2}\}\bigr) -2 +1\Bigr]
\langle\hspace{0.02cm}\bar{\xi}^{\,\prime}\hspace{0.02cm}|\,\xi\hspace{0.02cm}\rangle.
\]
As we can see from this expression, all terms of zeroth and second orders in the para-Grassmann variables are cancelled out and thus, we eventually obtain
\[
\langle\hspace{0.02cm}\bar{\xi}^{\,\prime}\hspace{0.02cm}|
\Bigl(\hspace{0.03cm}\frac{1}{2}\,\{a^{+}_{1},a^{-}_{1}\} - 1\Bigr)\!
\Bigl(\hspace{0.03cm}\frac{1}{2}\,\{a^{+}_{2},a^{-}_{2}\} - 1\Bigr)|\,\xi\hspace{0.03cm}\rangle
=
\]
\[
=
\frac{1}{4}\,\bigl(\hspace{0.03cm}\bar{\xi}^{\,\prime}_{1}
\hspace{0.03cm}\xi^{\phantom{\prime}}_{1}\hspace{0.03cm}
\bar{\xi}^{\,\prime}_{2}\hspace{0.03cm}\xi^{\phantom{\prime}}_{2}
-
\bar{\xi}^{\,\prime}_{1}\hspace{0.03cm}\bar{\xi}^{\,\prime}_{2}\hspace{0.03cm}
\xi^{\phantom{\prime}}_{2}\hspace{0.03cm}\xi^{\phantom{\prime}}_{1}
-
\bar{\xi}^{\,\prime}_{2}\hspace{0.03cm}\bar{\xi}^{\,\prime}_{1}\hspace{0.03cm}
\xi^{\phantom{\prime}}_{1}\hspace{0.03cm}\xi^{\phantom{\prime}}_{2}
+
\xi^{\phantom{\prime}}_{2}\hspace{0.03cm}\bar{\xi}^{\,\prime}_{2}\hspace{0.03cm}
\xi^{\phantom{\prime}}_{1}\hspace{0.03cm}\bar{\xi}^{\,\prime}_{1}
\hspace{0.03cm}\bigr)
\langle\hspace{0.02cm}\bar{\xi}^{\,\prime}\hspace{0.02cm}|\,\xi\hspace{0.02cm}\rangle.
\]
Making use of this expression and the formulae (\ref{eq:5s}) and (\ref{eq:6p}), we can now write the whole matrix element for the Harish-Chandra operator $\hat{\omega}^2$ in the representation (\ref{eq:6o})
\begin{equation}
\langle\hspace{0.02cm}\bar{\xi}^{\,\prime}\hspace{0.02cm}|\,
\hat{\omega}^{\,2}|\,\xi\hspace{0.03cm}\rangle
=
\Bigl\{\hspace{0.03cm}\frac{3}{2}\,x^{\hspace{0.02cm}2}\hspace{0.01cm}y^{\hspace{0.02cm}2}
-
\frac{3}{2}\,(x^{\hspace{0.02cm}2}\hspace{0.01cm}y + x\hspace{0.01cm}y^{\hspace{0.02cm}2})
+
\frac{3}{2}\,x\hspace{0.01cm}y + (x^{\hspace{0.02cm}2} + y^{\hspace{0.02cm}2})
- (x + y) + 1\,-
\label{eq:6g}
\end{equation}
\[
-\,\frac{\!1}{2^{\hspace{0.02cm}3}}\,
\bigl(\hspace{0.03cm}\bar{\xi}^{\,\prime}_{1}
\hspace{0.03cm}\xi^{\phantom{\prime}}_{1}\hspace{0.03cm}
\bar{\xi}^{\,\prime}_{2}\hspace{0.03cm}\xi^{\phantom{\prime}}_{1}
-
\bar{\xi}^{\,\prime}_{1}\hspace{0.03cm}\bar{\xi}^{\,\prime}_{2}\hspace{0.03cm}
\xi^{\phantom{\prime}}_{2}\hspace{0.03cm}\xi^{\phantom{\prime}}_{1}
-
\bar{\xi}^{\,\prime}_{2}\hspace{0.03cm}\bar{\xi}^{\,\prime}_{1}\hspace{0.03cm}
\xi^{\phantom{\prime}}_{1}\hspace{0.03cm}\xi^{\phantom{\prime}}_{2}
+
\xi^{\phantom{\prime}}_{2}\hspace{0.03cm}\bar{\xi}^{\,\prime}_{2}\hspace{0.03cm}
\xi^{\phantom{\prime}}_{1}\hspace{0.03cm}\bar{\xi}^{\,\prime}_{1}
\hspace{0.03cm}\bigr)\!\Bigr\}\hspace{0.03cm}
\langle\hspace{0.02cm}\bar{\xi}^{\,\prime}\hspace{0.02cm}|\,\xi\hspace{0.02cm}\rangle.
\hspace{0.01cm}
\]
If one takes into account the trivial identity
\[
[\hspace{0.03cm}\bar{\xi}^{\,\prime}_{1}, \bar{\xi}^{\,\prime}_{2}\hspace{0.03cm}]\hspace{0.03cm}
[\hspace{0.03cm}\xi^{\phantom{\prime}}_{1}, \xi^{\phantom{\prime}}_{2}\hspace{0.03cm}]
-
[\hspace{0.03cm}\bar{\xi}^{\,\prime}_{1}, \xi^{\phantom{\prime}}_{2}\hspace{0.03cm}]\hspace{0.03cm}
[\hspace{0.03cm}\bar{\xi}^{\,\prime}_{2}, \xi^{\phantom{\prime}}_{1}\hspace{0.03cm}]
=
-\bigl(\hspace{0.03cm}\bar{\xi}^{\,\prime}_{1}
\hspace{0.03cm}\xi^{\phantom{\prime}}_{1}\hspace{0.03cm}
\bar{\xi}^{\,\prime}_{2}\hspace{0.03cm}\xi^{\phantom{\prime}}_{1}
-
\bar{\xi}^{\,\prime}_{1}\hspace{0.03cm}\bar{\xi}^{\,\prime}_{2}\hspace{0.03cm}
\xi^{\phantom{\prime}}_{2}\hspace{0.03cm}\xi^{\phantom{\prime}}_{1}
-
\bar{\xi}^{\,\prime}_{2}\hspace{0.03cm}\bar{\xi}^{\,\prime}_{1}\hspace{0.03cm}
\xi^{\phantom{\prime}}_{1}\hspace{0.03cm}\xi^{\phantom{\prime}}_{2}
+
\xi^{\phantom{\prime}}_{2}\hspace{0.03cm}\bar{\xi}^{\,\prime}_{2}\hspace{0.03cm}
\xi^{\phantom{\prime}}_{1}\hspace{0.03cm}\bar{\xi}^{\,\prime}_{1}
\hspace{0.03cm}\bigr),
\]
then we see that the expression (\ref{eq:6g}) ideally reproduces the expression for the convolution 
$\Omega\hspace{0.03cm}\!*\Omega$, Eq.\,(\ref{eq:5a}). In the subsequent sections we will discuss in more detail why omitting the Harish-Chandra relation (\ref{eq:6q}) 
leads us to a correct result. It is precisely these expression (\ref{eq:6o}) that defines the square of the operator $a_{0}(\equiv-\hat{\omega})$, which in contrast to 
the Geyer symbol $a_{0}^{2}$ we will designate as $(a_{0})^{2}$.

%%%%%%%%%%%%%%%%%%%%%%%% section 7 %%%%%%%%%%%%%%%%%%%%%%%%%%%%

\section{The Casimir operators $\widehat{C}_{2}$ and $\widehat{C}^{\hspace{0.03cm}\prime}_{2}$}
\setcounter{equation}{0}
\label{section_7}

In the paper Omote {\it et al.} \cite{omote_1976} an explicit form of the quadratic Casimir operator for the group $SO(2M)$ was written out in terms of the generators
\[
\widehat{C}_{2} = \sum\limits_{k,l}\bigl(\hspace{0.03cm}2N_{kl}N_{lk}
+ L_{kl}M_{lk} + M_{kl}L_{lk}\bigr).
\]
Here, the indices $k, l$ run $1, 2, \ldots, M$. In our case $M = 2$ and therefore
\begin{equation}
\widehat{C}_{2} = 2\hspace{0.03cm}\bigl(\hspace{0.03cm}
-\{\hspace{0.02cm}L_{12},M_{12}\} + \{\hspace{0.02cm}N_{12},N_{21}\}
+ (N^{2}_{1} + N^{2}_{2})\bigr).
\label{eq:7q}
\end{equation}
For comparison, here we write out an explicit form of the operator $a_0$ obtained in section 6 of Part I
\begin{equation}
a_{0} = -\frac{1}{4}\,\bigl(\{\hspace{0.02cm}L_{12}, M_{12}\hspace{0.02cm}\}
+
\{\hspace{0.02cm}N_{12}, N_{21}\hspace{0.02cm}\}
-
\{\hspace{0.02cm}N_{1}, N_{2}\hspace{0.02cm}\}\bigr).
\label{eq:7w}
\end{equation}
For the group $SO(2M+1)$ the quadratic Casimir operator is
\begin{equation}
\widehat{C}^{\hspace{0.03cm}\prime}_{2} = \widehat{C}_{2} + \hat{\Lambda},
\label{eq:7e}
\end{equation}
where $\hat{\Lambda} \equiv \sum_{k=1}^{M} \{a_{k}^{+}, a_{k}^{-}\}$. In our case for $M = 2$ the operator $\hat{\Lambda}$ is
\begin{equation}
\hat{\Lambda} = \{a^{+}_{1},a^{-}_{1}\} + \{a^{+}_{2},a^{-}_{2}\}.
\label{eq:7r}
\end{equation}
From the form of Casimir's operator (\ref{eq:7q}) we may notice that the contribution $(N_{1}^{\hspace{0.02cm}2} + N_{\hspace{0.03cm}2}^{2})$ has its analogue in the expression for the Harish-Chandra operator (\ref{eq:6o}). This suggests that the product
\[
\Bigl(\hspace{0.03cm}\frac{1}{2}\,\{a^{+}_{1},a^{-}_{1}\} - 1\Bigr)\!
\Bigl(\hspace{0.03cm}\frac{1}{2}\,\{a^{+}_{2},a^{-}_{2}\} - 1\Bigr)
\]
from (\ref{eq:6o}) may have a connection with the difference
\begin{equation}
\{\hspace{0.02cm}N_{12},N_{21}\} - \{\hspace{0.02cm}L_{12},M_{12}\}
\label{eq:7t}
\end{equation}
from the definition of the Casimir operator (\ref{eq:7q}). We verify this assumption by a direct transformation of (\ref{eq:7t}).\\
\indent Let us consider the first term in (\ref{eq:7t}). By the definitions of the generators $N_{12}$ and $N_{21}$
\[
N_{12} = \displaystyle\frac{1}{2}\;[\hspace{0.03cm}a^{+}_{1}, a^{-}_{2}\hspace{0.02cm}],
\quad
N_{21} = \displaystyle\frac{1}{2}\;[\hspace{0.03cm}a^{+}_{2}, a^{-}_{1}\hspace{0.02cm}]
\]
and the algebra of para-Fermi operators of order $p = 2$, Eqs.\,(I.3.5)\,--\,(I.3.7), for this term we have
\[
\begin{split}
&N_{12}N_{21} = \frac{1}{4}\,\bigl[\hspace{0.03cm}-a^{+}_{1}a^{-}_{1}a^{+}_{2}a^{-}_{2} - a^{+}_{1}a^{-}_{2}a^{-}_{1}a^{+}_{2} - a^{-}_{2}a^{+}_{1}a^{+}_{2}a^{-}_{1}
- a^{-}_{1}a^{+}_{1}a^{-}_{2}a^{+}_{2} +
2\hspace{0.03cm}(\hspace{0.02cm}a^{+}_{1}a^{-}_{1} + a^{-}_{2}a^{+}_{2})\hspace{0.02cm}\bigr], \\[1ex]
&N_{21}N_{12} = \frac{1}{4}\,\bigl[\hspace{0.03cm}-a^{+}_{1}a^{-}_{1}a^{+}_{2}a^{-}_{2} -  a^{+}_{1}a^{-}_{2}a^{-}_{1}a^{+}_{2} - a^{-}_{2}a^{+}_{1}a^{+}_{2}a^{-}_{1}
- a^{-}_{1}a^{+}_{1}a^{-}_{2}a^{+}_{2} +
2\hspace{0.03cm}(\hspace{0.02cm}a^{-}_{1}a^{+}_{1} + a^{+}_{2}a^{-}_{2})\hspace{0.02cm}\bigr].
\end{split}
\]
Summing these two expressions, we get
\begin{equation}
\{\hspace{0.02cm}N_{12},N_{21}\} =
-\frac{1}{2}\,\bigl[\hspace{0.03cm}a^{+}_{1}a^{-}_{1}a^{+}_{2}a^{-}_{2} + a^{+}_{1}a^{-}_{2}a^{-}_{1}a^{+}_{2} + a^{-}_{2}a^{+}_{1}a^{+}_{2}a^{-}_{1}
+
a^{-}_{1}a^{+}_{1}a^{-}_{2}a^{+}_{2} \,-
\label{eq:7y}
\end{equation}
\[
-\,\bigl(\hspace{0.02cm}\{a^{+}_{1},a^{-}_{1}\} + \{a^{+}_{2},a^{-}_{2}\}\bigr)\hspace{0.02cm}\bigr].
\]
\indent In what follows we shall analyse the second term in (\ref{eq:7t}). Here, we recall that the generators $L_{12}$ and $M_{12}$ are given by the expressions
\[
L_{12} = \displaystyle\frac{1}{2}\;[\hspace{0.03cm}a^{+}_{1}, a^{+}_{2}\hspace{0.02cm}],
\quad
M_{12} = \displaystyle\frac{1}{2}\;[\hspace{0.03cm}a^{-}_{1}, a^{-}_{2}\hspace{0.02cm}].
\]
By virtue of the algebra (I.3.5)\,--\,(I.3.7) for this term we have
\[
\begin{split}
&L_{12}M_{12} = \frac{1}{4}\,\bigl[\hspace{0.03cm}-a^{+}_{1}a^{-}_{2}a^{-}_{1}a^{+}_{2} + a^{+}_{1}a^{-}_{1}a^{-}_{2}a^{+}_{2} + a^{-}_{1}a^{+}_{1}a^{+}_{2}a^{-}_{2}
- a^{-}_{2}a^{+}_{1}a^{+}_{2}a^{-}_{1} -
2\hspace{0.03cm}(\hspace{0.02cm}a^{+}_{1}a^{-}_{1} + a^{+}_{2}a^{-}_{2})\hspace{0.02cm}\bigr], \\[1ex]
&M_{12}L_{12} = \frac{1}{4}\,\bigl[\hspace{0.03cm}-a^{+}_{1}a^{-}_{2}a^{-}_{1}a^{+}_{2} + a^{+}_{1}a^{-}_{1}a^{-}_{2}a^{+}_{2} + a^{-}_{1}a^{+}_{1}a^{+}_{2}a^{-}_{2}
- a^{-}_{2}a^{+}_{1}a^{+}_{2}a^{-}_{1} -
2\hspace{0.03cm}(\hspace{0.02cm}a^{-}_{1}a^{+}_{1} + a^{-}_{2}a^{+}_{2})\hspace{0.02cm}\bigr],
\end{split}
\]
that in turn, gives
\[
\hspace{0.5cm}
\{\hspace{0.02cm}L_{12},M_{12}\} =
-\frac{1}{2}\,\bigl[\hspace{0.03cm}a^{+}_{1}a^{-}_{2}a^{-}_{1}a^{+}_{2} - a^{+}_{1}a^{-}_{1}a^{-}_{2}a^{+}_{2} - a^{-}_{1}a^{+}_{1}a^{+}_{2}a^{-}_{2}
+ a^{-}_{2}a^{+}_{1}a^{+}_{2}a^{-}_{1} \,+
\]
\[
+\,\bigl(\hspace{0.01cm}\{a^{+}_{1},a^{-}_{1}\} + \{a^{+}_{2},a^{-}_{2}\}\bigr)\hspace{0.02cm}\bigr].
\]
Substituting this expression and (\ref{eq:7y}) into (\ref{eq:7t}), further we obtain
\[
\{\hspace{0.02cm}N_{12},N_{21}\} - \{\hspace{0.02cm}L_{12},M_{12}\}
=
\]
\[
\begin{split}
&=
-\frac{1}{2}\,\bigl[\hspace{0.03cm}a^{+}_{1}a^{-}_{1}a^{+}_{2}a^{-}_{2} +
a^{-}_{1}a^{+}_{1}a^{-}_{2}a^{+}_{2} + a^{+}_{1}a^{-}_{1}a^{-}_{2}a^{+}_{2}
+ a^{-}_{1}a^{+}_{1}a^{+}_{2}a^{-}_{2}
-
2\hspace{0.03cm}\bigl(\hspace{0.01cm}\{a^{+}_{1},a^{-}_{1}\} + \{a^{+}_{2},a^{-}_{2}\}\bigr)\hspace{0.02cm}\bigr] = \\[1ex]
&=
-\frac{1}{2}\,\bigl[\hspace{0.03cm}\{a^{+}_{1},a^{-}_{1}\}\{a^{+}_{2},a^{-}_{2}\}
- 2\hspace{0.03cm}\{a^{+}_{1},a^{-}_{1}\} - 2\hspace{0.03cm}\{a^{+}_{2},a^{-}_{2}\}\hspace{0.02cm}\bigr]
= \\[1ex]
&=
-2\Bigl(\hspace{0.03cm}\frac{1}{2}\,\{a^{+}_{1},a^{-}_{1}\} - 1\Bigr)\!
\Bigl(\hspace{0.03cm}\frac{1}{2}\,\{a^{+}_{2},a^{-}_{2}\} - 1\Bigr) + 2.
\end{split}
\]
Thus, we finally have
\[
\Bigl(\hspace{0.03cm}\frac{1}{2}\,\{a^{+}_{1},a^{-}_{1}\} - 1\Bigr)\!
\Bigl(\hspace{0.03cm}\frac{1}{2}\,\{a^{+}_{2},a^{-}_{2}\} - 1\Bigr)
=
-\frac{1}{2}\,\bigl(\hspace{0.03cm}\{\hspace{0.02cm}N_{12},N_{21}\} - \{\hspace{0.02cm}L_{12},M_{12}\}\bigr) + 1.
\]
With allowance made for the last relation and the definition (\ref{eq:7q}), the Harish-Chandra operator (\ref{eq:6o}) can be cast in a rather compact and visual form
\begin{equation}
\hat{\omega}^{\,2} = \frac{3}{4}\,a^{\hspace{0.02cm}2}_{0}
+
\frac{1}{8}\;\bigl(\hspace{0.01cm}\widehat{C}_{2} - 6\hspace{0.02cm}\bigr)
\label{eq:7u}
\end{equation}
or making use of the representation for the $a_{0}^{2}$ in terms of the $G$-parity operator, Eq.\,(I.11.13), it can be also written as
\[
\hat{\omega}^{\,2} = \frac{3}{8}\,[\hspace{0.03cm}1\pm (-1)^{n}\hspace{0.03cm}]
+
\frac{1}{8}\,\bigl(\hspace{0.01cm}\widehat{C}_{2} - 6\hspace{0.02cm}\bigr).
\]
\indent The Casimir operator $\hat{C}_2$ (and correspondingly, $\hat{C}_{2}^{\prime}$) can be presented as a polynomial of the operator $\hat{\Lambda}$. For this purpose, it is easy to use the formulae (I.8.3)\,--\,(I.8.6) from the Part I without using the relation (\ref{eq:6q}). Based on these formulae we have
\begin{equation}
\hat{\omega}^{\,2} = \frac{3}{4}\,a^{\hspace{0.02cm}2}_{0}
+
\frac{\!1}{4^{\hspace{0.02cm}3}}\,\bigl[\hspace{0.01cm}-(2\hspace{0.01cm}\hat{B} - 4\hspace{0.02cm})^{\hspace{0.01cm}2} + 4^{\hspace{0.02cm}2}\hspace{0.03cm}\bigr].
\label{eq:7i}
\end{equation}
Here, the operator $\hat{B}\equiv\hat{\beta}_{\mu}\hat{\beta}_{\mu}$ by the relations of the (\ref{eq:6y}) type and the definition (\ref{eq:7r}) equals
\begin{equation}
\hat{B} = \frac{1}{2}\,\hat{\Lambda}.
\label{eq:7o}
\end{equation}
Comparing the last terms in (\ref{eq:7u}) and (\ref{eq:7i}), we can easily find the required representation
\begin{equation}
\widehat{C}_{2} = -\frac{1}{2}\,\hat{\Lambda}\bigl(\hspace{0.01cm}\hat{\Lambda} - 8\bigr),
\label{eq:7p}
\end{equation}
and as a consequence of (\ref{eq:7e}), we obtain
\begin{equation}
\hspace{0.2cm}
\widehat{C}^{\hspace{0.03cm}\prime}_{2} = -\frac{1}{2}\,\hat{\Lambda}\bigl(\hspace{0.01cm}\hat{\Lambda} - 10\bigr).
\label{eq:7a}
\end{equation}
We write out the rules of action of the Casimir operator $\hat{C}_2$ on the vector states (I.6.1). They can be easily obtained by calculations identical with those performed by us in section 6 of Part I for the operator $a_0$:
\begin{equation}
\begin{array}{lll}
&\widehat{C}_{2}|\hspace{0.03cm}0\rangle = 8|\hspace{0.03cm}0\rangle, \\[1ex]
&\widehat{C}_{2}|\hspace{0.03cm}1\rangle = 6 |\hspace{0.03cm}1\rangle,
&\widehat{C}_{2}|\hspace{0.03cm}2\rangle = 6 |\hspace{0.03cm}2\rangle,
\\[1ex]
&\widehat{C}_{2}|\hspace{0.03cm}11\rangle = 8 |\hspace{0.03cm}11\rangle,
&\widehat{C}_{2}|\hspace{0.03cm}22\rangle = 8 |\hspace{0.03cm}22\rangle,\\[1ex]
&\widehat{C}_{2}|\hspace{0.03cm}12\rangle = 8 |\hspace{0.03cm}12\rangle,
&\widehat{C}_{2}|\hspace{0.03cm}21\rangle = 8 |\hspace{0.03cm}21\rangle,
\\[1ex]
&\widehat{C}_{2}|\hspace{0.03cm}112\rangle = 6 |\hspace{0.03cm}112\rangle,
\hspace{-1cm}
&\widehat{C}_{2}|\hspace{0.03cm}221\rangle = 6 |\hspace{0.03cm}221\rangle,
\\[1ex]
&\widehat{C}_{2}|\hspace{0.03cm}1122\rangle = 8 |\hspace{0.03cm}1122\rangle.
\end{array}
\label{eq:7s}
\end{equation}
Making use of these rules\footnote{\hspace{0.03cm}The eigenvalues obtained in the right-hand side of (\ref{eq:7s}) are in perfect agreement with the conclusions of Bracken and Green \cite{bracken_1972}. In notations of the authors the quadratic Casimir operator $\widehat{\sigma}_2(2M)$ of the $SO(2M)$ group (which should be considered as a subgroup of the $SO(2M+1)$ group) has the following eigenvalues for arbitrary $M$ and $p$ 
\[
\widehat{\sigma}_{2}(2M) = \bigl[\hspace{0.03cm}p\hspace{0.03cm}M\bigl(M +\frac{1}{2}\,p - 1\bigr) 
- 
2\hspace{0.02cm}q^{\prime}\bigl(p - q^{\prime}\bigr)\bigr]\hspace{0.01cm}\hat{I}, 
\]
where $q^{\prime}=0, 1,\, \ldots\, , p$. For our specific case $M = 2$ and $p = 2$, it immediately follows that
\[
\widehat{\sigma}_{2}(4) = \left\{
\begin{array}{rl}
\!\!8,\! & \mbox{if } q^{\prime} = 0, 2\\
\!\!6,\! & \mbox{if } q^{\prime} = 1.
\end{array}
\right.
\]
}, it is not difficult to see that action of the Harish-Chandra operator $\hat{\omega}^2$ in the representation (\ref{eq:7u}) on the state vectors is exactly the same as the rules of action of the Geyer operator $a_0^2$, Eq.\,(I.6.2). By this means in the usual Fock space these two operators are fully equivalent. This equivalence breaks down if we introduce the so-called {\it generalized state-vector space} \cite{ohnuki_1980}. We will discuss this fact in the next section, but here, in the remainder of this section we analysed some properties of the Casimir operator $C_{2}^{\prime}$.\\
\indent Let us define a connection of the Casimir operator $C_{2}^{\prime}$ for the $SO(2M+1)$ group (for $M = 2$) with $\theta$-element of the centre of the DKP-algebra \cite{harish-chandra_1947}. The element of the centre, $\theta$, in the matrix representation is expressed in terms of the matrix $B$ with the help of relation
\begin{equation}
\theta = B\hspace{0.03cm}(\hspace{0.02cm}5 - B\hspace{0.02cm}).
\label{eq:7d}
\end{equation}
In the operator representation we have the connection (\ref{eq:7o}) and consequently $\theta$ takes the form
\[
\hat{\theta} = \frac{1}{4}\,\hat{\Lambda}\bigl(\hspace{0.01cm}10 - \hat{\Lambda}\bigr).
\]
Comparing this expression with (\ref{eq:7a}), we derive the required relation
\begin{equation}
\hspace{0.2cm}
\widehat{C}^{\hspace{0.03cm}\prime}_{2} = 2\hspace{0.03cm}\hat{\theta}.
\label{eq:7f}
\end{equation}
In accordance with the Harish-Chandra approach the matrix $\theta$ (for $D = 4$) has the eigenvalues 6 and therefore by virtue of the fact that $\theta$ is in the centre of the DKP-algebra, it must be put
\begin{equation}
\hspace{0.2cm}
\theta = 6\hspace{0.01cm}I \equiv \theta_{0},
\label{eq:7g}
\end{equation}
where $I$ is the unit matrix. In an operator formalism it goes in the relation
\[
\hat{\theta} = 6\hspace{0.01cm}\hat{I},
\]
where in turn $\hat{I}$ is the unit operator. As a consequence, in view of (\ref{eq:7f}) the Casimir operator for $SO(5)$ group is equal to
\begin{equation}
\hspace{0.2cm}
\widehat{C}^{\hspace{0.03cm}\prime}_{2} = 12\hspace{0.01cm}\hat{I}.
\label{eq:7h}
\end{equation}
In fact it is precisely this relation\footnote{\hspace{0.03cm}This relation is in agreement again with the conclusions of the paper \cite{bracken_1972}. The quadratic Casimir operator $\widehat{\sigma}_2(2M+1)$ for arbitrary $M$ and $p$ has the following eigenvalue:
\[
\widehat{\sigma}_{2}(2M + 1) = p\hspace{0.03cm}M\bigl(M +\frac{1}{2}\,p\bigr)\hspace{0.02cm}\hat{I}, 
\]
whence, for $M=2$ and $p=2$, it follows that
\[
\widehat{\sigma}_{2}(5) = 12\hspace{0.02cm}\hat{I}.
\]
In addition we note also that the invariant $\hat{q}$ introduced in the paper \cite{bracken_1972} in analysis of the reduction of the orthogonal group $SO(2M+1)$ with respect to the unitary group $U(M)$ is simply related to the operator $\hat{\Lambda}$. From a general formula \cite{bracken_1972}
\[
\widehat{\sigma}_{2}(2M + 1) - \widehat{\sigma}_{2}(2M) = 2\hspace{0.03cm}\hat{q}\hspace{0.03cm}
(p - \hat{q}) + p\hspace{0.03cm}M,
\]
in our specific case with allowance made for the representations (\ref{eq:7p}) and (\ref{eq:7a}) it follows that
\[
2\hspace{0.03cm}\hat{q}\hspace{0.03cm}(2 - \hat{q}) + 4 = \hat{\Lambda}.
\]
By this means the invariant $\hat{q}$ cannot be expressed as a rational function of $\hat{\Lambda}$. 
} that is the primary source of disagreement between two expressions (\ref{eq:5a}) and (\ref{eq:5s}).\\
\indent Let us consider action of the operator $\hat{C}_{2}^{\prime}=\hat{C}_2 + \hat{\Lambda}$ on the vector states. Action of the operator $\hat{C}_2 $ is defined by the formulae (\ref{eq:7s}), and action of the operator $\hat{\Lambda}$ is easily defined from its definition (\ref{eq:7r}):
\[
\hat{\Lambda}|\hspace{0.03cm}0\rangle = 4|\hspace{0.03cm}0\rangle,
\quad
\hat{\Lambda}|\hspace{0.03cm}1\rangle = 6|\hspace{0.03cm}1\rangle,
\quad
\hat{\Lambda}|\hspace{0.03cm}11\rangle = 4|\hspace{0.03cm}11\rangle,\,
\ldots\,,
\]
i.e. action of $\hat{\Lambda}$ is reduced to multiplying by 4 or 6 depending on evenness of the number of parafermions in a particular state. Then for the Casimir operator $\hat{C}_{2}^{\prime}= \hat{C}_2 + \hat{\Lambda}$ we  have
\[
\widehat{C}^{\hspace{0.03cm}\prime}_{2}|\hspace{0.03cm}0\rangle = 12|\hspace{0.03cm}0\rangle,
\quad
\widehat{C}^{\hspace{0.03cm}\prime}_{2}|\hspace{0.03cm}1\rangle = 12|\hspace{0.03cm}1\rangle,
\quad
\widehat{C}^{\hspace{0.03cm}\prime}_{2}|\hspace{0.03cm}11\rangle = 12|\hspace{0.03cm}11\rangle, \;\ldots\,,
\]
in a perfect agreement with the formula (\ref{eq:7h}).\\
\indent Let us calculate a matrix element of the operator $\hat{C}_{2}^{\prime}$ in the basis of parafermion coherent states. It is easy to determine the matrix element for the Casimir operator $\hat{C}_{2}$ based on the expression (\ref{eq:7u}). Taking into account that the matrix element of the Harish-Chandra operator $\hat{\omega}^{\hspace{0.02cm}2}$ is defined by the expression (\ref{eq:5a}), and the matrix element of the Geyer operator $a_{0}^{2}$ is defined by expression (\ref{eq:5s}), from (\ref{eq:7u}) we derive
\begin{equation}
\langle\hspace{0.02cm}\bar{\xi}^{\,\prime}\hspace{0.02cm}|\,
\widehat{C}_{2}|\,\xi\hspace{0.03cm}\rangle
 =
\label{eq:7j}
\end{equation}
\[
=
2\hspace{0.02cm}\biggl[\hspace{0.02cm}2\hspace{0.02cm}\biggl\{\!
\biggl(\displaystyle\frac{1}{2}\;[\hspace{0.03cm}\bar{\xi}^{\,\prime}_{1}, \xi^{\phantom{\prime}}_{2}\hspace{0.03cm}]\biggr)\!
\biggl(\displaystyle\frac{1}{2}\;[\hspace{0.03cm}\bar{\xi}^{\,\prime}_{2}, \xi^{\phantom{\prime}}_{1}\hspace{0.03cm}]\biggr)
-
\biggl(\displaystyle\frac{1}{2}\;[\hspace{0.03cm}\bar{\xi}^{\,\prime}_{1}, \bar{\xi}^{\,\prime}_{2}\hspace{0.03cm}]\biggr)\!
\biggl(\,\displaystyle\frac{1}{2}\;[\hspace{0.03cm}\xi^{\phantom{\prime}}_{1}, \xi^{\phantom{\prime}}_{2}\hspace{0.03cm}]\biggr)\!\biggr\}
+
(x^{\hspace{0.02cm}2} + y^{\hspace{0.02cm}2}) - (x + y) + 4\hspace{0.02cm}\biggr]
\langle\hspace{0.02cm}\bar{\xi}^{\,\prime}\hspace{0.02cm}|\,\xi\hspace{0.02cm}\rangle.
\]
Incidentally, this expression can be obtained by a direct calculation starting from the definition (\ref{eq:7q}) with the use of the results in section 7 of Part I. The matrix element of the operator $\hat{\Lambda}$ equals
\begin{equation}
\langle\hspace{0.02cm}\bar{\xi}^{\,\prime}\hspace{0.02cm}|\hspace{0.04cm}
\hat{\Lambda}|\,\xi\hspace{0.03cm}\rangle
 =
\bigl(\hspace{0.01cm}\{\hspace{0.02cm}
\bar{\xi}^{\,\prime}_{1},\xi^{\phantom{\prime}}_{1}\}
+
\{\hspace{0.02cm}\bar{\xi}^{\,\prime}_{2},\xi^{\phantom{\prime}}_{2}\} + 4\hspace{0.02cm}\bigr) \langle\hspace{0.02cm}\bar{\xi}^{\,\prime}\hspace{0.02cm}|\,\xi\hspace{0.02cm}\rangle.
\label{eq:7k}
\end{equation}
Adding the last two expressions, we obtain that in the basis of parafermion coherent states the following non-equality takes place
\[
\langle\hspace{0.02cm}\bar{\xi}^{\,\prime}\hspace{0.02cm}|\,
\widehat{C}^{\hspace{0.03cm}\prime}_{2}|\,\xi\hspace{0.03cm}\rangle
\neq 12\hspace{0.03cm}
\langle\hspace{0.02cm}\bar{\xi}^{\,\prime}\hspace{0.02cm}|\,\xi\hspace{0.02cm}\rangle,
\]
and in such a manner the relation (\ref{eq:7h}) is not true within the framework of generalized state-vector space.

%%%%%%%%%%%%%%%%%%%%%%%% section 8 %%%%%%%%%%%%%%%%%%%%%%%%%%%%

\section{Ohnuki and Kamefuchi's generalized state-vector space}
\setcounter{equation}{0}
\label{section_9}

As was noted for the first time by Ohnuki and Kamefuchi \cite{ohnuki_1980} {\it ``The introduction of para-Grassmann numbers into the framework of theory naturally necessitates a corresponding generalization of the state-vector space 
$\mathfrak{A}$.''} This state-vector space is usually spanned by state vectors such as
\[
{\cal M} (a_{i}^{+}, a_{j}^{+},\,\ldots\,) \vert\hspace{0.03cm}0\rangle,
\]
where ${\cal M}$ is a monomial in the creation operators $a_{k}^{+}$, which in our case is defined by the expressions (I.6.1). The use of the para-Grassmann numbers $\xi_k$ results in the fact that we have to allow such ${\cal M}^{\prime}$s to contain as well $\xi^{\prime}$s. We have already faced with this situation in section 11 of Part I in analysis of the structure of the parafermion coherent state $|\,\xi\hspace{0.02cm}\rangle$. There was shown that some terms of the expansion of the coherent state in powers of $[\hspace{0.02cm}\xi, a^{+}]$ under the action on the vacuum state in principle are not reduced to the expansion in the state vectors (I.6.1).\\
\indent Instead of the usual state-vector space $\mathfrak{A}$ now we should consider a generalized state-vector space $\mathfrak{A}_{\hspace{0.02cm}G}$, which is spanned by ket vectors such as
\[
{\cal M} (a_{i}^{+}, a_{j}^{+},\ldots,\xi_{k},\ldots)\vert\hspace{0.03cm}0\rangle.
\]
By this means the Harish-Chandra operator (\ref{eq:7u}) and the Geyer operator (I.B.17) are equivalent in the space 
$\mathfrak{A}$ and are not equivalent in the enlarged space $\mathfrak{A}_{\hspace{0.02cm}G}$. The accounting this circumstance results in an appreciable complication of the matrix elements of the operator expressions containing the Geyer operator $a_{0}^{2}$.\\
\indent In principle, the complication of this kind can be avoided if instead of the parafermion coherent state of the form
\[
|\,\xi\hspace{0.02cm}\rangle
=
{\rm e}^{\textstyle-\frac{1}{2}\sum_{l}\hspace{0.03cm}[\hspace{0.02cm}
\xi^{\phantom{+}\!}_{l}, a^{+}_{l}\hspace{0.02cm}]}
|\hspace{0.03cm}0\rangle
\]
one uses the coherent state admitting an expansion in the number basis as it takes place in the case of the usual Fermi (and Bose) statistics. An example of such a coherent state for the case of a single-mode parafermi system can be found in the paper by Jing and Nelson \cite{jing_1999}. 
In this paper the para-Fermi eigenstate of annihilation operator $a^{-}$ is presented in the form of the expansion in the number basis
\[
|\,n\hspace{0.02cm}\rangle
=
\frac{1}{\sqrt{\{n\}\hspace{0.02cm}!}}\,(a^{+})^{\hspace{0.02cm}n} |\hspace{0.03cm}0\rangle,
\]
such that $\hat{n}_f|\hspace{0.03cm}n\rangle = n|\hspace{0.03cm}n\rangle$, where the parafermi number operator is
\[
\hat{n}_{f\,} = \displaystyle\frac{1}{2}\;[\hspace{0.03cm}a^{+}, a^{-}\hspace{0.02cm}] +
\frac{1}{2}\,p
\]
and
\[
\{n\} \equiv n\hspace{0.02cm}(p+1-n), \quad \{n\}! = \{n\}\{n - 1\} \ldots \{1\} = \frac{n!\hspace{0.02cm} p\hspace{0.03cm}!}{(p-n)!}.
\]
Note that the coefficient $\{n\}$ coincides with the coefficient in the differentiation formula of para-Grassmann number $\xi^{n}$, Eq.\,(I.C.11). 
The para-Fermi coherent state has the following form:
\begin{equation}
|\,(\xi)_{p}\hspace{0.02cm}\rangle = \sum\limits^{p}_{n = 0} |\,n\hspace{0.02cm}\rangle
\frac{\xi^{\hspace{0.02cm}n}}{\sqrt{\{n\}\hspace{0.02cm}!}}.
\label{eq:8q}
\end{equation}
It obeys the relation $a^{-}|\,(\xi)_{p}\hspace{0.02cm}\rangle = |\,(\xi)_{p}\hspace{0.02cm}\rangle\hspace{0.03cm}\xi$ and has the overlap function
\[
\langle\hspace{0.025cm}(\bar{\xi}^{\,\prime})_{p}\hspace{0.02cm}
|\hspace{0.02cm}(\xi)_{p}\hspace{0.02cm}\rangle
=
\sum\limits^{p}_{n = 0}
\frac{\!1}{\{n\}\hspace{0.01cm}!}\,(\bar{\xi}^{\,\prime})^{\hspace{0.01cm}n}    \hspace{0.02cm}(\xi)^{\hspace{0.02cm}n}.
\]
In particular, for the special case $p = 2$ the expression (\ref{eq:8q}) takes the form
\[
|\,(\xi)_{2}\hspace{0.02cm}\rangle =  |\,0\hspace{0.02cm}\rangle +
\frac{1}{\sqrt{2}} |\,1\hspace{0.02cm}\rangle\hspace{0.03cm}\xi +
\frac{1}{2} |\,1\hspace{0.02cm}\rangle\hspace{0.03cm}\xi^{\hspace{0.02cm}2}.
\]
Making use of the parafermion state (\ref{eq:8q}) (and its generalization to two-mode para-Fermi system) enables one to avoid the introduction of the generalized 
state-vector space $\mathfrak{A}_{\hspace{0.02cm}G}$, however, in so doing the usual exponential representation both of the coherent state and, correspondingly of 
the overlap function rather convenient for the construction of the path integral representation, is lost.

%%%%%%%%%%%%%%%%%%%%%%%% section 9 %%%%%%%%%%%%%%%%%%%%%%%%%%%%

\section{Another representation of Harish-Chandra operator $\hat{\omega}^{2}$}
\setcounter{equation}{0}
\label{section_9}

We define the expression for Harish-Chandra operator $\hat{\omega}^2$, Eq.\,(\ref{eq:7u}) in terms of the Geyer operator $a_{0}^{2}$ and the Casimir operator $\hat{C}_2$ for the orthogonal $SO(4)$ group. Let us define a representation for this operator in terms of the Casimir operator $\hat{C}_2^{\prime}$ for $SO(5)$ group and operator $\hat{\Lambda}$. For this purpose, we will need the formula (68) from the paper \cite{harish-chandra_1947}. In the special case $M = 2$ this formula (in matrix representation) for the non-normalized pseudo-matrix $\omega^2$ gives us
\begin{equation}
\omega^{\hspace{0.02cm}2} = 4\hspace{0.02cm}\bigl\{2\hspace{0.02cm}B_{2} - 2\hspace{0.02cm}B_{3} + B_{4}\bigr\},
\label{eq:9q}
\end{equation}
where the matrices $B_k$ are recurrently defined by the matrix $B$ (see Eq.\,\eqref{ap:C8})
\begin{equation}
B_{2} = B\hspace{0.02cm}(B - 1),\quad  B_{3} = B_{2}\hspace{0.02cm}(B - 2),\quad  B_{4} = B_{3}\hspace{0.02cm}(B - 3).
\label{eq:9w}
\end{equation}
It should be noted that the formula (68) in \cite{harish-chandra_1947} was given with wrong number coefficients of the matrices $B_k$. In particular, for the case $M = 2$ the coefficient 2 of $B_{2}$ is absent (in (\ref{eq:9q}) we give the correct expression). The formula (68) in Harish-Chandra's original text further is not used with the exception of the following formula (69) and therefore all subsequent expressions are true. However, the following formula (69) from \cite{harish-chandra_1947} will be incorrect if one uses (68). All of these questions we discuss in detail in Appendix C, where a correct expression for (68) is given.\\
\indent Let us use the definition of the element of the centre $\theta$, Eq.\,(\ref{eq:7d}) to exclude in (\ref{eq:9w}) all terms nonlinear in the matrix B, then
\[
\begin{split}
&B_{2} = -\theta + 4\hspace{0.01cm}B, \\[1ex]
&B_{3} = -2\hspace{0.02cm}\theta - B\hspace{0.02cm}(\hspace{0.02cm}\theta - 12), \\[1ex]
&B_{4} = \theta^{2} - 6\hspace{0.03cm}\theta - B\hspace{0.02cm}(4\hspace{0.02cm}\theta - 24).
\end{split}
\]
Putting these expressions into (\ref{eq:9q}) yields
\begin{equation}
\omega^{\hspace{0.02cm}2} = 4\hspace{0.02cm}\bigl[P(\theta) + Q(\theta)B\hspace{0.02cm}\bigr],
\label{eq:9e}
\end{equation}
where $P(\theta) \equiv \theta^2 - 4\hspace{0.02cm}\theta$ and $Q(\theta) \equiv -2\hspace{0.02cm}(\theta - 4)$. We eliminate the term $\theta^{\hspace{0.02cm}2}$ from the function $P(\theta)$ using the minimal equation for $\theta$:
\[
(\hspace{0.02cm}\theta - 4)(\hspace{0.02cm}\theta - 6) = 0,
\]
then $P(\theta) = 6\hspace{0.02cm}(\theta - 4)$, and the expression (\ref{eq:9e}) takes the form
\begin{equation}
\omega^{\hspace{0.02cm}2} = \frac{1}{2}\,(\hspace{0.02cm}\theta - 4)\hspace{0.02cm}(\hspace{0.02cm}3 - B\hspace{0.02cm}).
\label{eq:9r}
\end{equation}
Here, we have come to the normalized pseudo-matrix in accordance with the rule: $\omega \rightarrow 4\hspace{0.03cm}\omega$. The expression (\ref{eq:9r}) correctly reproduces the formula (70) from \cite{harish-chandra_1947} suggested by Harish-Chandra from general reasoning. In particular, when we fix the element of the centre $\theta$, Eq.\,(\ref{eq:7g}), from (\ref{eq:9r}) follows (\ref{eq:6q}). The use of the original formula (68) (see \eqref{ap:C2} in Appendix C) leads to the improper expressions (\ref{eq:9r}) and (\ref{eq:6q}).\\
\indent For the operator formulation of relation (\ref{eq:9r}) we make use of the formulae of connection (\ref{eq:7f}) and (\ref{eq:7o}) that give us the desired representation for the Harish-Chandra operator
\[
\hat{\omega}^{\hspace{0.02cm}2} = \frac{1}{8}\,
(\hspace{0.02cm}\widehat{C}^{\hspace{0.03cm}\prime}_{2} - 8)\hspace{0.02cm} (\hspace{0.02cm}6 - \hat{\Lambda}\hspace{0.02cm}).
\]
If one recalls the representation of the Casimir operator  $\hat{C}_2^{\prime}$ in terms of $\hat{\Lambda}$, Eq.\,(\ref{eq:7a}), then the preceding expression can be put into another form
\begin{equation}
\hat{\omega}^{\hspace{0.02cm}2} = \frac{1}{16}\,
(\hspace{0.02cm}\hat{\Lambda} - 8)\hspace{0.02cm} (\hspace{0.02cm}\hat{\Lambda} - 6\hspace{0.02cm})(\hspace{0.02cm}\hat{\Lambda} - 2\hspace{0.02cm})
\label{eq:9t}
\end{equation}
and thus the operator $\hat{\omega}^{2}$ represents a polynomial of the third order in $\hat{\Lambda}$. As a consequence of (\ref{eq:9t}), (\ref{eq:7u}) and (\ref{eq:7p}), we obtain a similar representation for the Geyer operator $a_0^2$:
\begin{equation}
a^{\hspace{0.02cm}2}_{0} = \frac{1}{12}\,
(\hspace{0.02cm}\hat{\Lambda} - 7)\hspace{0.02cm} (\hspace{0.02cm}\hat{\Lambda} - 6\hspace{0.02cm})(\hspace{0.02cm}\hat{\Lambda} - 2\hspace{0.02cm}).
\label{eq:9y}
\end{equation}
By this mean for the $a^{\hspace{0.02cm}2}_{0}$ operator we have the third representation for this case in terms of $\hat{\Lambda}$. The first representation is the original expression of Geyer, Eq.\,(I.B.17), and the second one was written out in terms of the fermion number counter $(-1)^{n}$, Eq.\,(I.11.13).\\
\indent By virtue of the obtained expressions (\ref{eq:7o})\,--\.(\ref{eq:7a}), (\ref{eq:9t}) and (\ref{eq:9y}) one can state a question on the possibility of the representation of the operator $\hat{\omega}\hspace{0.03cm}(\equiv - a_{0})$ as a function of $\hat{\Lambda}$. However, analysis showed that the representation of the $\hat{\omega}$ as a polynomial of the second and the third orders in $\hat{\Lambda}$ is incompatible with the property
\[
\hat{\omega}^{\hspace{0.02cm}3} = \hat{\omega},
\]
i.e. we result either in a contradiction or in the trivial case $\hat{\omega} \equiv 0$. In principle, that should be expected since $\hat{\omega}$ is a {\it pseudo-scalar} operator as distinct from the other operators.\\
\indent In the remainder of this section, we would like to return to analysis of the integral convolution (\ref{eq:5w}). As was discussed in section 5, we proved the validity of (\ref{eq:5w}) in Part I (sections 7 and 12), where as the function $\tilde{\Omega}\hspace{0.02cm}(\bar{\xi}^{\prime}, \xi)$ we had taken an expression from the matrix element for the Geyer operator $a_0^2$, Eqs.\,(\ref{eq:2e}) and (\ref{eq:2r}). However, in the extended state-vector space 
$\mathfrak{A}_{\hspace{0.02cm}G}$ instead of the operator $a_{0}^{2}$ we must use the operator $\hat{\omega}^2\equiv (a_{0})^{2}$, whose matrix element by virtue of (\ref{eq:7u}) has the form
\begin{equation}
\langle\hspace{0.02cm}\bar{\xi}^{\,\prime}\hspace{0.02cm}|\,
\hat{\omega}^{\hspace{0.02cm}2}|\,\xi\hspace{0.02cm}\rangle
=
\widetilde{\widetilde{\Omega}}\hspace{0.03cm}(\bar{\xi}^{\,\prime},\xi)\hspace{0.03cm}
\langle\hspace{0.02cm}\bar{\xi}^{\,\prime}\hspace{0.02cm}|\,\xi\hspace{0.02cm}\rangle,
\label{eq:9u}
\end{equation}
where now
\begin{equation}
\widetilde{\widetilde{\Omega}}\hspace{0.03cm}(\bar{\xi}^{\,\prime},\xi)
=
\frac{3}{4}\:\widetilde{\Omega}\hspace{0.03cm}(\bar{\xi}^{\,\prime},\xi)
+
\frac{1}{8}\,(\hspace{0.02cm}{\cal C}_{2}\hspace{0.02cm}(\bar{\xi}^{\,\prime},\xi) - 6\hspace{0.02cm}).
\label{eq:9i}
\end{equation}
We define the function ${\cal C}_{2}(\bar{\xi}^{\prime},\xi)$ as usually by the relation
\[
\langle\hspace{0.02cm}\bar{\xi}^{\,\prime}\hspace{0.02cm}|\,
\widehat{C}_{2}|\,\xi\hspace{0.03cm}\rangle
=
{\cal C}_{2}\hspace{0.02cm}(\bar{\xi}^{\,\prime},\xi)\hspace{0.03cm}
\langle\hspace{0.02cm}\bar{\xi}^{\,\prime}\hspace{0.02cm}|\,\xi\hspace{0.02cm}\rangle.
\]
An explicit form of ${\cal C}_2(\bar{\xi}, \xi)$ is given by the expression (\ref{eq:7j}). Instead of (\ref{eq:5w}) we need to prove the validity of the relation
\begin{equation}
\Omega\hspace{0.03cm}(\bar{\xi}^{\,\prime},\xi)
=
\Omega\hspace{0.03cm}(\bar{\xi}^{\,\prime},\xi)\! *\hspace{0.02cm}
\widetilde{\widetilde{\Omega}}\hspace{0.03cm}(\bar{\xi}^{\,\prime},\xi).
\label{eq:9o}
\end{equation}
For the first term on the right-hand side of (\ref{eq:9i}) the relation (\ref{eq:5w}) is still true and therefore instead of (\ref{eq:9o}) we can write
\[
\Omega\hspace{0.03cm}(\bar{\xi}^{\,\prime},\xi)
=
\frac{1}{2}\;\Omega\hspace{0.03cm}(\bar{\xi}^{\,\prime},\xi)\! *\hspace{0.03cm}
(\hspace{0.02cm}{\cal C}_{2}\hspace{0.02cm}(\bar{\xi}^{\,\prime},\xi) - 6\hspace{0.02cm})
\]
or
\begin{equation}
\Omega\hspace{0.03cm}(\bar{\xi}^{\,\prime},\xi)
=
\frac{1}{8}\;\Omega\hspace{0.03cm}(\bar{\xi}^{\,\prime},\xi)\! *\hspace{0.03cm}
{\cal C}_{2}\hspace{0.02cm}(\bar{\xi}^{\,\prime},\xi).
\label{eq:9p}
\end{equation}
However, a straightforward calculation of the convolution in (\ref{eq:9p}) is too cumbersome, consequently we consider the proof of (\ref{eq:9p}) within the operator formalism which is somewhat easier.\\
\indent We exploit  the fact that the right-hand side of (\ref{eq:9p}) can be written as
\begin{equation}
\frac{1}{8}\,\langle\hspace{0.02cm}\bar{\xi}^{\,\prime}\hspace{0.02cm}|\,
a_{0}\hspace{0.03cm}\widehat{C}_{2}|\,\xi\hspace{0.03cm}\rangle
\hspace{0.03cm}
\langle\hspace{0.02cm}\bar{\xi}^{\,\prime}\hspace{0.02cm}|\,\xi\hspace{0.02cm}\rangle^{-1}.
\label{eq:9a}
\end{equation}
In the matrix representation the following relation (see \eqref{ap:C7}) was proved:
\begin{equation}
\omega B = 2\hspace{0.03cm}\omega.
\label{eq:9s}
\end{equation}
In the operator representation it takes the form
\begin{equation}
a_{0}\hspace{0.03cm}\hat{\Lambda} = 4\hspace{0.02cm}a_{0}.
\label{eq:9d}
\end{equation}
Further we use the representation of the Casimir operator $\hat{C}_{2}$ in terms of the operator $\hat{\Lambda}$, Eq.\,(\ref{eq:7p}). Then by virtue of (\ref{eq:9d}) the following is true
\[
a_{0}\hspace{0.03cm}\widehat{C}_{2} = -\frac{1}{2}\,a_{0}\hspace{0.03cm}\hat{\Lambda}^{2}
+
4\hspace{0.02cm}a_{0}\hspace{0.03cm}\hat{\Lambda}
=
\Bigl[-\frac{1}{2}\;4^{\hspace{0.02cm}2} + 4^{\hspace{0.02cm}2}\Bigr]a_{0} = 8\hspace{0.02cm}a_{0}.
\]
The expression (\ref{eq:9a}) in view of the last relation is really equals $\Omega(\bar{\xi}^{\prime}, \xi)$. However, we will need clearly to show that the operator equality (\ref{eq:9d}) indeed takes place.\\
\indent We rewrite the equality (\ref{eq:9d}) in an identity form
\[
\frac{1}{2}\,\{\hspace{0.01cm}a_{0},\hat{\Lambda}\}
+
\frac{1}{2}\,[\hspace{0.03cm}a_{0},\hat{\Lambda}\hspace{0.02cm}]  = 4\hspace{0.02cm}a_{0}.
\]
Let us prove that the following relations are valid
\begin{equation}
\{\hspace{0.01cm}a_{0},\hat{\Lambda}\} = 8\hspace{0.02cm}a_{0},
\quad
[\hspace{0.03cm}a_{0},\hat{\Lambda}\hspace{0.02cm}] = 0.
\label{eq:9f}
\end{equation}
Here, we consider the proof of the first relation, the proof of the second one is given in Appendix D. Substituting an explicit form of the operators $a_{0}$ and $\hat{\Lambda}$ into the first relation in (\ref{eq:9f}), we get
\[
\bigl(\hspace{0.02cm}\{\!\hspace{0.04cm}\{\hspace{0.03cm}L_{12}, M_{12}\hspace{0.02cm}\},\{\hspace{0.02cm}a^{+}_{1},a^{-}_{1}\}\!\hspace{0.04cm}\}
+
\{\!\hspace{0.04cm}\{\hspace{0.03cm}N_{12}, N_{21}\hspace{0.02cm}\},\{\hspace{0.02cm}a^{+}_{1},a^{-}_{1}\}\!\hspace{0.04cm}\}
-
\{\!\hspace{0.04cm}\{\hspace{0.03cm}N_{1}, N_{2}\hspace{0.02cm}\},\{\hspace{0.02cm}a^{+}_{1},a^{-}_{1}\} \!\hspace{0.04cm}\}\hspace{0.01cm}\bigr) + (1 \rightleftarrows 2) =
\]
\begin{equation}
=
8\hspace{0.02cm}\bigl(\hspace{0.02cm}\{\hspace{0.03cm}L_{12}, M_{12}\hspace{0.02cm}\}
+
\{\hspace{0.03cm}N_{12}, N_{21}\hspace{0.02cm}\}
-
\{\hspace{0.03cm}N_{1}, N_{2}\hspace{0.02cm}\}\bigr).
\label{eq:9g}
\end{equation}
We will need two operator identities
\begin{align}
\{C,\{A,B\hspace{0.02cm}\}\!\hspace{0.03cm}\} &= [A,[B,C\hspace{0.03cm}]\hspace{0.03cm}] + \{B,\{A,C\hspace{0.02cm}\}\!\hspace{0.03cm}\},
\label{eq:9h}  \\[1ex]
[\hspace{0.04cm}C,\{A,B\hspace{0.02cm}\}] &= \{B,[\hspace{0.02cm}C,A\hspace{0.02cm}]\} - \{A,[B,C\hspace{0.03cm}]\hspace{0.015cm}\}.
\label{eq:9j}
\end{align}
We consider the first term on the left-hand side of (\ref{eq:9g}). By using the identity (\ref{eq:9h}), we obtain
\begin{equation}
\{\!\hspace{0.04cm}\{\hspace{0.03cm}L_{12}, M_{12}\hspace{0.02cm}\},\{\hspace{0.02cm}a^{+}_{1},a^{-}_{1}\}\!\hspace{0.04cm}\}
=
[\hspace{0.03cm}L_{12},[M_{12},\{\hspace{0.02cm}a^{+}_{1},a^{-}_{1}\}\hspace{0.03cm}]
\hspace{0.03cm}]
+
\{M_{12},\{\!\hspace{0.04cm}\{\hspace{0.02cm}a^{+}_{1},a^{-}_{1}\},L_{12}\hspace{0.02cm}\}
\!\hspace{0.03cm}\}.
\label{eq:9k}
\end{equation}
Further we take advantage of the identity (\ref{eq:9j}). The internal commutator in the first term in (\ref{eq:9k}) is \[
[\hspace{0.03cm}M_{12},\{\hspace{0.02cm}a^{+}_{1},a^{-}_{1}\}\hspace{0.03cm}]
=
\{\hspace{0.02cm}a^{-}_{1},[\hspace{0.02cm}M_{12},a^{+}_{1}\hspace{0.02cm}]\} - \{\hspace{0.02cm}a^{+}_{1},[\hspace{0.04cm}a^{-}_{1},M_{12}\hspace{0.03cm}] \hspace{0.015cm}\}
=
-\{\hspace{0.01cm}a^{-}_{1},a^{-}_{2}\}.
\]
Here, we have used the commutation rules (I.6.15). Then the first term in (\ref{eq:9k})  takes the form
\begin{equation}
-\hspace{0.02cm}[\hspace{0.03cm}L_{12},\{\hspace{0.02cm}a^{-}_{1},a^{-}_{2}\}
\hspace{0.03cm}]
=
-\hspace{0.02cm}\{\hspace{0.02cm}a^{-}_{2},[\hspace{0.02cm}L_{12},a^{-}_{1}\hspace{0.02cm}]
\hspace{0.03cm}\}
+ \{\hspace{0.02cm}a^{-}_{1},[\hspace{0.04cm}a^{-}_{2},L_{12}\hspace{0.03cm}]
\hspace{0.015cm}\}
=
\{\hspace{0.01cm}a^{+}_{2},a^{-}_{2}\} - \{\hspace{0.01cm}a^{+}_{1},a^{-}_{1}\},
\label{eq:9l}
\end{equation}
where we again have used the identity (\ref{eq:9j}) and the rules (I.6.15).\\
\indent Let us analyse now the second term on the right-hand side of (\ref{eq:9k}). We have the following chain of equalities:
\[
\{\!\hspace{0.04cm}\{\hspace{0.02cm}a^{+}_{1},a^{-}_{1}\},L_{12}\hspace{0.02cm}\}
=
[\hspace{0.03cm}a^{+}_{1},[\hspace{0.02cm}a^{-}_{1},L_{12}\hspace{0.03cm}]\hspace{0.04cm}] +
\{\hspace{0.02cm}a^{-}_{1},\{L_{12},a^{+}_{1}\hspace{0.01cm}\}\!\hspace{0.03cm}\}
=
\]
\[
=
[\hspace{0.03cm}a^{+}_{1},a^{+}_{2}\hspace{0.03cm}]
+
\{\hspace{0.02cm}a^{-}_{1},\{L_{12},a^{+}_{1}\hspace{0.01cm}\}\!\hspace{0.03cm}\}
=
2\hspace{0.02cm}L_{12} - \frac{1}{2}\,
\{\hspace{0.02cm}a^{-}_{1},[\hspace{0.03cm}a^{+}_{2},(a^{+}_{1})^{2}\hspace{0.03cm}] \hspace{0.02cm}\}.
\]
Here, at the last step we made transformation with the use of the definition of the generator $L_{12}$, Eq.\,(I.6.9),
\[
\{L_{12},a^{+}_{1}\hspace{0.01cm}\} =
\frac{1}{2}\,\{[\hspace{0.03cm}a^{+}_{1},a^{+}_{2}\hspace{0.03cm}],a^{+}_{1}\hspace{0.01cm}\}
\equiv
-\frac{1}{2}\,[\hspace{0.03cm}a^{+}_{2},(a^{+}_{1})^{2}\hspace{0.03cm}] .
\]
By this means the second term in (\ref{eq:9k}) can be presented in  the following form:
\[
\{M_{12},\{\!\hspace{0.04cm}\{\hspace{0.02cm}a^{+}_{1},a^{-}_{1}\},L_{12}\hspace{0.02cm}\}
\!\hspace{0.03cm}\}
=
2\hspace{0.03cm}\{M_{12}, L_{12}\hspace{0.02cm}\}
-
\frac{1}{2}\,\{M_{12},\{a^{-}_{2}, [\hspace{0.03cm}a^{+}_{2},(a^{+}_{1})^{2}\hspace{0.03cm}]
\hspace{0.02cm}\}\!\hspace{0.03cm}\}.
\]
With the use of this expression and (\ref{eq:9l}) the first contribution on the left-hand side of (\ref{eq:9g}) takes the form
\begin{equation}
\{\!\hspace{0.04cm}\{\hspace{0.03cm}L_{12}, M_{12},\hspace{0.02cm}\},\{\hspace{0.02cm}a^{+}_{1},a^{-}_{1}\}\!\hspace{0.04cm}\}
=
\label{eq:9z}
\end{equation}
\[
=
\bigl(\{\hspace{0.01cm}a^{+}_{2},a^{-}_{2}\} - \{\hspace{0.01cm}a^{+}_{1},a^{-}_{1}\}\bigr)
+
2\hspace{0.03cm}\{M_{12}, L_{12}\hspace{0.02cm}\}
-
\frac{1}{2}\,\{M_{12},\{a^{-}_{1}, [\hspace{0.03cm}a^{+}_{2},(a^{+}_{1})^{2}\hspace{0.03cm}]
\hspace{0.02cm}\}\!\hspace{0.03cm}\}.
\]
\indent Analysis of the second and the third contributions in (\ref{eq:9g}) is performed in accordance with the same scheme, because of this, here we write out only the final expressions:
\vspace{-0.3cm}
\begin{flushleft}
for the second term
\end{flushleft}
\vspace{-0.7cm}
\begin{equation}
\{\!\hspace{0.04cm}\{\hspace{0.03cm}N_{12}, N_{21}\hspace{0.02cm}\},\{\hspace{0.02cm}a^{+}_{1},a^{-}_{1}\}\!\hspace{0.04cm}\}
=
\label{eq:9x}
\end{equation}
\[
=
\bigl(-\{\hspace{0.01cm}a^{+}_{2},a^{-}_{2}\} + \{\hspace{0.01cm}a^{+}_{1},a^{-}_{1}\}\bigr)
+
2\hspace{0.03cm}\{N_{21}, N_{12}\hspace{0.02cm}\}
-
\frac{1}{2}\,\{N_{21},\{a^{-}_{1}, [\hspace{0.03cm}a^{-}_{2},(a^{+}_{1})^{2}\hspace{0.03cm}]
\hspace{0.02cm}\}\!\hspace{0.03cm}\},
\]
\vspace{-0.3cm}
\begin{flushleft}
for the third term
\end{flushleft}
\vspace{-0.9cm}
\begin{equation}
\{\!\hspace{0.04cm}\{\hspace{0.03cm}N_{1}, N_{2}\hspace{0.02cm}\},\{\hspace{0.02cm}a^{+}_{1},a^{-}_{1}\} \!\hspace{0.04cm}\}
=
4\hspace{0.03cm}\{\hspace{0.03cm}N_{1}, N_{2}\hspace{0.02cm}\}.
\label{eq:9c}
\end{equation}
\indent It remains for us to analyse the last terms in (\ref{eq:9z}) and (\ref{eq:9x}). Let us consider the first of them. Making use of the operator identity (\ref{eq:9j}) for the internal anticommutator we have
\[
\{a^{-}_{1}, [\hspace{0.03cm}a^{+}_{2},(a^{+}_{1})^{2}\hspace{0.03cm}]
\hspace{0.02cm}\}
=
\{(a^{+}_{1})^{2}, [\hspace{0.03cm}a^{-}_{1},a^{+}_{2}\hspace{0.03cm}]
\hspace{0.02cm}\}
+
[\hspace{0.03cm}a^{+}_{2},\{(a^{+}_{1})^{2},a^{-}_{1}\hspace{0.02cm}\}\hspace{0.02cm}]
=
-2\hspace{0.03cm}\{(a^{+}_{1})^{2}, N_{21}\} - 4\hspace{0.02cm}L_{12}.
\]
Then this term takes the form
\[
2\hspace{0.03cm}\{M_{12},L_{12}\} + \{M_{12},\{(a^{+}_{1})^{2}, N_{21}\}\!\hspace{0.04cm}\}
=
2\hspace{0.03cm}\{M_{12},L_{12}\} +
[\hspace{0.03cm}(a^{+}_{1})^{2},
[\hspace{0.03cm}N_{21},M_{12}\hspace{0.03cm}]\hspace{0.04cm}]
+
\{N_{21},\{M_{12},(a^{+}_{1})^{2}\hspace{0.02cm}\}\!\hspace{0.025cm}\}.
\]
Here, we have used the identity (\ref{eq:9h}). The second term on the right-hand side of the above expression vanishes by virtue of the algebra (I.6.11) and as a result, we obtain
\[
-
\frac{1}{2}\,\{M_{12},\{a^{-}_{2}, [\hspace{0.03cm}a^{+}_{2},(a^{+}_{1})^{2}\hspace{0.03cm}]
\hspace{0.02cm}\}\!\hspace{0.03cm}\}
=
2\hspace{0.03cm}\{M_{12},L_{12}\}
+
\{N_{21},\{M_{12},(a^{+}_{1})^{2}\hspace{0.02cm}\}\!\hspace{0.03cm}\}.
\]
Completely similar analysis for the last term in (\ref{eq:9x}) leads us to
\[
-
\frac{1}{2}\,\{N_{21},\{a^{-}_{2}, [\hspace{0.03cm}a^{-}_{2},(a^{+}_{1})^{2}\hspace{0.03cm}]
\hspace{0.02cm}\}\!\hspace{0.03cm}\}
=
2\hspace{0.03cm}\{N_{21},N_{12}\}
-
\{N_{21},\{(a^{+}_{1})^{2},M_{12}\hspace{0.02cm}\}\!\hspace{0.03cm}\}
\]
and thus, instead of (\ref{eq:9z}) and (\ref{eq:9x}), now we can write, correspondingly,
\[
\begin{split}
\{\!\hspace{0.04cm}\{\hspace{0.03cm}L_{12}, M_{12}\hspace{0.02cm}\},\{\hspace{0.02cm}a^{+}_{1},a^{-}_{1}\}\!\hspace{0.04cm}\}
&=
\bigl(\hspace{0.01cm}\{\hspace{0.01cm}a^{+}_{2},a^{-}_{2}\} - \{\hspace{0.01cm}a^{+}_{1},a^{-}_{1}\}\bigr)
+
4\hspace{0.03cm}\{M_{12},L_{12}\}
+
\{N_{21},\{M_{12},(a^{+}_{1})^{2}\hspace{0.02cm}\}\!\hspace{0.03cm}\}, \\[1ex]
\{\!\hspace{0.04cm}\{\hspace{0.03cm}N_{12}, N_{21}\hspace{0.02cm}\},\{\hspace{0.02cm}a^{+}_{1},a^{-}_{1}\}\!\hspace{0.04cm}\}
&=
\bigl(\hspace{0.01cm}\{\hspace{0.01cm}a^{+}_{1},a^{-}_{1}\} - \{\hspace{0.01cm}a^{+}_{2},a^{-}_{2}\}\bigr)
+
4\hspace{0.03cm}\{N_{21},N_{12}\}
-
\{N_{21},\{M_{12},(a^{+}_{1})^{2}\hspace{0.02cm}\}\!\hspace{0.03cm}\}.
\end{split}
\]
Putting the last expressions together and subtracting (\ref{eq:9c}) from them, we obtain that the contribution of the first three terms on the left-hand side of (\ref{eq:9g}) equals
\[
4\hspace{0.02cm}\bigl(\hspace{0.01cm}\{L_{12},M_{12}\}  + \{N_{21},N_{12}\} -
\{N_{1},N_{2}\} \bigr).
\]
The terms with the replacement $(1 \rightleftarrows 2)$ give us the remaining half on the left-hand side of (\ref{eq:9g}). In this way we really reproduce the first relation in (\ref{eq:9f}). In fact we have proved even a more weak statement. Let us present the operator $\hat{\Lambda}$ as a sum
\[
\hat{\Lambda} = \hat{\Lambda}_{1} + \hat{\Lambda}_{2},
\]
where $\hat{\Lambda}_1 \equiv \{ a_{1}^{+}, a_{1}^{-} \},\,\hat{\Lambda}_2 \equiv \{ a_{2}^{+}, a_{2}^{-} \}$. Then by virtue of the aforementioned we proved the validity of two independent relations
\begin{equation}
\{\hspace{0.01cm}a_{0},\hat{\Lambda}_{1}\} = 4\hspace{0.02cm}a_{0},
\quad
\{\hspace{0.01cm}a_{0},\hat{\Lambda}_{2}\} = 4\hspace{0.02cm}a_{0}.
\label{eq:9v}
\end{equation}
However, if we recall an existence of the second relation in (\ref{eq:9f}), then in addition to (\ref{eq:9v}) we have to verify the validity of the equalities
\begin{equation}
[\hspace{0.03cm}a_{0},\hat{\Lambda}_{1}\hspace{0.02cm}] = 0,
\quad
[\hspace{0.03cm}a_{0},\hat{\Lambda}_{2}\hspace{0.02cm}] = 0.
\label{eq:9b}
\end{equation}
As we have shown in Appendix D, the relations (\ref{eq:9b}) do not hold separately, and they hold only in a sum.\\
\indent The necessity of using the operator $\hat{\omega}^{2}$ instead of $a_{0}^{2}$ leads to a modification of some expressions with $a_0^2$, which were obtained in Part I. In particular, it is concerned with the matrix elements $\langle\hspace{0.02cm}\bar{\xi}^{\,\prime}\hspace{0.02cm}|\,
[\hspace{0.03cm}a_{0}^{\hspace{0.02cm}2}, a^{\pm}_{n}\hspace{0.02cm}]|\,\xi\hspace{0.02cm} \rangle$ derived in section 9 of part I. It is not difficult to show that now, instead of (I.9.21) and (I.9.22), we must use the equalities
\[
\begin{split}
\langle\hspace{0.02cm}\bar{\xi}^{\,\prime}\hspace{0.02cm}|\,
[\hspace{0.03cm}\hat{\omega}^{\hspace{0.02cm}2}, a^{+}_{n}\hspace{0.02cm}]|\,\xi\hspace{0.02cm}\rangle
&=
-\biggl(\frac{\partial\hspace{0.04cm}\widetilde{\widetilde{\Omega}}}
{\partial\hspace{0.02cm}\xi_{n}}\biggr)
\hspace{0.02cm}
\langle\hspace{0.02cm}\bar{\xi}^{\,\prime}\hspace{0.02cm}|\,\xi\hspace{0.02cm}\rangle \\[1ex]
\langle\hspace{0.02cm}\bar{\xi}^{\,\prime}\hspace{0.02cm}|\,
[\hspace{0.03cm}\hat{\omega}^{\hspace{0.02cm}2}, a^{-}_{n}\hspace{0.02cm}]|\,\xi\hspace{0.02cm}\rangle
&=
-\biggl(\frac{\partial\hspace{0.04cm}\widetilde{\widetilde{\Omega}}}
{\partial\hspace{0.02cm}\bar{\xi}^{\,\prime}_{n}}\biggr) \hspace{0.02cm}
\langle\hspace{0.02cm}\bar{\xi}^{\,\prime}\hspace{0.02cm}|\,\xi\hspace{0.02cm}\rangle,
\end{split}
\]
with the function $\widetilde{\widetilde{\Omega}}(\bar{\xi}^{\hspace{0.02cm}\prime},\xi)$ defined by the formula 
(\ref{eq:9i}). Some other modifications of the formalism under consideration will be discussed below.

%%%%%%%%%%%%%%%%%%%%%%%% section 10 %%%%%%%%%%%%%%%%%%%%%%%%%%%%

\section{Calculation of the commutators $[\hspace{0.02cm}\hat{A}, a^{+}_{k}\hspace{0.02cm}]$
and $[\hspace{0.02cm}\hat{A},[\hspace{0.03cm}a_{0},a^{+}_{k}\hspace{0.02cm}]\hspace{0.03cm}]$}
\setcounter{equation}{0}
\label{section_10}

We determine the commutators of the operator $\hat{A}$ with the creation operator $a_{k}^{+}$ and with the commutator $[\hspace{0.03cm}a_{0},a^{+}_{k}\hspace{0.02cm}]$. The aim of these calculations is to obtain a more compact and visual representation for the matrix element $\langle\hspace{0.02cm}(k)^{\prime}_{p}\hspace{0.02cm}|\hspace{0.03cm} \hat{A}\hspace{0.03cm}\hat{\eta}_{\mu}(z)\hspace{0.02cm} \hat{D}_{\mu}\hspace{0.01cm}|\hspace{0.02cm}(k - 1)_{x}\hspace{0.02cm}\rangle$, instead of the expression (I.10.14). We recall that the awkwardness of (I.10.14) is caused by the necessity to move the operator $a_{k}^{+}$ and the commutator   
$[\hspace{0.03cm}a_{0},a^{+}_{k}\hspace{0.02cm}]$ towards the left of the operator $\hat{A}$, Eqs.\,(I.9.2), (I.10.10). For the calculation of the required commutators we use the exponential representation of the operator $\hat{A}$
\begin{equation}
\hat{A} = \alpha\hspace{0.04cm}
{\rm e}^{\textstyle-i\hspace{0.02cm}\frac{2\pi}{3}\,a^{\phantom{+}}_{0}}.
\label{eq:10q}
\end{equation}
We need the expression for the trilinear commutator (I.B.9), which for convenience of reference, we also give here\footnote{\hspace{0.02cm}We recall that in the case under consideration for parastatistics of order $p = 2$ we have redefined the operator $a_{0}:\,a_{0} \rightarrow 2\hspace{0.02cm}a_{0}.$}
\begin{equation}
[\hspace{0.03cm}a^{\phantom{+}}_{0}\!, [\hspace{0.03cm}a^{\phantom{+}}_{0}\!,a^{+}_{k}\hspace{0.02cm}]\hspace{0.02cm}] = a^{+}_{k}.
\label{eq:10w}
\end{equation}
\indent First we consider the commutator $[\hspace{0.02cm}\hat{A}, a^{+}_{k}\hspace{0.02cm}]$. We rearrange the operator $\hat{A}$ to the right
\[
 \alpha\hspace{0.04cm}[\hspace{0.04cm}{\rm e}^{\textstyle-i\hspace{0.02cm}\frac{2\pi}{3}\, a^{\phantom{+}}_{0}}\!,a^{+}_{k}\hspace{0.02cm}]
 =
\alpha\hspace{0.01cm}\Bigl({\rm e}^{\textstyle-i\hspace{0.02cm}\frac{2\pi}{3}\,a^{\phantom{+}}_{0}}
\hspace{0.02cm}a^{+}_{k}\hspace{0.02cm}
{\rm e}^{\textstyle i\hspace{0.02cm}\frac{2\pi}{3}\,a^{\phantom{+}}_{0}}
-
a^{+}_{k}\Bigr)\hspace{0.02cm}{\rm e}^{\textstyle-i\hspace{0.02cm}\frac{2\pi}{3}\,a^{\phantom{+}}_{0}}
=
\]
\[
\begin{split}
&=
\biggl(\hspace{0.02cm}a^{+}_{k} - \biggl(\frac{2\pi\hspace{0.01cm} i}{3}\biggr)\hspace{0.02cm}
[\hspace{0.03cm}a^{\phantom{+}}_{0}\!,a^{+}_{k}\hspace{0.02cm}]
\,+
\frac{\!1}{2\hspace{0.02cm}!}\hspace{0.02cm}\biggl(\frac{2\pi\hspace{0.01cm} i}{3}\biggr)^{\!\!2\,}
[\hspace{0.03cm}a^{\phantom{+}}_{0}\!, [\hspace{0.03cm}a^{\phantom{+}}_{0}\!,a^{+}_{k}\hspace{0.02cm}]\hspace{0.03cm}]
-\,\ldots\,- a^{+}_{k}\!\biggr)\hspace{0.01cm}\hat{A}
= \\[1ex]
&=
\biggl(- \hspace{0.02cm}\biggl\{\!\biggl(\frac{2\pi\hspace{0.01cm} i}{3}\biggr)
+ \frac{\!1}{3\hspace{0.02cm}!}\hspace{0.02cm}\biggl(\frac{2\pi\hspace{0.01cm} i}{3}\biggr)^{\!\!3\,}
+ \ldots\,\biggr\}\hspace{0.03cm}
[\hspace{0.03cm}a^{\phantom{+}}_{0}\!,a^{+}_{k}\hspace{0.02cm}]
\,+\,
\biggl\{\frac{\!1}{2\hspace{0.02cm}!}\hspace{0.02cm}\biggl(\frac{2\pi\hspace{0.01cm} i}{3}\biggr)^{\!\!2}
+
\frac{\!1}{4\hspace{0.02cm}!}\hspace{0.02cm}\biggl(\frac{2\pi\hspace{0.01cm} i}{3}\biggr)^{\!\!4}
+\,\ldots\,\biggr\}\hspace{0.03cm}a^{+}_{k}\!\biggr)\hspace{0.01cm}\hat{A}
= \\[1ex]
&=
\biggl\{-i\hspace{0.02cm}\sin\frac{2\pi\hspace{0.01cm} i}{3}\:[\hspace{0.03cm}a^{\phantom{+}}_{0}\!,a^{+}_{k}\hspace{0.02cm}]
+
\cos\frac{2\pi\hspace{0.01cm} i}{3}\:a^{+}_{k}\biggr\}\hat{A}
=
\biggl\{\biggl(\!-\frac{i\sqrt{3}}{2}\biggr)\hspace{0.02cm}
[\hspace{0.03cm}a^{\phantom{+}}_{0}\!,a^{+}_{k}\hspace{0.02cm}]
+
\biggl(-\frac{3}{2}\biggr)\hspace{0.02cm}a^{+}_{k}\biggr\}\hspace{0.02cm}\hat{A}.
\end{split}
\]
Here, we have used the operator identity (I.5.3) and the commutation rule (\ref{eq:10w}). The commutator $[\hspace{0.02cm}\hat{A},[\hspace{0.03cm}a_{0},a^{+}_{k}\hspace{0.02cm}]\hspace{0.03cm}]$ is calculated in the same way\footnote{\hspace{0.02cm}It is easiest to calculate the double commutator by an {\it algebraic} differentiation, i.e. by the commutation of the expression for $[\hspace{0.02cm}\hat{A}, a^{+}_{k}\hspace{0.02cm}]$ with the operator $a_0$, in view of (\ref{eq:10w}).} and as a result, we can write the desired commutators
\begin{subequations}
\begin{align}
&[\hspace{0.02cm}\hat{A}, a^{+}_{k}\hspace{0.02cm}]
=
\biggl\{\!\biggl(-\frac{3}{2}\hspace{0.02cm}\biggr)\hspace{0.02cm}a^{+}_{k}
+
\biggl(\!-\frac{i\sqrt{3}}{2}\hspace{0.03cm}\biggr)\hspace{0.02cm}
[\hspace{0.03cm}a^{\phantom{+}}_{0}\!,a^{+}_{k}\hspace{0.02cm}]\biggr\}
\hspace{0.02cm}\hat{A},
\label{eq:10ea} \\[1ex]
&[\hspace{0.02cm}\hat{A},[\hspace{0.03cm}a^{\phantom{+}}_{0}\!,a^{+}_{k}\hspace{0.02cm}]
\hspace{0.03cm}]
=
\biggl\{\!\biggl(\!-\frac{i\sqrt{3}}{2}\hspace{0.03cm}\biggr)\hspace{0.02cm}a^{+}_{k}
+
\biggl(-\frac{3}{2}\hspace{0.02cm}\biggr)\hspace{0.02cm}
[\hspace{0.03cm}a^{\phantom{+}}_{0}\!,a^{+}_{k}\hspace{0.02cm}]\biggr\}
\hspace{0.02cm}\hat{A}.
\label{eq:10eb}
\end{align}
\label{eq:10e}
\end{subequations}
\indent We will analyse these relations by calculating their matrix elements in the basis of parafermion coherent states. Let us consider the second relation. For the left-hand side we make use of the relation\footnote{\hspace{0.02cm}Recall that
\begin{equation}
\beta = \biggl(\frac{i\sqrt{3}}{2}\biggr)\hspace{0.02cm}\alpha,
\quad
\gamma = \biggl(-\frac{3}{2}\biggr)\hspace{0.02cm}\alpha,
\quad
\alpha^{3} = \frac{1}{m}\,.
\label{eq:10r}
\end{equation}}
(I.10.12)
\begin{equation}
[\hspace{0.02cm}\hat{A},[\hspace{0.03cm}a^{\phantom{+}}_{0}\!,a^{+}_{k}\hspace{0.02cm}]
\hspace{0.03cm}]
=
-\beta\hspace{0.03cm}a^{+}_{k} + 2\hspace{0.02cm}\gamma\hspace{0.03cm}a^{+}_{k}a^{\phantom{+}}_{0}\! + \gamma\hspace{0.03cm}[\hspace{0.03cm}a^{\phantom{+}}_{0}\!,a^{+}_{k}\hspace{0.02cm}].
\label{eq:10t}
\end{equation}
The matrix elements of two terms on the right-hand side (\ref{eq:10eb}) can be written as
\begin{equation}
\begin{split}
&\langle\hspace{0.02cm}\bar{\xi}^{\,\prime}\hspace{0.02cm}|\,
a^{+}_{k}\hat{A}|\,\xi\hspace{0.02cm}\rangle
=
\bar{\xi}^{\,\prime}_{k}\hspace{0.04cm}
{\cal A}\hspace{0.03cm}(\bar{\xi}^{\,\prime},\xi)\hspace{0.02cm}
\langle\hspace{0.02cm}\bar{\xi}^{\,\prime}\hspace{0.02cm}|\,\xi\hspace{0.02cm}\rangle,
\\[1ex]
&\langle\hspace{0.02cm}\bar{\xi}^{\,\prime}\hspace{0.02cm}|\,
[\hspace{0.03cm}a^{\phantom{+}}_{0}\!,a^{+}_{k}\hspace{0.02cm}]\hat{A}|\,\xi\hspace{0.02cm}
\rangle
=
-\hspace{0.02cm}\biggl(\!\frac{\partial\hspace{0.03cm}\Omega
\hspace{0.03cm}(\bar{\xi}^{\,\prime},\xi)}
{\partial\hspace{0.03cm}\xi^{\phantom{\prime}}_{k}}\biggr)
\!*
{\cal A}\hspace{0.03cm}(\bar{\xi}^{\,\prime},\xi)\hspace{0.02cm}
\langle\hspace{0.02cm}\bar{\xi}^{\,\prime}\hspace{0.02cm}|\,\xi\hspace{0.02cm}\rangle.
\end{split}
\label{eq:10y}
\end{equation}
Here, we have used the definitions of the function ${\cal A}(\hat{\xi}^{\prime}, \xi)$, Eq.\,(I.9.23), of the derivative $\partial\hspace{0.02cm}\Omega/ \partial\hspace{0.01cm}\xi_k$, Eq.\,(I.9.12), and of the star product (\ref{eq:4t}). Based on (\ref{eq:10t}) and (\ref{eq:10y}) the matrix element of (\ref{eq:10eb}) can be presented in the form
\begin{equation}
(\alpha/\gamma)\hspace{0.02cm}\biggl\{\bar{\xi}^{\,\prime}_{k}\hspace{0.03cm}
\Bigl(\hspace{0.02cm}-\beta + 2\hspace{0.02cm}\gamma\hspace{0.04cm}\Omega
+ (\beta/\alpha){\cal A}\hspace{0.02cm}\Bigr)
-
\gamma\hspace{0.02cm}\biggl(\frac{\partial\hspace{0.03cm}\Omega}
{\partial\hspace{0.03cm}\xi^{\phantom{\prime}}_{k}}\biggr)\!\biggr\}
=
-\hspace{0.02cm}\biggl(\frac{\partial\hspace{0.03cm}\Omega}
{\partial\hspace{0.03cm}\xi^{\phantom{\prime}}_{k}}\biggr)\!*{\cal A}.
\label{eq:10u}
\end{equation}
On the left- and right-hand sides we cancelled by the overlap function $\langle\hspace{0.02cm}\bar{\xi}^{\,\prime}\hspace{0.02cm}|\,\xi\hspace{0.02cm}\rangle$.\\
\indent Further we consider the relation (\ref{eq:10ea}). Taking into account that
\[
\langle\hspace{0.02cm}\bar{\xi}^{\,\prime}\hspace{0.02cm}|\,
[\hspace{0.02cm}\hat{A}, a^{+}_{k}\hspace{0.02cm}]|\,\xi\hspace{0.02cm}\rangle
=
-\frac{\partial\hspace{0.01cm}{\cal A}\hspace{0.03cm}(\bar{\xi}^{\,\prime},\xi)}
{\partial\hspace{0.03cm}\xi^{\phantom{\prime}}_{k}}
\hspace{0.04cm}
\langle\hspace{0.02cm}\bar{\xi}^{\,\prime}\hspace{0.02cm}|\,\xi\hspace{0.02cm}\rangle,
\]
we can write the matrix element of the operator relation (\ref{eq:10ea}) in the following form:
\begin{equation}
(\alpha/\beta)\hspace{0.02cm}\biggl\{(\gamma/\alpha)\hspace{0.04cm}\bar{\xi}^{\,\prime}_{k}
\hspace{0.04cm}{\cal A} \,+\, \frac{\partial\hspace{0.01cm}{\cal A}}
{\partial\hspace{0.03cm}\xi^{\phantom{\prime}}_{k}}\biggr\}
=
-\hspace{0.02cm}\biggl(\frac{\partial\hspace{0.03cm}\Omega}
{\partial\hspace{0.03cm}\xi^{\phantom{\prime}}_{k}}\biggr)\!*{\cal A}.
\label{eq:10i}
\end{equation}
Equating the left-hand sides of (\ref{eq:10u}) and (\ref{eq:10i}), we obtain a condition for consistency of the operator relations (\ref{eq:10ea}) and (\ref{eq:10eb}) in the enlarged state-vector space $\mathfrak{A}_{\hspace{0.02cm}G}$:
\begin{equation}
\frac{1}{\gamma}\,\biggl\{\bar{\xi}^{\,\prime}_{k}\hspace{0.03cm}
\Bigl(\hspace{0.02cm}-\beta + 2\hspace{0.02cm}\gamma\hspace{0.04cm}\Omega
+ (\beta/\alpha){\cal A}\hspace{0.02cm}\Bigr)
-
\gamma\hspace{0.02cm}\biggl(\frac{\partial\hspace{0.03cm}\Omega}
{\partial\hspace{0.03cm}\xi^{\phantom{\prime}}_{k}}\biggr)\!\biggr\}
=
\frac{1}{\beta}\,\biggl\{(\gamma/\alpha)\hspace{0.04cm}\bar{\xi}^{\,\prime}_{k}
\hspace{0.04cm}{\cal A} \,+\, \frac{\partial\hspace{0.01cm}{\cal A}}
{\partial\hspace{0.03cm}\xi^{\phantom{\prime}}_{k}}\biggr\}.
\label{eq:10o}
\end{equation}
This relation must be fulfilled identically. Let us verify this circumstance. The function
${\cal A}(\bar{\xi}^{\prime}, \xi)$ by virtue of the definitions (\ref{eq:5r}) and (\ref{eq:9u}) has the following form:
\begin{equation}
{\cal A}\hspace{0.03cm}(\bar{\xi}^{\,\prime},\xi)
=
\alpha -
\beta\hspace{0.05cm}\Omega\hspace{0.03cm}(\bar{\xi}^{\,\prime},\xi)
+
\gamma\hspace{0.03cm}\widetilde{\widetilde{\Omega}}\hspace{0.03cm}(\bar{\xi}^{\,\prime},\xi),
\label{eq:10p}
\end{equation}
where, in turn, the function $\widetilde{\widetilde{\Omega}}$ is given by the expression (\ref{eq:9i}). Making use of the representation (I.11.17) for the function $\widetilde{\Omega}$, we have
\begin{equation}
\frac{\partial\hspace{0.03cm}\widetilde{\Omega}}{\partial\hspace{0.03cm}
\xi^{\phantom{\prime}}_{k}}
=
\bar{\xi}^{\,\prime}_{k}\hspace{0.04cm}
(\hspace{0.02cm}2\hspace{0.03cm}\widetilde{\Omega} - 1).
\label{eq:10a}
\end{equation}
Taking into account the last expression and (\ref{eq:9i}), we obtain the derivative of the function ${\cal A}$ with respect to $\xi_{k}$:
\[
\frac{\partial\hspace{0.01cm}{\cal A}}{\partial\hspace{0.03cm}\xi^{\phantom{\prime}}_{k}}
=
-\beta\hspace{0.03cm}\frac{\partial\hspace{0.03cm}\Omega}{\partial\hspace{0.03cm}
\xi^{\phantom{\prime}}_{k}}
\,+\,
\gamma\hspace{0.03cm}\frac{3}{4}\,\bar{\xi}^{\,\prime}_{k}\hspace{0.04cm}
(\hspace{0.02cm}2\hspace{0.03cm}\widetilde{\Omega} - 1)
+
\gamma\hspace{0.03cm}\frac{1}{8}\,\frac{\partial\hspace{0.04cm}
{\cal C}_{2}}{\partial\hspace{0.03cm}\xi^{\phantom{\prime}}_{k}}.
\]
Substituting this derivative into (\ref{eq:10o}), by using numerical values of the parameters $\beta$ and $\gamma$, Eq.\,(\ref{eq:10r}), one can show that eventually the consistency condition (\ref{eq:10o}) is reduced to the simple relation
\begin{equation}
\frac{\partial\hspace{0.04cm}{\cal C}_{2}\hspace{0.03cm}(\bar{\xi}^{\,\prime},\xi)}
{\partial\hspace{0.03cm}\xi^{\phantom{\prime}}_{k}}
=
2\hspace{0.02cm}\bar{\xi}^{\,\prime}_{k}\hspace{0.04cm}
\bigl(\hspace{0.02cm}{\cal C}_{2}\hspace{0.03cm}(\bar{\xi}^{\,\prime},\xi) - 7\bigr).
\label{eq:10s}
\end{equation}
Finally, by using an explicit form of the function ${\cal C}_2(\bar{\xi}^{\prime}, \xi)$, Eq.\,(\ref{eq:7j}), it is not difficult to verify that the consistency condition does not turn into identity. Let us define the reason of such a contradiction.\\
\indent In deriving the formulae (\ref{eq:10e}) and (\ref{eq:10t}) we used the quantization rule (\ref{eq:10w}). In a special case of parastatistics $p = 2$ this rule is a consequence of two relations
\begin{equation}
a^{\phantom{+}}_{0}\!\hspace{0.02cm}a^{+}_{k}a^{\phantom{+}}_{0} = 0,
\quad
(a^{\phantom{+}}_{0}\!)^{2}\hspace{0.02cm}a^{+}_{k} +
a^{+}_{k}(a^{\phantom{+}}_{0}\!)^{2} = a^{+}_{k},
\label{eq:10d}
\end{equation}
which, in turn, is the operator representation of matrix relations (see Appendix A in Part I)
\begin{equation}
\omega\beta_{\mu\,}\omega = 0,
\qquad
\omega^{2\!}\hspace{0.035cm}\beta_{\mu} + \beta_{\mu\,}\omega^2 = \beta_{\mu}.
\hspace{0.8cm}
\label{eq:10f}
\end{equation}
Let us define the matrix element in the basis of parafermion coherent states of the second operator relation in (\ref{eq:10d}). For the left-hand side we have
\[
\langle\hspace{0.02cm}\bar{\xi}^{\,\prime}\hspace{0.02cm}|\,
\{(a^{\phantom{+}}_{0}\!)^{2}\!,a^{+}_{k}\}|\,\xi\hspace{0.02cm}
\rangle
\equiv
\langle\hspace{0.02cm}\bar{\xi}^{\,\prime}\hspace{0.02cm}|\,
2\hspace{0.02cm}a^{+}_{k}(a^{\phantom{+}}_{0}\!)^{2} +
[\hspace{0.03cm}(a^{\phantom{+}}_{0}\!)^{2}\!,a^{+}_{k}\hspace{0.02cm}]|\,\xi\hspace{0.02cm}
\rangle
=
2\hspace{0.02cm}\bar{\xi}^{\,\prime}_{k}\hspace{0.03cm}
\langle\hspace{0.02cm}\bar{\xi}^{\,\prime}\hspace{0.02cm}|\,
(a^{\phantom{+}}_{0}\!)^{2}|\,\xi\hspace{0.02cm}
\rangle
+
\langle\hspace{0.02cm}\bar{\xi}^{\,\prime}\hspace{0.02cm}|\,
[\hspace{0.03cm}(a^{\phantom{+}}_{0}\!)^{2}\!,a^{+}_{k}\hspace{0.02cm}]|\,\xi\hspace{0.02cm}
\rangle.
\]
At first we take as the operator $(a_{0})^2$ the Geyer operator (I.B.17). Then making use of (I.9.19) and (I.9.21), the desired matrix element after cancelling by the overlap function, takes the form
\begin{equation}
2\hspace{0.04cm}\bar{\xi}^{\,\prime}_{k}\hspace{0.05cm}\widetilde{\Omega}
-
\frac{\partial\hspace{0.03cm}\widetilde{\Omega}}{\partial\hspace{0.03cm}
\xi^{\phantom{\prime}}_{k}}
=
\bar{\xi}^{\,\prime}_{k},
\label{eq:10g}
\end{equation}
and thus we result in (\ref{eq:10a}), i.e. there is no contradiction. However, as we already known from the preceding consideration, instead of the Geyer operator $a_0^2$, we must use the Harish-Chandra operator $\hat{\omega}^{2} \equiv (a_{0})^{2}$ and therefore, instead of (\ref{eq:10g}), we have to write
\[
2\hspace{0.04cm}\bar{\xi}^{\,\prime}_{k}\hspace{0.05cm}\widetilde{\widetilde{\Omega}}
-
\frac{\partial\hspace{0.03cm}\widetilde{\widetilde{\Omega}}}{\partial\hspace{0.03cm}
\xi^{\phantom{\prime}}_{k}}
=
\bar{\xi}^{\,\prime}_{k}.
\]
Substituting (\ref{eq:9i}) into this equality, we result in the relation (\ref{eq:10s}), which is contradictory.  This circumstance provides a hint that the second matrix relation in (\ref{eq:10d}), and as a consequence, its operator formulation (\ref{eq:10d}) are not entirely correct.\\
\indent In Appendix A of Part I we have given all basic formulae of the $\omega$\hspace{0.02cm}-\hspace{0.02cm}$\beta_{\mu}$ algebra. All of them except the formula (I.A.3) were proved in \cite{harish-chandra_1947} without fixing the element of centre (\ref{eq:7g}). Let us discard this fixing in deriving the formula (I.A.3). We will need the relation (\ref{eq:9e}) in which we come to the normalized matrix $\omega$ according to the rule $\omega\rightarrow 2^{2}\hspace{0.03cm} \omega$:
\[
\omega^{\hspace{0.02cm}2} = \frac{1}{4}\,\bigl[P(\theta) + Q(\theta)B\hspace{0.02cm}\bigr].
\]
Here we recall $P(\theta) = 6\hspace{0.02cm}(\theta-4)$ and $Q(\theta) = -2\hspace{0.02cm}(\theta - 4)$. By virtue of the fact that $\theta$ is element of the centre of the DKP-algebra and the equality $\{B,\beta_{\mu}\} = 5\hspace{0.02cm}\beta_{\mu}$, instead of (I.A.3), we have
\[
\{\hspace{0.02cm}\omega^{\hspace{0.02cm}2},\beta_{\mu}\}
=
\frac{1}{4}\,\bigl(2\hspace{0.02cm}P(\theta)\hspace{0.02cm}\beta_{\mu} + Q(\theta)\{B,\beta_{\mu}\}\bigr)
=
\frac{1}{4}\,(2\hspace{0.02cm}\theta - 8)\beta_{\mu}.
\]
To pass to the operator representation it is sufficient to perform the following replacements: $\omega \rightarrow \hat{\omega},\,\beta_{\mu} \rightarrow a_{k}^{\pm},\,2\hspace{0.02cm} \theta\rightarrow \hat{C}^{\hspace{0.02cm}\prime}_2$. Thus we finally arrive at the desired generalization of the second relation in (\ref{eq:10d})
\[
\hat{\omega}^{\hspace{0.02cm}2}\hspace{0.02cm}a^{\pm}_{k}
+
a^{\pm}_{k}\hspace{0.02cm}\hat{\omega}^{\hspace{0.02cm}2}
= \frac{1}{4}\,(\hspace{0.02cm}\widehat{C}^{\hspace{0.03cm}\prime}_{2} - 8)\hspace{0.03cm} a^{\pm}_{k}.
\]
In fixing the Casimir operator $\hat{C}^{\hspace{0.02cm} \prime}_{2} = 12\hspace{0.02cm}  \hat{I}$ we return to the original relation (\ref{eq:10f}). With allowance made for this generalization, the double commutator (\ref{eq:10w}) takes the form
\begin{equation}
[\hspace{0.03cm}a^{\phantom{+}}_{0}\!, [\hspace{0.03cm}a^{\phantom{+}}_{0}\!,a^{+}_{k}\hspace{0.02cm}]\hspace{0.02cm}]
=
\frac{1}{4}\,(\hspace{0.02cm}\widehat{C}^{\hspace{0.03cm}\prime}_{2} - 8)\hspace{0.03cm} a^{+}_{k},
\label{eq:10h}
\end{equation}
and now instead of (I.10.12) we obtain
\[
[\hspace{0.03cm}\hat{A}, [\hspace{0.03cm}a^{\phantom{+}}_{0}\!,a^{+}_{k}\hspace{0.02cm}]\hspace{0.02cm}]
=
-\frac{1}{4}\,\beta\hspace{0.02cm}(\hspace{0.02cm}\widehat{C}^{\hspace{0.03cm}\prime}_{2} - 8)\hspace{0.03cm} a^{+}_{k}
+
\frac{1}{4}\,\gamma\hspace{0.03cm}
\{\hspace{0.02cm}a_{0}\hspace{0.01cm}(\hspace{0.02cm}
\widehat{C}^{\hspace{0.03cm}\prime}_{2} - 8),a^{+}_{k}\}.
\]
Further, based on the representation (\ref{eq:7a}) and the property (\ref{eq:9d}) we can put the last expression in the final form
\begin{equation}
[\hspace{0.03cm}\hat{A}, [\hspace{0.03cm}a^{\phantom{+}}_{0}\!,a^{+}_{k}\hspace{0.02cm}]\hspace{0.02cm}]
=
-\frac{1}{4}\,\beta\hspace{0.02cm}(\hspace{0.02cm}\widehat{C}^{\hspace{0.03cm}\prime}_{2} - 8)\hspace{0.03cm} a^{+}_{k}
+
2\hspace{0.02cm}\gamma\hspace{0.03cm}a^{+}_{k}a^{\phantom{+}}_{0}\! + \gamma\hspace{0.03cm}[\hspace{0.03cm}a^{\phantom{+}}_{0}\!,a^{+}_{k}\hspace{0.02cm}].
\label{eq:10j}
\end{equation}
Comparing this expression with (\ref{eq:10t}), we see that change occurred only in the first term on the right-hand side.\\
\indent Further, we need to modify the relations (\ref{eq:10e}), since now for their deriving we should make use of the commutation rules (\ref{eq:10h}) instead of (\ref{eq:10w}). It is not difficult to show that in this case, instead of (\ref{eq:10e}), we get
\begin{subequations}
\begin{align}
&[\hspace{0.02cm}\hat{A}, a^{+}_{k}\hspace{0.02cm}]
=
\biggl\{\!\biggl(-\frac{3}{2}\hspace{0.02cm}\biggr)\hspace{0.02cm}a^{+}_{k}
+
\biggl(\!-\frac{i\sqrt{3}}{2}\hspace{0.03cm}\biggr)\hspace{0.02cm}
[\hspace{0.03cm}a^{\phantom{+}}_{0}\!,a^{+}_{k}\hspace{0.02cm}]\biggr\}
\hspace{0.02cm}\hat{A}
+
\frac{1}{4}\,\gamma\hspace{0.04cm} a^{+}_{k}\hspace{0.02cm}
(\hspace{0.02cm}\widehat{C}^{\hspace{0.03cm}\prime}_{2} - 12),
\label{eq:10ka} \\[1ex]
&[\hspace{0.02cm}\hat{A},[\hspace{0.03cm}a^{\phantom{+}}_{0}\!,a^{+}_{k}\hspace{0.02cm}]
\hspace{0.03cm}]
=
\biggl\{\!\biggl(\!-\frac{i\sqrt{3}}{2}\hspace{0.03cm}\biggr)\hspace{0.02cm}a^{+}_{k}
+
\biggl(-\frac{3}{2}\hspace{0.02cm}\biggr)\hspace{0.02cm}
[\hspace{0.03cm}a^{\phantom{+}}_{0}\!,a^{+}_{k}\hspace{0.02cm}]\biggr\}
\hspace{0.02cm}\hat{A}
-
\frac{1}{4}\,\beta\hspace{0.04cm} a^{+}_{k}\hspace{0.02cm}
(\hspace{0.02cm}\widehat{C}^{\hspace{0.03cm}\prime}_{2} - 12).
\label{eq:10kb}
\end{align}
\label{eq:10k}
\end{subequations}
Here, we see appearing additional terms in comparison with (\ref{eq:10e}). At this stage on the left-hand side of (10.18b) we have to use the relation (\ref{eq:10j}).\\
\indent Let us consider now the matrix elements of the generalized operator relations (\ref{eq:10k}). The change in the first term of the right-hand side of (\ref{eq:10j}) leads to the following replacement on the left-hand side of (\ref{eq:10u}):
\[
\bar{\xi}^{\,\prime}_{k}\hspace{0.03cm}\beta \rightarrow
\bar{\xi}^{\,\prime}_{k}\hspace{0.04cm}\beta\,
\frac{1}{4}\,\bigl(\hspace{0.025cm}
{\cal C}^{\hspace{0.02cm}\prime}_{2}\hspace{0.03cm}(\bar{\xi}^{\,\prime},\xi) - 8\bigr),
\]
and additional terms in (\ref{eq:10ka}) and (\ref{eq:10kb}) result in appearance of additional terms on the left-hand side of the matrix elements (\ref{eq:10u}) and (\ref{eq:10i}), namely the expression
\[
-\hspace{0.02cm}\bar{\xi}^{\,\prime}_{k}\hspace{0.04cm}\beta\,
\frac{1}{4}\,\bigl(\hspace{0.025cm}
{\cal C}^{\hspace{0.02cm}\prime}_{2}\hspace{0.03cm}(\bar{\xi}^{\,\prime},\xi) - 12\hspace{0.01cm}\bigr),
\]
should be added to the left-hand side of (\ref{eq:10u}) and the expression
\[
\hspace{0.5cm}
\bar{\xi}^{\,\prime}_{k}\hspace{0.04cm}\gamma\,
\frac{1}{4}\,\bigl(\hspace{0.025cm}
{\cal C}^{\hspace{0.02cm}\prime}_{2}\hspace{0.03cm}(\bar{\xi}^{\,\prime},\xi) - 12\hspace{0.01cm}\bigr)
\]
should be added to the the left-hand side of (\ref{eq:10i}). The consistency condition of thus obtained two matrix elements in view of the equality
\[
{\cal C}^{\hspace{0.02cm}\prime}_{2}\hspace{0.03cm}(\bar{\xi}^{\,\prime},\xi)
=
{\cal C}_{2}\hspace{0.03cm}(\bar{\xi}^{\,\prime},\xi)
+
\Lambda\hspace{0.03cm}(\bar{\xi}^{\,\prime},\xi),
\]
results us, instead of (\ref{eq:10s}), in a completely different relation, namely
\begin{equation}
\frac{\partial\hspace{0.04cm}{\cal C}_{2}\hspace{0.03cm}(\bar{\xi}^{\,\prime},\xi)}
{\partial\hspace{0.03cm}\xi^{\phantom{\prime}}_{k}}
=
-\hspace{0.02cm}2\hspace{0.03cm}\bar{\xi}^{\,\prime}_{k}\hspace{0.04cm}
\bigl(\hspace{0.005cm}\Lambda\hspace{0.03cm}(\bar{\xi}^{\,\prime},\xi) - 5\hspace{0.01cm}\bigr).
\label{eq:10l}
\end{equation}
By using an explicit form of the functions ${\cal C}_{2}\hspace{0.03cm}(\bar{\xi}^{\,\prime},\xi)$ and $\Lambda\hspace{0.03cm}(\bar{\xi}^{\,\prime},\xi)$, Eqs.\,(\ref{eq:7j}) 
and (\ref{eq:7k}), it is easy to verify that (\ref{eq:10l}) turns into identity.

%%%%%%%%%%%%%%%%%%%%%%%% section 11 %%%%%%%%%%%%%%%%%%%%%%%%%%%%

\section{Matrix element $\langle\hspace{0.02cm}(k)^{\prime}_{p}\hspace{0.02cm}|\hspace{0.02cm}
\hat{A}\hspace{0.02cm}\hat{\eta}_{\mu}(z)\hat{D}_{\mu}\hspace{0.01cm}|\hspace{0.02cm}
(k - 1)_{x}\hspace{0.02cm}\rangle$}
\setcounter{equation}{0}
\label{section_11}

We return to the matrix element (I.4.2), more precisely, to the most problematic part containing the creation operator $a_{n}^{+}$. We write out this part once again
\begin{align}
-\frac{i}{2}\,\sum\limits^{2}_{n\hspace{0.02cm}=\hspace{0.02cm}1}
\,&\biggl\{\!\hspace{0.01cm}\biggl(1 + \frac{1}{2}\,z\!\hspace{0.03cm}\biggr)
\langle\hspace{0.02cm}\bar{\xi}^{\,\prime(k)}\hspace{0.02cm}|\,
\hat{A}\hspace{0.03cm}a^{+}_{n}|\,\xi^{(k - 1)}\hspace{0.02cm}\rangle
+
z\hspace{0.03cm}\biggl(\!-\frac{i\sqrt{3}}{2}\hspace{0.02cm}\biggr)\hspace{0.02cm}
\langle\hspace{0.02cm}\bar{\xi}^{\,\prime(k)}\hspace{0.02cm}|\hspace{0.02cm}
\hat{A}\hspace{0.03cm}[\hspace{0.03cm}a^{\phantom{+}}_{0}, a^{+}_{n}\hspace{0.02cm}]
\hspace{0.02cm}|\,\xi^{(k - 1)}\hspace{0.02cm}\rangle\biggr\}\hspace{0.03cm}\times
\label{eq:11q} \\[1ex]
&\times\!
\bigl(p^{(k)}_{n} - e\hspace{0.02cm}A_{n}(x^{(k-1)})\bigr)\hspace{0.02cm}
\langle\hspace{0.02cm}p^{(k)}|\,x^{(k-1)\hspace{0.02cm}}\rangle.
\notag
\end{align}
As in sections 9 and 10 of Part I the first step is to shift the operators $a_{n}^{+}$ and $[\hspace{0.03cm}a^{\phantom{+}}_{0}, a^{+}_{n}\hspace{0.02cm}]$ to the left of the operator $\hat{A}$, i.e. we rewrite the matrix elements in (\ref{eq:11q}) as follows:
\[
\begin{split}
&\langle\hspace{0.02cm}\bar{\xi}^{\,\prime}\hspace{0.02cm}|\,
\hat{A}\hspace{0.04cm}a^{+}_{n}|\,\xi\hspace{0.02cm}\rangle
=
\langle\hspace{0.02cm}\bar{\xi}^{\,\prime}\hspace{0.02cm}|\,
a^{+}_{n}\hat{A}|\,\xi\hspace{0.02cm}\rangle
+
\langle\hspace{0.02cm}\bar{\xi}^{\,\prime}\hspace{0.02cm}|\,
[\hspace{0.03cm}\hat{A}, a^{+}_{n}\hspace{0.02cm}]\hspace{0.02cm}|\,\xi\hspace{0.02cm}\rangle,
\\[1ex]
&\langle\hspace{0.02cm}\bar{\xi}^{\,\prime}\hspace{0.02cm}|\hspace{0.02cm}
\hat{A}\hspace{0.04cm}[\hspace{0.03cm}a^{\phantom{+}}_{0}, a^{+}_{n}\hspace{0.02cm}]|\,\xi\hspace{0.02cm}\rangle
=
\langle\hspace{0.02cm}\bar{\xi}^{\,\prime}\hspace{0.02cm}|\,
[\hspace{0.03cm}a^{\phantom{+}}_{0}, a^{+}_{n}\hspace{0.02cm}]\hspace{0.02cm}\hat{A}\hspace{0.02cm}|\,\xi\hspace{0.02cm}\rangle
+
\langle\hspace{0.02cm}\bar{\xi}^{\,\prime}\hspace{0.02cm}|\,
[\hspace{0.02cm}\hat{A},[\hspace{0.03cm}a^{\phantom{+}}_{0}, a^{+}_{n}\hspace{0.02cm}]\hspace{0.015cm}]\hspace{0.015cm}|\,\xi\hspace{0.02cm}\rangle.
\end{split}
\]
Here, for the sake of simplicity we have omitted iteration numbers $(k)$ and $(k - 1)$. Further, for the second terms on the right-hand side of these expressions we use the commutation relations (\ref{eq:10k}) that gives us
\[
\begin{split}
&\langle\hspace{0.02cm}\bar{\xi}^{\,\prime}\hspace{0.02cm}|\,
\hat{A}\hspace{0.04cm}a^{+}_{n}|\,\xi\hspace{0.02cm}\rangle
=
\biggl(\!-\frac{1}{2}\biggr)\langle\hspace{0.02cm}\bar{\xi}^{\,\prime}\hspace{0.02cm}|\,
a^{+}_{n}\hat{A}|\,\xi\hspace{0.02cm}\rangle
+
\biggl(\!-\frac{i\sqrt{3}}{2}\hspace{0.02cm}\biggr)
\langle\hspace{0.02cm}\bar{\xi}^{\,\prime}\hspace{0.02cm}|\,
[\hspace{0.03cm}a^{\phantom{+}}_{0}, a^{+}_{n}\hspace{0.02cm}]\hspace{0.02cm}\hat{A}\hspace{0.02cm}|\,\xi\hspace{0.02cm}\rangle
+
\frac{1}{4}\,\gamma\hspace{0.04cm}\langle\hspace{0.02cm}\bar{\xi}^{\,\prime}\hspace{0.02cm}|\, a^{+}_{n}\hspace{0.02cm}
(\hspace{0.02cm}\widehat{C}^{\hspace{0.03cm}\prime}_{2} - 12)|\,\xi\hspace{0.02cm}\rangle,
\\[1ex]
&\langle\hspace{0.02cm}\bar{\xi}^{\,\prime}\hspace{0.02cm}|\hspace{0.02cm}
\hat{A}\hspace{0.04cm}[\hspace{0.03cm}a^{\phantom{+}}_{0}, a^{+}_{n}\hspace{0.02cm}]|\,\xi\hspace{0.02cm}\rangle
=
\biggl(\!-\frac{i\sqrt{3}}{2}\hspace{0.02cm}\biggr)\langle\hspace{0.02cm}\bar{\xi}^{\,\prime}
\hspace{0.02cm}|\,
a^{+}_{n}\hat{A}|\,\xi\hspace{0.02cm}\rangle
+
\biggl(\!-\frac{1}{2}\biggr)
\langle\hspace{0.02cm}\bar{\xi}^{\,\prime}\hspace{0.02cm}|\,
[\hspace{0.03cm}a^{\phantom{+}}_{0}, a^{+}_{n}\hspace{0.02cm}]\hspace{0.02cm}\hat{A}\hspace{0.02cm}|\,\xi\hspace{0.02cm}\rangle
-\!
\frac{1}{4}\,\beta\hspace{0.04cm}\langle\hspace{0.02cm}\bar{\xi}^{\,\prime}\hspace{0.02cm}|\, a^{+}_{n}\hspace{0.02cm}
(\hspace{0.02cm}\widehat{C}^{\hspace{0.03cm}\prime}_{2} - 12)|\,\xi\hspace{0.02cm}\rangle.
\end{split}
\]
Substituting the above expressions into (\ref{eq:11q}) and collecting similar terms, we get
\begin{align}
-\frac{i}{2}\,\sum\limits^{2}_{n\hspace{0.02cm}=\hspace{0.02cm}1}
\,\biggl\{\!\hspace{0.01cm}&\biggl[\biggl(\!-\frac{1}{2}\biggr)
\biggl(1 + \frac{1}{2}\,z\biggr)
+
z\hspace{0.03cm}\biggl(\!-\frac{i\sqrt{3}}{2}\hspace{0.02cm}\biggr)^{\!\!2}\,\biggr]
\langle\hspace{0.02cm}\bar{\xi}^{\,\prime}\hspace{0.02cm}|\,
\hat{A}\hspace{0.03cm}a^{+}_{n}|\,\xi\hspace{0.02cm}\rangle
\,+
\label{eq:11w}\\[1ex]
+
&\biggl[\biggl(\!-\frac{i\sqrt{3}}{2}\hspace{0.02cm}\biggr)\biggl(1 + \frac{1}{2}\,z\biggr)
+
\biggl(\!-\frac{1}{2}\biggr)\biggl(\!-\frac{i\sqrt{3}}{2}\hspace{0.02cm}\biggr)\biggr]
\langle\hspace{0.02cm}\bar{\xi}^{\,\prime}\hspace{0.02cm}|\hspace{0.02cm}
\hat{A}\hspace{0.03cm}[\hspace{0.03cm}a^{\phantom{+}}_{0}, a^{+}_{n}\hspace{0.02cm}]
\hspace{0.02cm}|\,\xi\hspace{0.02cm}\rangle\,+
\notag \\[1ex]
+&\,\alpha\hspace{0.03cm}\frac{1}{4}\hspace{0.03cm}\biggl[\biggl(\!-\frac{3}{2}\biggr)\biggl(1 + \frac{1}{2}\,z\biggr)
+
z\hspace{0.03cm}\biggl(\!-\frac{i\sqrt{3}}{2}\hspace{0.02cm}\biggr)^{\!\!2}\,\biggr]
\langle\hspace{0.02cm}\bar{\xi}^{\,\prime}\hspace{0.02cm}|\,a^{+}_{n}\hspace{0.02cm}
(\hspace{0.02cm}\widehat{C}^{\hspace{0.03cm}\prime}_{2} - 12)|\,\xi\hspace{0.02cm}\rangle
\hspace{0.01cm}\!\biggr\}\hspace{0.02cm}
\bigl(p_{\hspace{0.02cm}n} - e\hspace{0.02cm}A_{n}(x)\bigr)\hspace{0.02cm}
\langle\hspace{0.02cm}p\hspace{0.02cm}|\,x\hspace{0.02cm}\rangle.
\notag
\end{align}
Here, in the last term we have taken into account the values of parameters $\beta$ and $\gamma$, Eq.\,(\ref{eq:10r}). By this means the use of the commutation relations (\ref{eq:10k}) enables us to present (\ref{eq:11q}) in a form without additional asymmetric contributions, which appear in a straightforward calculation of the matrix elements (see Eqs.\,(I.9.2), (I.10.10) and (I.10.13)). However, in (\ref{eq:11w}) we have a change of number coefficients of the first two matrix elements in comparison with similar terms containing the annihilation operator $a_{n}^{-}$, Eq.\,(I.4.2). Moreover, the contribution with the matrix element of the Casimir operator $\hat{C}_{2}^{\prime}$ appears. Let us discuss these two circumstances in more detail.\\
\indent Let us consider the limit $z \rightarrow q$, where $q$ is a primitive cubic root of unity (I.2.5). The coefficients of three matrix elements in (\ref{eq:11w}) in the limit $z \rightarrow q$ can be presented in the following form, correspondingly,
\begin{equation}
\biggl(1 + \frac{1}{2}\,q\!\hspace{0.03cm}\biggr)\hspace{0.02cm}q^{2},
\quad
\biggl[q\hspace{0.03cm}\biggl(\!-\frac{i\sqrt{3}}{2}\hspace{0.02cm}\biggr)\biggr]q^{2},
\quad
-\frac{1}{4}\,\gamma\hspace{0.03cm}q^{2}.
\label{eq:11e}
\end{equation}
Further, taking into account
\[
\begin{array}{lll}
&\langle\hspace{0.02cm}\bar{\xi}^{\,\prime}\hspace{0.02cm}|\,
[\hspace{0.03cm}a^{\phantom{+}}_{0}, a^{+}_{n}\hspace{0.02cm}]\hspace{0.02cm}\hat{A}\hspace{0.02cm}|\,\xi\hspace{0.02cm}\rangle
=
-\hspace{0.02cm}\biggl(\displaystyle\frac{\partial\hspace{0.03cm}\Omega}
{\partial\hspace{0.03cm}\xi^{\phantom{\prime}}_{n}}\biggr)\!*{\cal A}\,
\langle\hspace{0.02cm}\bar{\xi}^{\,\prime}\hspace{0.02cm}|\,\xi\hspace{0.02cm}\rangle,
&\langle\hspace{0.02cm}\bar{\xi}^{\,\prime}\hspace{0.02cm}|\,
a^{+}_{n}\hat{A}|\,\xi\hspace{0.02cm}\rangle
=
\bar{\xi}^{\,\prime}_{n}\hspace{0.04cm}{\cal A}\,
\langle\hspace{0.02cm}\bar{\xi}^{\,\prime}\hspace{0.02cm}|\,\xi\hspace{0.02cm}\rangle,
\\[3ex]
&\langle\hspace{0.02cm}\bar{\xi}^{\,\prime}\hspace{0.02cm}|\hspace{0.02cm}
\hat{A}\hspace{0.04cm}[\hspace{0.03cm}a^{\phantom{+}}_{0}, a^{-}_{n}\hspace{0.02cm}]|\,\xi\hspace{0.02cm}\rangle
=
-\hspace{0.02cm}
{\cal A}\!\hspace{0.03cm}*\hspace{0.03cm}\!
\biggl(\displaystyle\frac{\partial\hspace{0.03cm}\Omega}
{\partial\hspace{0.03cm}\bar{\xi}^{\prime}_{n}}\biggr)
\langle\hspace{0.02cm}\bar{\xi}^{\,\prime}\hspace{0.02cm}|\,\xi\hspace{0.02cm}\rangle,
&\langle\hspace{0.02cm}\bar{\xi}^{\,\prime}\hspace{0.02cm}|\,
\hat{A}\hspace{0.03cm}a^{-}_{n}|\,\xi\hspace{0.02cm}\rangle
=
\xi_{n}\hspace{0.03cm}{\cal A}\,
\langle\hspace{0.02cm}\bar{\xi}^{\,\prime}\hspace{0.02cm}|\,\xi\hspace{0.02cm}\rangle
\end{array}
\]
and using the expression (\ref{eq:11w}) with coefficients (\ref{eq:11e}), we rewrite the matrix element (I.4.2) in the limit $z \rightarrow q$ in the following form:
\begin{equation}
\langle\hspace{0.02cm}(k)^{\prime}_{p}\hspace{0.02cm}|\hspace{0.02cm}
\hat{A}\hspace{0.02cm}\hat{\eta}_{\mu}(q)\hat{D}_{\mu}\hspace{0.01cm}|\hspace{0.02cm}
(k - 1)_{x}\hspace{0.02cm}\rangle
=
\label{eq:11r}
\end{equation}
\[
\begin{split}
=&
-\frac{i}{2}\,\sum\limits^{2}_{n\hspace{0.02cm}=\hspace{0.02cm}1}
\,\biggl\{\!\hspace{0.02cm}\biggl[\biggl(1 + \frac{1}{2}\,q\!\hspace{0.03cm}\biggr)
{\cal A}\hspace{0.04cm}\xi^{(k-1)}_{\bar{n}}
+
q\hspace{0.03cm}\biggl(\frac{i\sqrt{3}}{2}\hspace{0.02cm}\biggr)\hspace{0.02cm}
\biggl({\cal A}\!\hspace{0.03cm}*\hspace{0.03cm}\!
\frac{\!\!\partial\hspace{0.03cm}\Omega}
{\partial\hspace{0.03cm}\bar{\xi}^{\hspace{0.03cm}\prime(k)}_{n}}\biggr)\biggr]
\bigl(p^{(k)}_{\bar{n}} - e\hspace{0.02cm}A_{\bar{n}}(x^{(k-1)})\bigr)
- \\[1ex]
&-q^{2}
\biggl[\biggl(1 + \frac{1}{2}\,q\!\hspace{0.03cm}\biggr)
\bar{\xi}^{\hspace{0.03cm}\prime(k)}_{n}\!{\cal A}
+
q\hspace{0.03cm}\biggl(\frac{i\sqrt{3}}{2}\hspace{0.02cm}\biggr)\hspace{0.02cm}
\biggl(\frac{\partial\hspace{0.03cm}\Omega}
{\partial\hspace{0.03cm}\xi^{(k-1)}_{n}}\!*{\cal A}\biggr)\biggr]
\bigl(p^{(k)}_{n} - e\hspace{0.02cm}A_{n}(x^{(k-1)})\bigr) -
\\[1ex]
&-q^{2}\gamma\hspace{0.04cm}\frac{1}{4}\;
({\cal C}^{\hspace{0.02cm}\prime}_{2} - 12)
\,\bar{\xi}^{\hspace{0.03cm}\prime(k)}_{n}\hspace{0.01cm}
\bigl(p^{(k)}_{n} - e\hspace{0.02cm}A_{n}(x^{(k-1)})\bigr)\!\hspace{0.02cm}\biggr\}
\langle\hspace{0.02cm}\bar{\xi}^{\,\prime(k)}\hspace{0.02cm}|\,\xi^{(k-1)}
\hspace{0.02cm}\rangle
\langle\hspace{0.02cm}p^{(k)}|\,x^{(k-1)\hspace{0.02cm}}\rangle,
\end{split}
\]
where ${\cal A} \equiv {\cal A}(\hat{\xi}^{\hspace{0.02cm}\prime (k)}, \xi^{(k-1)}),\,\Omega \equiv \Omega (\hat{\xi}^{\hspace{0.02cm}\prime (k)}, \xi^{(k-1)})$ and ${\cal C}_{2}^{\prime}\equiv {\cal C}_{2}^{\prime}(\hat{\xi}^{\hspace{0.02cm}\prime(k)}, \xi^{(k-1)})$. One can write this expression in a more compact form.  Bearing in mind the definition of the star product, Eq.\,(\ref{eq:4t}), we obtain
\begin{equation}
{\cal A}\hspace{0.03cm}(\bar{\xi}^{\,\prime},\xi)\!*\xi_{n}
=
{\cal A}\hspace{0.03cm}(\bar{\xi}^{\,\prime},\xi)\hspace{0.04cm}\xi_{n},
\qquad
\bar{\xi}^{\hspace{0.03cm}\prime}_{n}\!*{\cal A}\hspace{0.03cm}(\bar{\xi}^{\,\prime},\xi)
=
\bar{\xi}^{\hspace{0.03cm}\prime}_{n}\hspace{0.04cm}
{\cal A}\hspace{0.03cm}(\bar{\xi}^{\,\prime},\xi).
\label{eq:11t}
\end{equation}
Further, making use of the definition of the functions $\Xi_{\hspace{0.03cm}\bar{n}}(z)$ and $\bar{\Xi}_{\hspace{0.03cm}n}(z)$:
\begin{subequations}
\begin{align}
&\Xi^{(\hspace{0.02cm}k-1,\hspace{0.03cm}k)}_{\bar{n}}(z)
=
\biggl(1 + \frac{1}{2}\,z\!\hspace{0.03cm}\biggr)\hspace{0.02cm}\xi^{(k-1)}_{\bar{n}}
+
z\hspace{0.03cm}\biggl(\frac{i\sqrt{3}}{2}\hspace{0.02cm}\biggr)\!\hspace{0.02cm}
\biggl(\displaystyle\frac{\partial\hspace{0.03cm}\Omega}
{\partial\hspace{0.03cm}\bar{\xi}^{\prime (k)}_{n}}\biggr),
\label{eq:11ya} \\[1ex]
&\bar{\Xi}^{(\hspace{0.02cm}k-1,\hspace{0.03cm}k)}_{n}(z)
=
\biggl(1 + \frac{1}{2}\,z\!\hspace{0.03cm}\biggr)\hspace{0.02cm}
\bar{\xi}^{\hspace{0.03cm}\prime(k)}_{n}
+
z\hspace{0.03cm}\biggl(\frac{i\sqrt{3}}{2}\hspace{0.02cm}\biggr)\!\hspace{0.02cm}
\biggl(\displaystyle\frac{\partial\hspace{0.03cm}\Omega}
{\partial\hspace{0.03cm}\xi^{(k-1)}_{n}}\biggr),
\label{eq:11yb}
\end{align}
\label{eq:11y}
\end{subequations}
instead of (\ref{eq:11r}), we get
\begin{equation}
\langle\hspace{0.02cm}(k)^{\prime}_{p}\hspace{0.02cm}|\hspace{0.02cm}
\hat{A}\hspace{0.02cm}\hat{\eta}_{\mu}(q)\hat{D}_{\mu}\hspace{0.01cm}|\hspace{0.02cm}
(k - 1)_{x}\hspace{0.02cm}\rangle
=
\label{eq:11u}
\end{equation}
\[
\begin{split}
=&
-\frac{i}{2}\,\biggl\{\hspace{0.03cm}\sum\limits^{2}_{n\hspace{0.02cm}=\hspace{0.02cm}1}
\,\Bigl[\hspace{0.01cm}{\cal A}\hspace{0.02cm}\!*\Xi^{(\hspace{0.02cm}k-1,\hspace{0.03cm}k)}_{\bar{n}}(q)
\bigl(p^{(k)}_{\bar{n}} - e\hspace{0.02cm}A_{\bar{n}}(x^{(k-1)})\bigr)
-
q^{2\,}\bar{\Xi}^{(\hspace{0.02cm}k-1,\hspace{0.03cm}k)}_{n}(q)\!*{\cal A}
\hspace{0.04cm}\bigl(p^{(k)}_{n} - e\hspace{0.02cm}A_{n}(x^{(k-1)})\bigr)\Bigr] +
\\[1ex]
&-q^{2}
\gamma\hspace{0.04cm}\frac{1}{4}\;
({\cal C}^{\hspace{0.02cm}\prime}_{2} - 12)
\biggl[\,\sum\limits^{2}_{n\hspace{0.02cm}=\hspace{0.02cm}1}
\,\bar{\xi}^{\hspace{0.03cm}\prime(k)}_{n}\hspace{0.02cm}
\bigl(p^{(k)}_{n} - e\hspace{0.02cm}A_{n}(x^{(k-1)})\bigr)\biggr]\!\hspace{0.02cm}\biggr\}
\hspace{0.02cm}
\langle\hspace{0.02cm}\bar{\xi}^{\,\prime(k)}\hspace{0.02cm}|\,\xi^{(k-1)}
\hspace{0.02cm}\rangle
\langle\hspace{0.02cm}p^{(k)}|\,x^{(k-1)\hspace{0.02cm}}\rangle.
\end{split}
\]
However, here it is impossible to collect the first two terms in one vector expression (as it takes place in the original operator expression), since in the second term the functions $\bar{\Xi}_{n}(z)$ and ${\cal A}$ are interchange, and moreover, there is an additional factor $q^2$. It is unclear also the presence of the last contribution on the right-hand side of (\ref{eq:11u}). Below we will show how one can solve these two problems simultaneously.\\
\indent At first we consider some technical details. Based on the general rule of calculating the matrix elements of commutators containing the operators $a_{n}^{\pm}$, we can immediately write
\begin{equation}
\langle\hspace{0.02cm}\bar{\xi}^{\,\prime}\hspace{0.02cm}|\,
[\hspace{0.03cm}\hat{A}, a^{+}_{n}\hspace{0.02cm}]\hspace{0.02cm}|\,\xi\hspace{0.02cm}\rangle
=
-\frac{\partial\hspace{0.01cm}{\cal A}\hspace{0.03cm}(\bar{\xi}^{\,\prime},\xi)}
{\partial\hspace{0.03cm}\xi^{\phantom{\prime}}_{n}}\,
\langle\hspace{0.02cm}\bar{\xi}^{\,\prime}\hspace{0.02cm}|\,\xi\hspace{0.02cm}\rangle.
\label{eq:11i}
\end{equation}
On the other hand, making use of the notion of the star product, we can present the left-hand side of the preceding expression by the so-called Moyal bracket and as the result we have
\begin{equation}
{\cal A}\hspace{0.03cm}(\bar{\xi}^{\,\prime},\xi)\!\hspace{0.03cm}*\hspace{0.03cm}
\bar{\xi}^{\hspace{0.03cm}\prime}_{n}
-
\bar{\xi}^{\hspace{0.03cm}\prime}_{n}\!\hspace{0.01cm}*\hspace{0.01cm}
{\cal A}\hspace{0.03cm}(\bar{\xi}^{\,\prime},\xi)
\equiv
[\hspace{0.03cm}{\cal A}\hspace{0.03cm}(\bar{\xi}^{\,\prime},\xi), \bar{\xi}^{\hspace{0.03cm}\prime}_{n}\hspace{0.02cm}]^{\phantom{A}}_{\hspace{0.03cm}*}
=
-\frac{\partial\hspace{0.01cm}{\cal A}\hspace{0.03cm}(\bar{\xi}^{\,\prime},\xi)}
{\partial\hspace{0.03cm}\xi^{\phantom{\prime}}_{n}}\,.
\label{eq:11o}
\end{equation}
Here, we have cancelled by the common factor $\langle\hspace{0.02cm}\bar{\xi}^{\,\prime}\hspace{0.02cm}|\,\xi\hspace{0.02cm}\rangle$. Further, we need the relations of the (\ref{eq:10k}) type, only now we rearrange the operator $\hat{A}$ to the left 
\begin{subequations}
\begin{align}
&[\hspace{0.02cm}\hat{A}, a^{+}_{k}\hspace{0.02cm}]
=
\hat{A}\hspace{0.02cm}
\biggl\{\frac{3}{2}\;a^{+}_{k}
+\,
\biggl(\!-\frac{i\sqrt{3}}{2}\hspace{0.03cm}\biggr)\hspace{0.02cm}
[\hspace{0.03cm}a^{\phantom{+}}_{0}\!,a^{+}_{k}\hspace{0.02cm}]\biggr\}
-
\frac{1}{4}\,\gamma\hspace{0.04cm} a^{+}_{k}\hspace{0.02cm}
(\hspace{0.02cm}\widehat{C}^{\hspace{0.03cm}\prime}_{2} - 12),
\label{eq:11pa} \\[1ex]
&[\hspace{0.02cm}\hat{A},[\hspace{0.03cm}a^{\phantom{+}}_{0}\!,a^{+}_{k}\hspace{0.02cm}]
\hspace{0.03cm}]
=
\hat{A}\hspace{0.02cm}
\biggl\{\!\biggl(\!-\frac{i\sqrt{3}}{2}\hspace{0.03cm}\biggr)\hspace{0.02cm}a^{+}_{k}
+\,
\frac{3}{2}\;
[\hspace{0.03cm}a^{\phantom{+}}_{0}\!,a^{+}_{k}\hspace{0.02cm}]\biggr\}
-
\frac{1}{4}\,\beta\hspace{0.04cm} a^{+}_{k}\hspace{0.02cm}
(\hspace{0.02cm}\widehat{C}^{\hspace{0.03cm}\prime}_{2} - 12).
\label{eq:11pb}
\end{align}
\label{eq:11p}
\end{subequations}
We equate the right-hand sides of the operator expressions (\ref{eq:10ka}) and (\ref{eq:11pa}) and calculate the matrix element of thus obtained expression. As a result, we have
\[
\begin{split}
&-\frac{3}{2}\,
\bar{\xi}^{\hspace{0.03cm}\prime}_{n}\!\hspace{0.01cm}*\hspace{0.01cm}{\cal A}
+
\biggl(\frac{i\sqrt{3}}{2}\hspace{0.02cm}\biggr)\hspace{0.02cm}
\frac{\partial\hspace{0.03cm}\Omega}
{\partial\hspace{0.03cm}\xi^{\phantom{\prime}}_{n}}\!*{\cal A}
+
\frac{1}{4}\,\gamma\hspace{0.04cm}\bar{\xi}^{\hspace{0.03cm}\prime}_{n}\hspace{0.02cm}
({\cal C}^{\hspace{0.02cm}\prime}_{2} - 12)
= \\[0.5ex]
&= \frac{3}{2}\,{\cal A}\!\hspace{0.03cm}*\hspace{0.03cm}
\bar{\xi}^{\hspace{0.03cm}\prime}_{n}
+
\biggl(\frac{i\sqrt{3}}{2}\hspace{0.02cm}\biggr)\hspace{0.02cm}
{\cal A}\!*\frac{\partial\hspace{0.03cm}\Omega}
{\partial\hspace{0.03cm}\xi^{\phantom{\prime}}_{n}}
-
\frac{1}{4}\,\gamma\hspace{0.04cm}\bar{\xi}^{\hspace{0.03cm}\prime}_{n}\hspace{0.02cm}
({\cal C}^{\hspace{0.02cm}\prime}_{2} - 12).
\end{split}
\]
Further, on the right-hand side for the first term we use the relation (\ref{eq:11o}). On rearrangement of the terms, we get
\begin{equation}
\biggl(\frac{i\sqrt{3}}{2}\hspace{0.02cm}\biggr)\hspace{0.02cm}
{\cal A}\hspace{0.03cm}\!*\frac{\partial\hspace{0.03cm}\Omega}
{\partial\hspace{0.03cm}\xi^{\phantom{\prime}}_{n}}
=
\biggl(\frac{i\sqrt{3}}{2}\hspace{0.02cm}\biggr)\hspace{0.02cm}
\frac{\partial\hspace{0.03cm}\Omega}
{\partial\hspace{0.03cm}\xi^{\phantom{\prime}}_{n}}\hspace{0.03cm}\!*{\cal A}
-
3\hspace{0.04cm}
\bar{\xi}^{\hspace{0.03cm}\prime}_{n}\!\hspace{0.01cm}*\hspace{0.01cm}{\cal A}
+
\frac{3}{2}\,\frac{\partial\hspace{0.01cm}{\cal A}}
{\partial\hspace{0.03cm}\xi^{\phantom{\prime}}_{n}}
+
\frac{1}{2}\,\gamma\hspace{0.04cm}\bar{\xi}^{\hspace{0.03cm}\prime}_{n}\hspace{0.02cm}
({\cal C}^{\hspace{0.02cm}\prime}_{2} - 12).
\label{eq:11a}
\end{equation}
Finally we eliminate the derivative $\partial{\cal A}/\partial \xi_{n}$ on the right-hand sides of (\ref{eq:11o}) and (\ref{eq:11a}) by using the formula
\[
\frac{\partial\hspace{0.01cm}{\cal A}}
{\partial\hspace{0.03cm}\xi^{\phantom{\prime}}_{n}}
=
\biggl(-\frac{i\sqrt{3}}{2}\hspace{0.02cm}\biggr)\hspace{0.02cm}
\frac{\partial\hspace{0.03cm}\Omega}
{\partial\hspace{0.03cm}\xi^{\phantom{\prime}}_{n}}\hspace{0.03cm}\!*{\cal A}
+
\frac{3}{2}\,
\bar{\xi}^{\hspace{0.03cm}\prime}_{n}\!\hspace{0.01cm}*\hspace{0.01cm}{\cal A}
-
\frac{1}{4}\,\gamma\hspace{0.04cm}\bar{\xi}^{\hspace{0.03cm}\prime}_{n}\hspace{0.02cm}
({\cal C}^{\hspace{0.02cm}\prime}_{2} - 12),
\]
which is a consequence of (\ref{eq:10ka}) and (\ref{eq:11i}). Then, instead of (\ref{eq:11o}) and (\ref{eq:11a}), we can write, correspondingly,
\begin{equation}
\begin{split}
{\cal A}\!\hspace{0.03cm}*\hspace{0.03cm}
\bar{\xi}^{\hspace{0.03cm}\prime}_{n}
&=
\biggl(\frac{i\sqrt{3}}{2}\hspace{0.02cm}\biggr)\hspace{0.02cm}
\frac{\partial\hspace{0.03cm}\Omega}
{\partial\hspace{0.03cm}\xi^{\phantom{\prime}}_{n}}\hspace{0.03cm}\!*{\cal A}
-
\frac{1}{2}\;\bar{\xi}^{\hspace{0.03cm}\prime}_{n}\!\hspace{0.01cm}*\hspace{0.01cm}{\cal A}
+
\frac{1}{4}\,\gamma\hspace{0.04cm}\bar{\xi}^{\hspace{0.03cm}\prime}_{n}\hspace{0.02cm}
({\cal C}^{\hspace{0.02cm}\prime}_{2} - 12),
\\[1ex]
\biggl(\frac{i\sqrt{3}}{2}\hspace{0.02cm}\biggr)\hspace{0.02cm}
{\cal A}\hspace{0.03cm}\!*\frac{\partial\hspace{0.03cm}\Omega}
{\partial\hspace{0.03cm}\xi^{\phantom{\prime}}_{n}}
&=
-\frac{1}{2}\,\biggl(\frac{i\sqrt{3}}{2}\hspace{0.02cm}\biggr)\hspace{0.02cm}
\frac{\partial\hspace{0.03cm}\Omega}
{\partial\hspace{0.03cm}\xi^{\phantom{\prime}}_{n}}\hspace{0.03cm}\!*{\cal A}
-
\frac{3}{4}\;
\bar{\xi}^{\hspace{0.03cm}\prime}_{n}\!\hspace{0.01cm}*\hspace{0.01cm}{\cal A}
+
\frac{1}{8}\,\gamma\hspace{0.04cm}\bar{\xi}^{\hspace{0.03cm}\prime}_{n}\hspace{0.02cm}
({\cal C}^{\hspace{0.02cm}\prime}_{2} - 12).
\end{split}
\label{eq:11s}
\end{equation}
\indent Let us consider the star product ${\cal A}\hspace{0.03cm}\!*\bar{\Xi}_{n}$. Based on the definition (\ref{eq:11yb}) we have
\[
{\cal A}\hspace{0.02cm}\!*\bar{\Xi}_{\hspace{0.02cm}n}(q)
=
\biggl(1 + \frac{1}{2}\,q\!\hspace{0.03cm}\biggr)\hspace{0.02cm}
{\cal A}\!\hspace{0.03cm}*\hspace{0.03cm}\bar{\xi}^{\hspace{0.03cm}\prime}_{n}
+
q\hspace{0.03cm}\biggl(\frac{i\sqrt{3}}{2}\hspace{0.02cm}\biggr)\hspace{0.02cm}
{\cal A}\hspace{0.03cm}\!*\frac{\partial\hspace{0.03cm}\Omega}
{\partial\hspace{0.03cm}\xi^{\phantom{\prime}}_{n}}.
\]
Substituting the expressions (\ref{eq:11s}) into the right-hand side and collecting similar terms, we arrive at the rearrangement rule between the functions ${\cal A}$ and $\bar{\Xi}_{\hspace{0.02cm}n}$
\begin{equation}
{\cal A}\hspace{0.02cm}\!*\bar{\Xi}_{\hspace{0.02cm}n}(q)
=
q^{\hspace{0.01cm}2}\,
\bar{\Xi}_{\hspace{0.02cm}n}(q)\hspace{0.02cm}\!*{\cal A}
-
\frac{1}{4}\,q^{2}\hspace{0.01cm}\gamma\hspace{0.04cm}
\bar{\xi}^{\hspace{0.03cm}\prime}_{n}\hspace{0.02cm}
({\cal C}^{\hspace{0.02cm}\prime}_{2} - 12).
\label{eq:11d}
\end{equation}
In our early paper \cite{markov_2015} we have obtained the rearrangement rule between the matrix $A$ and the deformed matrices $\eta_{\mu}(z)$, Eq.\,(I.2.10), at the fixing point $z = q$
\begin{equation}
A\hspace{0.025cm}\eta_{\mu}(q) = q^{\hspace{0.01cm}2}\hspace{0.01cm}\eta_{\mu}(q)\hspace{0.01cm}A.
\label{eq:11f}
\end{equation}
The relation (\ref{eq:11d}) should be considered as analogue of (\ref{eq:11f}) in the algebra of para-Grassmann numbers equipped with the star product (\ref{eq:4t}). The absence of an additional contribution on the right-hand side of (\ref{eq:11f}) means that this matrix relation was obtained by fixing the element of centre $\theta$, Eq.\,(\ref{eq:7g}), and therefore, it should be corrected.\\
\indent The substitution of (\ref{eq:11d}) into the matrix element (\ref{eq:11u}) gives us finally
\begin{equation}
\langle\hspace{0.02cm}(k)^{\prime}_{p}\hspace{0.02cm}|\hspace{0.02cm}
\hat{A}\hspace{0.02cm}\hat{\eta}_{\mu}(q)\hat{D}_{\mu}\hspace{0.01cm}|\hspace{0.02cm}
(k - 1)_{x}\hspace{0.02cm}\rangle
=
\label{eq:11g}
\end{equation}
\[
=
-\frac{i}{2}\,\biggl\{\,\sum\limits^{2}_{n\hspace{0.02cm}=\hspace{0.02cm}1}
\,\Bigl[\hspace{0.01cm}{\cal A}\!\hspace{0.02cm}*\Xi^{(\hspace{0.02cm}k-1,\hspace{0.03cm}k)}_{\bar{n}}(q)
\bigl(p^{(k)}_{\bar{n}} - e\hspace{0.02cm}A_{\bar{n}}(x^{(k-1)})\bigr)
-
{\cal A}\hspace{0.02cm}\!*\bar{\Xi}^{(\hspace{0.02cm}k,\hspace{0.03cm}k-1)}
_{\hspace{0.02cm}n}(q)
\hspace{0.04cm}\bigl(p^{(k)}_{n} - e\hspace{0.02cm}A_{n}(x^{(k-1)})\bigr)\Bigr] \biggr\}
\hspace{0.02cm}\times
\]
\[
\times
\hspace{0.02cm}
\langle\hspace{0.02cm}\bar{\xi}^{\,\prime(k)}\hspace{0.02cm}|\,\xi^{(k-1)}
\hspace{0.02cm}\rangle
\langle\hspace{0.02cm}p^{(k)}|\,x^{(k-1)\hspace{0.02cm}}\rangle,
\]
where, we recall, the function ${\cal A}$ is expressed in terms of $\Omega$ with the help of the star exponential
\[
{\cal A}\hspace{0.03cm}(\bar{\xi}^{\,\prime},\xi)
=
\alpha\hspace{0.03cm}{\rm e}^{\textstyle-i\hspace{0.02cm}\frac{2\pi}{3}\,
\Omega\hspace{0.03cm}(\bar{\xi}^{\,\prime},\xi)}_{\hspace{0.04cm}*}.
\]
Hence, by virtue of the definitions of the functions $\Xi_{\hspace{0.02cm}\bar{n}}$ and $\bar{\Xi}_{\hspace{0.02cm}n}$, Eq.\,(\ref{eq:11y}), we see that the matrix element (\ref{eq:11g}) is expressed formally through one function $\Omega$. The function should be rewritten (and the matrix element as a whole) in terms of a para-Grassmann 4-vector $\zeta_{\mu}$ with the components
\[
\begin{array}{lll}
&\zeta_{1} = \displaystyle\frac{1}{2}\,(\bar{\xi}^{\prime}_{1} + \xi^{\phantom{\prime}}_{1}),\;
&\zeta_{3} = \displaystyle\frac{1}{2}\,(\bar{\xi}^{\prime}_{2} + \xi^{\phantom{\prime}}_{2}),
\\[3ex]
&\zeta_{2} = \displaystyle\frac{1}{2}\,(\bar{\xi}^{\prime}_{1} - \xi^{\phantom{\prime}}_{1}),\;
&\zeta_{4} = \displaystyle\frac{1}{2}\,(\bar{\xi}^{\prime}_{2} - \xi^{\phantom{\prime}}_{2}).
\end{array}
\]
This question will be considered in Part III \cite{part_III} after calculating the matrix elements of the remaining terms of generalized Hamiltonian (I.3.13).\\
\indent Note that formally we could write out the expression (\ref{eq:11g}) at once based on (I.4.2). However, the main problem is deriving in an explicit form the star product between the function ${\cal A}$ and the functions $\Xi_{\hspace{0.02cm}\bar{n}}$ and $\bar{\Xi}_{\hspace{0.02cm}n}$. For the second term in the square brackets on the right-hand side of (\ref{eq:11g}), as it is written down here, a direct calculation is rather cumbersome in comparison with the star product $\bar{\Xi}_{n}\hspace{0.02cm}\!*{\cal A}$, since even for the star product ${\cal A}\hspace{0.02cm}\!*\bar{\xi}_{n}^{\prime}$ we have ${\cal A}\hspace{0.02cm}\!*\bar{\xi}_{n}^{\prime} \neq {\cal A}\cdot\bar{\xi}_{n}^{\prime}$ (compare with the second expression in (\ref{eq:11t})), but according to (\ref{eq:11o}),
\[
{\cal A}\!\hspace{0.03cm}*\hspace{0.03cm}\bar{\xi}^{\hspace{0.03cm}\prime}_{n}
=
\bar{\xi}^{\hspace{0.04cm}\prime}_{n}\hspace{0.03cm}{\cal A}
-
\frac{\partial\hspace{0.01cm}{\cal A}}
{\partial\hspace{0.03cm}\xi^{\phantom{\prime}}_{n}}.
\]
Further, the integrand of the convolution
\[
{\cal A}\hspace{0.02cm}\!*\frac{\partial\hspace{0.03cm}\Omega}
{\partial\hspace{0.03cm}\xi^{\phantom{\prime}}_{n}}
\,=\!
\iint\!\!{\cal A}\hspace{0.03cm}(\bar{\xi}^{\,\prime},\xi + \mu)\hspace{0.03cm}
\frac{\partial\hspace{0.03cm}\Omega\hspace{0.03cm}(\bar{\xi}^{\,\prime} + \bar{\mu},\xi)}
{\partial\hspace{0.03cm}\xi^{\phantom{\prime}}_{n}}\;
{\rm e}^{\textstyle-\frac{1}{2}\,[\hspace{0.03cm}\bar{\mu},\mu\hspace{0.03cm}]}
\,(d\mu)_{2}\hspace{0.03cm}(d\bar{\mu})_{2}
\]
on expanding in powers of $\mu$ and $\bar{\mu}$, will contain the large number of non-trivial contributions in comparison with the convolution $(\partial\hspace{0.02cm}\Omega/\partial\hspace{0.02cm}\xi_{n})\hspace{0.02cm}\!*\!\hspace{0.02cm}{\cal A}$. Thus, the rearrangement rule (\ref{eq:11d})  in fact gives us the most ``economical'' 
way of deriving the star product ${\cal A}\hspace{0.02cm}\!*\bar{\Xi}_n$ and thereby of the matrix element (\ref{eq:11g}) as a whole.

%%%%%%%%%%%%%%%%%%%%%%%% section 12 %%%%%%%%%%%%%%%%%%%%%%%%%%%%

\section{Conclusion}
\setcounter{equation}{0}
\label{section_12}

In this paper and in the preceding one we have set up the most part of the formalism needed to construct the path integral representation for the inverse third order wave operator within the framework of the Duffin-Kemmer-Petiau theory. An important element of this formalism is the notion of a star product of para-Grassmann-valued functions, which enabled to put the matrix elements into the elegant and compact form. Two different techniques of the calculation of the star product were considered. This made it possible to verify independently the results of the calculations. As an important by-product of the $*$ product we have established isomorphism between the algebra of parafermion creation and annihilation operators of order $p = 2$ and the para-Grassmann algebra of the same order.\\
\indent Another important element of our formalism is the notion of an extended Fock space $\mathfrak{A}_{\hspace{0.02cm}G}$. In fact, this notion is a key element of our approach. We have come to recognize that we must work within the framework of Ohnuki and Kamefuchi's generalized state-vector space $\mathfrak{A}_{\hspace{0.02cm}G}$ rather than in the context of the usual state-vector space $\mathfrak{A}$ that allowed us finally to resolve all the contradictions involved. As a particular consequence, a number of the trilinear matrix and operator relations obtained earlier in the papers by Harish-Chandra \cite{harish-chandra_1947} and by Geyer \cite{geyer_1968} have been refined. These improved relations include such objects as the element $\theta$ of the centre of the Duffin-Kemmer-Petiau algebra and the quadratic Casimir operator $\hat{C}^{\prime}_{2}$ of the Lie group $SO(2M+1)$.\\
\indent All mathematical technique of the calculations proposed by us will be used in Part III \cite{part_III} for analysis of considerably more complicated contributions of the second and the third orders with respect to the covariant derivative $\hat{D}_{\mu}$ in the generalized Hamilton operator (I.3.13). Here, qualitatively new features in calculating the matrix elements of these contributions connected primarily with non-linearity of the expression in $\hat{D}_{\mu}$, arise. Thus, for example, for calculating the matrix element of the term cubic in the covariant derivative the essential point is a necessity of preliminary symmetrization in $\hat{D}_{\mu}$ of the initial expression. Another nontrivial problem is the calculation of the kernel of the transition operator from the matrix elements in the parafermion coherent basis to the representation in which the matrices $\beta_{\mu}$ of the original DKP algebra have a given form. For this purpose the system of 20 algebraic equations will be written out and solved (for the case of spin 1/2 the number of such equations equals 8 \cite{borisov_1982}). The coefficients of this system represent para-Grassmann numbers of order 2.\\
\indent Finally, the use of the obtained matrix elements of the contributions linear, quadratic and cubic in the covariant derivative $\hat{D}_{\mu}$, which arise in the construction of the scheme of finite-multiplicity approximation, enables us when we go to the limit $N \rightarrow \infty$ to written out a complete expression of the required path integral representation of the inverse third order wave operator in an external Maxwell field.

%%%%%%%%%%%%%%%%%%%%%%%%%%%%%%%%%%%%%%%%%%%%%%%%%%%%%%%%%

\section*{\bf Acknowledgments}

The research of Yu.A.M. was supported by the program for improving competitiveness of the Tomsk National Research University among leading world scientific and educational centres.

%%%%%%%%%%%%%%%%%%%%%%% section 13 %%%%%%%%%%%%%%%%%%%%%%%%%%%%

%\newpage

\begin{appendices}
\numberwithin{equation}{section}

\section{Integration of para-Grassmann-valued functions}
%\numberwithin{equation}{section}
\label{appendix_A}

The general formulae of integration with respect to a single para-Grassmann variable $\mu$ of arbitrary order $p$ are given in \cite{ohnuki_1980}. Here, we write out only  the necessary formulae for the special case $p = 2$. The integrals different from zero have the form:
\begin{align}
&\int\!\mu^{2}\hspace{0.02cm}  d^{\hspace{0.02cm} 2\!}\hspace{0.02cm}\mu = 2\hspace{0.03cm} i^{\hspace{0.03cm} 2},  &\label{ap:A1}\\[0.5ex]
&\int[\hspace{0.03cm}\mu, \xi\hspace{0.02cm}]\,\mu\,d^{\hspace{0.02cm} 2\!}\hspace{0.02cm}\mu = -2\hspace{0.02cm} i^{\hspace{0.03cm} 2}\xi,
\label{ap:A2}\\[0.5ex]
&\int[\hspace{0.03cm}\mu, \xi\hspace{0.02cm}]\,[\hspace{0.03cm}\mu, \xi^{\prime}\hspace{0.02cm}]\,d^{\hspace{0.02cm} 2\!}\hspace{0.02cm}\mu = 2\hspace{0.02cm} i^{\hspace{0.03cm} 2}\{\hspace{0.03cm}\xi, \xi^{\prime}\hspace{0.02cm}\},
\label{ap:A3}\\[0.5ex]
&\int[\hspace{0.03cm}\mu, \xi\hspace{0.02cm}]\hspace{0.03cm}\{\hspace{0.03cm}\mu, \xi^{\prime}\hspace{0.02cm}\}\,d^{\hspace{0.02cm} 2\!}\hspace{0.02cm}\mu
= 2\hspace{0.02cm} i^{\hspace{0.03cm} 2}[\hspace{0.03cm}\xi, \xi^{\prime}\hspace{0.02cm}],
\label{ap:A4}
\end{align}
where $\xi$ and $\xi^{\prime}$ are arbitrary para-Grassmann numbers of the same order. The integrals containing zeroth and first powers of $\mu$ vanish by the definition.
Let us write out also the definition of the $\delta$-function
\begin{align}
\delta(\xi - \xi^{\prime}) \equiv \frac{1}{ 2\hspace{0.02cm} i^{\hspace{0.03cm} 2}}\,
(\xi - \xi^{\prime})^{2} =\!
\int\!{\rm e}^{\textstyle-\frac{1}{2}\,[\hspace{0.03cm}\xi - \xi^{\prime},\mu\hspace{0.03cm}]}\,
d^{\hspace{0.02cm} 2\!}\hspace{0.02cm}\mu.
\label{ap:A5}
\end{align}

%%
%%%%%%%%%%%%%%%%%%%%%%%%% Appendix B %%%%%%%%%%%%%%%%%%%%%%%%%%
%%

\section{\bf Differentiation of para-Grassmann-valued functions}
\label{appendix_B}
\numberwithin{equation}{section}

In this Appendix, in addition to the general formulae of differentiation (I.C.9)\,--\,(I.C.11), we derive a number of particular differentiation rules. For this purpose we will need the following formulae \cite{omote_1979, ohnuki_1980}
\begin{equation}
\frac{\overrightarrow{\partial}(\xi_{i})^{m}}{\partial\hspace{0.03cm}\xi_{i}}
=
m\hspace{0.03cm}(p + 1 - m)\hspace{0.02cm}(\xi_{i})^{m - 1},
\quad
\frac{\overrightarrow{\partial}}{\partial\hspace{0.03cm}\xi_{i}}\,
[\hspace{0.03cm}\xi_{i}, \bar{\mu}_{j}\hspace{0.02cm}]
= 2\hspace{0.01cm}\bar{\mu}_{j},
\quad
\frac{\overrightarrow{\partial}}{\partial\hspace{0.03cm}\xi_{i}}\,
\{\hspace{0.03cm}\xi_{i}, \bar{\mu}_{j}\hspace{0.02cm}\}
= 2\hspace{0.01cm}(p - 1)\hspace{0.02cm}\bar{\mu}_{j}.
\label{ap:B1}
\end{equation}
In the special case $p = 2$ from these formulae the expression for the left derivative follows
$$
\frac{\overrightarrow{\partial}}{\partial\hspace{0.03cm}\xi_{k}}\,
(\hspace{0.03cm} \bar{\mu}_{j}\hspace{0.03cm}\xi_{i}\hspace{0.02cm})
\equiv
-\frac{1}{2}\,\frac{\overrightarrow{\partial}}{\partial\hspace{0.03cm}\xi_{k}}\,
[\hspace{0.03cm}\xi_{i}, \bar{\mu}_{j}\hspace{0.02cm}]
+
\frac{1}{2}\,\frac{\overrightarrow{\partial}}{\partial\hspace{0.03cm}\xi_{k}}\,
\{\hspace{0.03cm}\xi_{i}, \bar{\mu}_{j}\hspace{0.02cm}\}
=
-\delta_{ki}\hspace{0.03cm}\bar{\mu}_{j} + \delta_{ki}\hspace{0.03cm}\bar{\mu}_{j} = 0
$$
and similarly for the right derivative we get
$$
(\hspace{0.03cm}\xi_{i}\hspace{0.03cm} \bar{\mu}_{j}\hspace{0.02cm})
\frac{\overleftarrow{\partial}}{\partial\hspace{0.03cm}\xi_{k}}
\equiv
\frac{1}{2}\,[\hspace{0.03cm}\xi_{i}, \bar{\mu}_{j}\hspace{0.02cm}]\,
\frac{\overleftarrow{\partial}}{\partial\hspace{0.03cm}\xi_{k}}
+
\frac{1}{2}\,\{\hspace{0.03cm}\xi_{i}, \bar{\mu}_{j}\hspace{0.02cm}\}\,
\frac{\overleftarrow{\partial}}{\partial\hspace{0.03cm}\xi_{k}}
=
-\bar{\mu}_{j}\hspace{0.03cm}\delta_{ki} + \bar{\mu}_{j}\hspace{0.03cm}\delta_{ki} = 0.
$$
By this mean we have
\begin{equation}
\frac{\overrightarrow{\partial}}{\partial\hspace{0.03cm}\xi_{k}}\,
(\hspace{0.03cm} \bar{\mu}_{j}\hspace{0.03cm}\xi_{i}\hspace{0.02cm}) = 0,
\quad
(\hspace{0.03cm}\xi_{i}\hspace{0.03cm} \bar{\mu}_{j}\hspace{0.02cm})
\frac{\overleftarrow{\partial}}{\partial\hspace{0.03cm}\xi_{k}} = 0.
\label{ap:B2}
\end{equation}
The fact that these derivatives vanish is a distinguishing feature of the case $p = 2$. For $p\neq2$ these derivatives always are different from zero, for example,
\[
\begin{split}
&\mbox{for}\;\; p = 1:\quad \frac{\overrightarrow{\partial}}{\partial\hspace{0.03cm}\xi_{1}}\,
(\hspace{0.03cm}\bar{\mu}_{j}\hspace{0.04cm}\xi_{1}\hspace{0.02cm}) =
-\bar{\mu}_{j}, \\[1ex]
&\mbox{for}\;\; p = 3:\quad \frac{\overrightarrow{\partial}}{\partial\hspace{0.03cm}\xi_{1}}\,
(\hspace{0.03cm}\bar{\mu}_{j}\hspace{0.04cm}\xi_{1}\hspace{0.02cm}) =
\bar{\mu}_{j}
\end{split}
\]
etc. In addition to \eqref{ap:B2} we also can write for $p = 2$
\begin{equation}
\frac{\overrightarrow{\partial}}{\partial\hspace{0.03cm}\xi_{k}}\,
(\hspace{0.03cm}\xi_{i}\hspace{0.03cm}\bar{\mu}_{j}\hspace{0.02cm})
= 2\hspace{0.02cm}\delta_{ik}\hspace{0.02cm}\bar{\mu}_{j},
\quad
(\hspace{0.03cm}\bar{\mu}_{j}\hspace{0.03cm}\xi_{i}\hspace{0.02cm})
\frac{\overleftarrow{\partial}}{\partial\hspace{0.03cm}\xi_{k}} = 2\hspace{0.02cm}\delta_{ik}\hspace{0.02cm}\bar{\mu}_{j}
\label{ap:B3}
\end{equation}
and as a consequence of \eqref{ap:B2} and \eqref{ap:B3} we have
\begin{equation}
\frac{\overrightarrow{\partial}}{\partial\hspace{0.03cm}\xi_{k}}\,
(\hspace{0.03cm}\xi_{j}\hspace{0.04cm}\xi_{i}\hspace{0.02cm})
= 2\hspace{0.03cm}\delta_{kj}\hspace{0.03cm}\xi_{i},
\quad
(\hspace{0.03cm}\xi_{i}\hspace{0.04cm}\xi_{j}\hspace{0.02cm})
\frac{\overleftarrow{\partial}}{\partial\hspace{0.03cm}\xi_{k}} = 2\hspace{0.03cm}\delta_{kj}\hspace{0.03cm}\xi_{i}.
\label{ap:B4}
\end{equation}

%
%%%%%%%%%%%%%%%%%%%%%%%%% Appendix C %%%%%%%%%%%%%%%%%%%%%%%%%%
%

\section{\bf Corrected Harish-Chandra's formula (68)}
\label{appendix_C}
\numberwithin{equation}{section}

Here, we give a derivation of the formula (68) from the paper \cite{harish-chandra_1947} with a proper number coefficients of the matrices
\begin{equation}
B_{r} =\! \! \sum_{(\hspace{0.02cm}k_{1}\,\ldots\;k_{r})}\!\!
\beta^{\hspace{0.01cm}2}_{k_{1}}\,\ldots\,\beta^{\hspace{0.01cm}2}_{k_{1}},
\label{ap:C1}
\end{equation}
where the summation is to be taken over all possible values of $k_{1},\ldots,k_{r}$ such that no two of $k^{\prime}$s are the same. In this Appendix we follow the notations accepted  in \cite{harish-chandra_1947}.\\
\indent The original formula (68) in \cite{harish-chandra_1947} has the following form:
\begin{align}
&\omega^{\hspace{0.02cm}2} = (-1)^{\bigl[\textstyle\frac{1}{2}\,s\hspace{0.02cm}(s - 1)\bigr]}
\biggl[\frac{s + 1}{2}\biggr]\!\biggl[\frac{s}{2}\biggr]\,\ldots\,\biggl[\frac{2}{2}\biggr]
\hspace{0.02cm}\biggl\{B_{[\frac{1}{2}\,(s + 1)]} - \biggl[\frac{s}{2}\biggr]
B_{[\frac{1}{2}\,(s + 3)]}\,+
\label{ap:C2} \\[1ex]
&+
\frac{\!1}{2\hspace{0.02cm}!}\biggl[\frac{s}{2}\biggr]\!
\biggl[\frac{s - 2}{2}\biggr]B_{[\frac{1}{2}\,(s + 5)]}
-
\frac{\!1}{3\hspace{0.02cm}!}\biggl[\frac{s}{2}\biggr]\!\biggl[\frac{s - 2}{2}\biggr]\!
\biggl[\frac{s - 4}{2}\biggr]B_{[\frac{1}{2}\,(s + 7)]}\,+\,\ldots\,+
(-1)^{\bigl[\textstyle\frac{1}{2}\,s\bigr]}B_{s}\!\hspace{0.02cm}\biggr\}.
\notag
\end{align}
Here, $s$ is a space-time dimensions, $[n]$ denotes the integral part of $n$. This expression is derived from the following formula (Eq.\,(67) in \cite{harish-chandra_1947})
\begin{equation}
\omega^{\hspace{0.02cm}2} = (-1)^{\bigl[\textstyle\frac{1}{2}\,s\hspace{0.02cm}(s - 1)\bigr]}
\biggl[\frac{s + 1}{2}\biggr]\!\biggl[\frac{s}{2}\biggr]\ldots\biggl[\frac{2}{2}\biggr]
P_{(\hspace{0.02cm}k_{1}\,\ldots\;k_{r})}\hspace{0.02cm}
\beta^{\hspace{0.01cm}2}_{k_{s}}\hspace{0.02cm}\beta^{\hspace{0.01cm}2}_{k_{s - 2}}\ldots
(1 - \beta^{\hspace{0.01cm}2}_{k_{s - 1}})(1 - \beta^{\hspace{0.01cm}2}_{k_{s - 3}})\,\ldots,
\label{ap:C3}
\end{equation}
where $P_{(\hspace{0.02cm}k_{1}\,\ldots\;k_{r})}\hspace{0.02cm}$ denotes a sum over all permutations of $(1, 2,\ldots,s)$ and $k_{1}, \ldots,k_{s}$ are all different. In turn, \eqref{ap:C3} is a result of a product of two (nonnormalized) matrices $\omega = \epsilon^{ k_{1}\ldots\, k_{s}} \beta_{k_1}\ldots \beta_{k_{s}}$.\\
\indent Let us remove the parentheses on the right-hand side of \eqref{ap:C3}:
\begin{align}
\omega^{\hspace{0.02cm}2} &= (-1)^{\bigl[\textstyle\frac{1}{2}\,s\hspace{0.02cm}(s - 1)\bigr]}
\biggl[\frac{s + 1}{2}\biggr]\!\biggl[\frac{s}{2}\biggr]\ldots\biggl[\frac{2}{2}\biggr]
P_{(\hspace{0.02cm}k_{1}\,\ldots\;k_{r})}\,\times \notag\\[1ex]
\times\Bigl\{\hspace{0.02cm}&C^{[s/2]}_{0}
\beta^{\hspace{0.01cm}2}_{k_{s}}\hspace{0.02cm}\beta^{\hspace{0.02cm}2}_{k_{s - 2}}\,\ldots\,
\beta^{\hspace{0.02cm}2}_{k_{2}}\hspace{0.02cm}
1_{k_{s - 1}}\hspace{0.02cm}1_{k_{s - 3}}\,\ldots\,1_{k_{1}}
- \notag \\[1ex]
-\;&C^{[s/2]}_{1}
\beta^{\hspace{0.01cm}2}_{k_{s}}\hspace{0.02cm}\beta^{\hspace{0.02cm}2}_{k_{s - 2}}\,\ldots\,
\beta^{\hspace{0.02cm}2}_{k_{2}}\beta^{\hspace{0.02cm}2}_{k_{s - 1}}\hspace{0.02cm}
1_{k_{s - 3}}\hspace{0.02cm}1_{k_{s - 5}}\,\ldots\,1_{k_{1}}
+ \label{ap:C4} \\[1ex]
+\;&C^{[s/2]}_{2}
\beta^{\hspace{0.01cm}2}_{k_{s}}\hspace{0.02cm}\beta^{\hspace{0.02cm}2}_{k_{s - 2}}\,\ldots\,
\beta^{\hspace{0.02cm}2}_{k_{2}}\beta^{\hspace{0.02cm}2}_{k_{s - 1}} \beta^{\hspace{0.02cm}2}_{k_{s - 3}}\hspace{0.02cm}
1_{k_{s - 5}}\hspace{0.02cm}1_{k_{s - 7}}\,\ldots\,1_{k_{1}}
- \notag \\[1ex]
-\;&C^{[s/2]}_{3}
\beta^{\hspace{0.01cm}2}_{k_{s}}\hspace{0.02cm}\beta^{\hspace{0.02cm}2}_{k_{s - 2}}\,\ldots\,
\beta^{\hspace{0.02cm}2}_{k_{2}}\beta^{\hspace{0.02cm}2}_{k_{s - 1}} \beta^{\hspace{0.02cm}2}_{k_{s - 3}}\beta^{\hspace{0.02cm}2}_{k_{s - 5}}\hspace{0.02cm}
1_{k_{s - 7}}\hspace{0.02cm}1_{k_{s - 9}}\,\ldots\,1_{k_{1}}
\,+
\notag \\[1ex]
+\,&\ldots\,+ (-1)^{\bigl[\textstyle\frac{1}{2}\,s\bigr]}C^{[s/2]}_{[s/2]}
\beta^{\hspace{0.01cm}2}_{k_{s}}\hspace{0.02cm}\beta^{\hspace{0.02cm}2}_{k_{s - 2}}\,\ldots\,
\beta^{\hspace{0.02cm}2}_{k_{2}}\beta^{\hspace{0.02cm}2}_{k_{s - 1}} \beta^{\hspace{0.02cm}2}_{k_{s - 3}}\,\ldots\,\beta^{\hspace{0.02cm}2}_{k_{1}}\Bigr\}\,.
\notag
\end{align}
Here, $C_{k}^{n}$ are binomial coefficients. Writing out their explicit form
\[
C^{[s/2]}_{0} = 1,\;\, C^{[s/2]}_{1} = \frac{\!1}{1\hspace{0.01cm}!}\biggl[\frac{s}{2}\biggr],
\;\,C^{[s/2]}_{2} = \frac{\!1}{2\hspace{0.02cm}!}\biggl[\frac{s}{2}\biggr]\!\biggl[\frac{s-2}{2}\biggr],
\;\,C^{[s/2]}_{3} = \frac{\!1}{3\hspace{0.02cm}!}\biggl[\frac{s}{2}\biggr]\!\biggl[\frac{s-2}{2}\biggr]\!
\biggl[\frac{s-4}{2}\biggr],\,\ldots\,,
\]
using the definition of matrices $B_{r}$, Eq.\,\eqref{ap:C1}, and omitting a product of the unity matrices $1_{k_{s}}$, we reproduce the original formula of Harish-Chandra \eqref{ap:C2}. However, indeed we have to take into account a possibility of rearrangement of the unity matrices $1_{k_{s-l}}, l = 1, 3, 5, \ldots$.  We explicitly set the markers $k_{s-l}$ for the unity matrices to emphasize the importance of accounting their rearrangements. This gives additional factors. So in the first term in braces in \eqref{ap:C4} the rearrangement of the unity matrices gives an additional factor $\bigl[s/2\bigr]!$, in the second term it gives the factor $\bigl[(s - 2)/2\bigr]!$ etc. By this mean, as a result, instead of \eqref{ap:C2} now we have
\begin{align}
&\omega^{\hspace{0.02cm}2} = (-1)^{\bigl[\textstyle\frac{1}{2}\,s\hspace{0.02cm}(s - 1)\bigr]}
\biggl[\frac{s + 1}{2}\biggr]\!\biggl[\frac{s}{2}\biggr]\,\ldots\,\biggl[\frac{2}{2}\biggr]
\hspace{0.02cm}
\biggl\{\!\biggl(\biggl[\frac{s}{2}\biggr]!\biggr)B_{[\frac{1}{2}\,(s + 1)]} - \biggl[\frac{s}{2}\biggr]
\biggl(\biggl[\frac{s - 2}{2}\biggr]!\biggr)B_{[\frac{1}{2}\,(s + 3)]}\,+
\notag \\[1ex]
&+
\frac{\!1}{2\hspace{0.02cm}!}\biggl[\frac{s}{2}\biggr]\!
\biggl[\frac{s - 2}{2}\biggr]\biggl(\biggl[\frac{s - 4}{2}\biggr]!\biggr)B_{[\frac{1}{2}\,(s + 5)]}
-
\frac{\!1}{3\hspace{0.02cm}!}\biggl[\frac{s}{2}\biggr]\!\biggl[\frac{s - 2}{2}\biggr]\!
\biggl[\frac{s - 4}{2}\biggr]
\biggl(\biggl[\frac{s - 6}{2}\biggr]!\biggr)B_{[\frac{1}{2}\,(s + 7)]}\,+\,\ldots\,+
\notag \\[1ex]
&+
(-1)^{\bigl[\textstyle\frac{1}{2}\,s\bigr]}B_{s}\!\hspace{0.02cm}\biggr\}
\notag
\end{align}
or taking the common factorial $\bigl(\bigl[s/2\bigr]!\bigr)$ outside braces, we get
\begin{align}
&\omega^{\hspace{0.02cm}2} = (-1)^{\bigl[\textstyle\frac{1}{2}\,s\hspace{0.02cm}(s - 1)\bigr]}
\biggl[\frac{s + 1}{2}\biggr]\!\Biggl[\frac{s}{2}\biggr]\,\ldots\,\biggl[\frac{2}{2}\biggr]
\biggl(\biggl[\frac{s}{2}\biggr]!\biggr)\hspace{0.02cm}
\biggl\{B_{[\frac{1}{2}\,(s + 1)]}
-
\frac{1}{1!}\,B_{[\frac{1}{2}\,(s + 3)]}\,+
\label{ap:C5} \\[1ex]
&+
\frac{\!1}{2\hspace{0.02cm}!}\,B_{[\frac{1}{2}\,(s + 5)]}
-
\frac{\!1}{3\hspace{0.02cm}!}\,B_{[\frac{1}{2}\,(s + 7)]}\,+\,\ldots\,+
(-1)^{\bigl[\textstyle\frac{1}{2}\,s\bigr]}
\biggl(\biggl[\frac{s}{2}\biggr]!\biggr)^{\!-1}\!B_{s}\!\hspace{0.02cm}\biggr\}.
\notag
\end{align}
For the special case $s = 4$ from this expression it follows
\[
\omega^{\hspace{0.02cm}2} = 2^{\hspace{0.02cm}2}\bigl\{2\hspace{0.02cm}B_{2} - 2\hspace{0.02cm}B_{3} + B_{4}\bigr\},
\]
while the original formula \eqref{ap:C2} gives us
\[
\omega^{\hspace{0.02cm}2} = 2^{\hspace{0.02cm}2}\bigl\{B_{2} - 2\hspace{0.02cm}B_{3} + B_{4}\bigr\}.
\]
\indent As previously discussed, in section 9, the formula \eqref{ap:C2} has never been used further in the text of the paper \cite{harish-chandra_1947}, except deriving the next formula (69) having the following form:
\begin{equation}
\omega^{\hspace{0.02cm}3} = (-1)^{\bigl[\textstyle\frac{1}{2}\,s\hspace{0.02cm}(s - 1)\bigr]}
\biggl\{\!\hspace{0.02cm}\biggl[\frac{s + 1}{2}\biggr]\!\hspace{0.02cm}\biggl[\frac{s}{2}\biggr]
\!\hspace{0.02cm}\biggl[\frac{s - 1}{2}\biggr]
\,\ldots\,\biggl[\frac{2}{2}\biggr]
\biggr\}^{\!2}\omega.
\label{ap:C6}
\end{equation}
Let us show that this expression can not be reproduced by using the original formula  \eqref{ap:C2}, whereas the revised formula  \eqref{ap:C5} do this.\\
\indent We will need the following matrix relation \cite{harish-chandra_1947}:
\begin{equation}
\omega B = \biggl[\frac{s + 1}{2}\biggr]\omega.
\label{ap:C7}
\end{equation}
We have used the special case of this formula in section 9, Eq.\,(\ref{eq:9s}). Besides, instead of the initial definition of matrices $B_{r}$, Eq.\,\eqref{ap:C1}, we make use of the following representation:
\begin{equation}
B_{r} =B\hspace{0.02cm}(B - 1)\,\ldots\,(B - r +1).
\label{ap:C8}
\end{equation}
Let us multiply \eqref{ap:C8} by $\omega$ from the left. Taking into account \eqref{ap:C7}, we find
\[
\omega B_{r} = \biggl[\frac{s + 1}{2}\biggr]\biggl(\biggl[\frac{s + 1}{2}\biggr] - 1\biggr)
\biggl(\biggl[\frac{s + 1}{2}\biggr] - 2\biggr)\,\ldots\,
\biggl(\biggl[\frac{s + 1}{2}\biggr] - r + 1\biggr)\omega.
\]
At a certain value $r \equiv r_{*}$ this expression for $r \geq r_{*}$ will vanish. This value equals
\begin{equation}
r_{*} = \biggl[\frac{s + 1}{2}\biggr] + 1 = \biggl[\frac{s + 3}{2}\biggr].
\label{ap:C9}
\end{equation}
Multiply the expression \eqref{ap:C2} by $\omega$. We see that all terms except the first one, by virtue of the condition \eqref{ap:C9} vanish and consequently we get
\begin{equation}
\omega^{\hspace{0.02cm}3} = (-1)^{\bigl[\textstyle\frac{1}{2}\,s\hspace{0.02cm}(s - 1)\bigr]}
\biggl(\biggl[\frac{s + 1}{2}\biggr]\!\hspace{0.02cm}\biggl[\frac{s}{2}\biggr]
\!\hspace{0.02cm}\biggl[\frac{s - 1}{2}\biggr]\,\ldots\,\biggl[\frac{2}{2}\biggr]\biggr)
\omega B_{[\frac{1}{2}\,(s + 1)]} =
\label{ap:C10}
\end{equation}
\[
=
(-1)^{\bigl[\textstyle\frac{1}{2}\,s\hspace{0.02cm}(s - 1)\bigr]}
\biggl(\biggl[\frac{s + 1}{2}\biggr]\!\hspace{0.02cm}\biggl[\frac{s}{2}\biggr]
\!\hspace{0.02cm}\biggl[\frac{s - 1}{2}\biggr]\,\ldots\,\biggl[\frac{2}{2}\biggr]\biggr)
\biggl(\biggl[\frac{s + 1}{2}\biggr]\!\hspace{0.02cm}\biggl[\frac{s - 1}{2}\biggr]\!\hspace{0.02cm}\biggl[\frac{s - 3}{2}\biggr]\,\ldots\,\biggl[\frac{2}{2}\biggr]\biggr)
\omega.
\]
Obviously, the formula (C.6) is not reproduced, because  the factors $\bigl[s/2\big]$, $\bigl[(s - 2)/2\bigr]$, $\bigl[(s - 4)/2\bigr],\;\ldots$ are lacking.\\
\indent Now we multiply the revised formula \eqref{ap:C5} by the matrix $\omega$. Then instead of \eqref{ap:C10}, we  have
\[
\omega^{\hspace{0.02cm}3} \!=\!
(-1)^{\bigl[\textstyle\frac{1}{2}\,s\hspace{0.02cm}(s - 1)\bigr]}
\biggl(\biggl[\frac{s + 1}{2}\biggr]\!\hspace{0.02cm}\biggl[\frac{s}{2}\biggr]
\!\hspace{0.02cm}\biggl[\frac{s - 1}{2}\biggr]\!\ldots\!\biggl[\frac{2}{2}\biggr]\biggr)
\!\biggl(\biggl[\frac{s}{2}\biggr]!\biggr)\!
\biggl(\biggl[\frac{s + 1}{2}\biggr]\!\hspace{0.02cm}\biggl[\frac{s - 1}{2}\biggr]\!\hspace{0.02cm}\biggl[\frac{s - 3}{2}\biggr]\!\ldots\!\biggl[\frac{2}{2}\biggr]\biggr)
\omega.
\]
Here, in contrast to \eqref{ap:C10}, an additional factor appears
\[
\biggl[\frac{s}{2}\biggr]! =
\biggl[\frac{s}{2}\biggr]\!\hspace{0.02cm}\biggl[\frac{s - 2}{2}\biggr]
\!\hspace{0.02cm}\biggl[\frac{s - 4}{2}\biggr]\,\ldots\,\biggl[\frac{2}{2}\biggr],
\]
which gives us the missing multiplies in \eqref{ap:C4}. Thereby the formula \eqref{ap:C5} reproduces \eqref{ap:C6}.
\newpage

%
%%%%%%%%%%%%%%%%%%%%%%%%% Appendix D %%%%%%%%%%%%%%%%%%%%%%%%%%
%

\section{Proof of the relation $[\hspace{0.03cm}a_{0},\hat{\Lambda}\hspace{0.02cm}] = 0$}
\numberwithin{equation}{section}

In this Appendix we will give a proof of the second relation in (\ref{eq:9f}). Let us present it as a sum of two terms
\begin{equation}
[\hspace{0.03cm}a_{0},\hat{\Lambda}_{1}\hspace{0.02cm}]
+
[\hspace{0.03cm}a_{0},\hat{\Lambda}_{2}\hspace{0.02cm}] = 0,
\label{ap:D1}
\end{equation}
where $\hat{\Lambda} = \{\hspace{0.02cm}a_{k}^{+}, a_{k}^{-}\},\,k = 1, 2$. Substituting an explicit form of the operator $a_{0}$, Eq.\,(I.6.17), we present the first term in \eqref{ap:D1} in the following form:
\begin{equation}
-\hspace{0.02cm}\frac{1}{4}\,\bigl(\hspace{0.02cm}[\hspace{0.03cm}\{\hspace{0.03cm}L_{12}, M_{12}\hspace{0.02cm}\},\{\hspace{0.02cm}a^{+}_{1},a^{-}_{1}\}\hspace{0.03cm}]
+
[\hspace{0.03cm}\{\hspace{0.03cm}N_{12}, N_{21}\hspace{0.02cm}\},\{\hspace{0.02cm}a^{+}_{1},a^{-}_{1}\}\hspace{0.03cm}]
-
[\hspace{0.03cm}\{\hspace{0.03cm}N_{1}, N_{2}\hspace{0.02cm}\},\{\hspace{0.02cm}a^{+}_{1},a^{-}_{1}\}\hspace{0.03cm}]
\hspace{0.02cm}\bigr).
\label{ap:D2}
\end{equation}
Let us consider the first contribution in \eqref{ap:D2}. Making use of the operator identity (\ref{eq:9j}), we present it in the form:
\[
[\hspace{0.03cm}\{\hspace{0.03cm}L_{12}, M_{12}\hspace{0.02cm}\},\{\hspace{0.02cm}a^{+}_{1},a^{-}_{1}\}\hspace{0.03cm}]
=
\{L_{12},[\hspace{0.02cm}M_{12},\{\hspace{0.02cm}a^{+}_{1},a^{-}_{1}\}\hspace{0.03cm}]\}
+
\{M_{12},[\hspace{0.02cm}L_{12},\{\hspace{0.02cm}a^{+}_{1},a^{-}_{1}\}\hspace{0.03cm}]
\hspace{0.015cm}\},
\]
where in turn
\[
[\hspace{0.02cm}M_{12},\{\hspace{0.02cm}a^{+}_{1},a^{-}_{1}\}\hspace{0.03cm}]
=
-\hspace{0.02cm}\{a^{+}_{1},[\hspace{0.04cm}a^{-}_{1},M_{12}\hspace{0.03cm}]\}
+
\{a^{-}_{1},[\hspace{0.02cm}M_{12},a^{+}_{1}\hspace{0.03cm}]\hspace{0.015cm}\}
=
-\hspace{0.02cm}\{\hspace{0.02cm}a^{-}_{1},a^{-}_{2}\}
\]
and
\[
[\hspace{0.02cm}L_{12},\{\hspace{0.02cm}a^{+}_{1},a^{-}_{1}\}\hspace{0.03cm}]
=
-\hspace{0.02cm}\{a^{+}_{1},[\hspace{0.04cm}a^{-}_{1},L_{12}\hspace{0.03cm}]\}
+
\{a^{-}_{1},[\hspace{0.02cm}L_{12},a^{+}_{1}\hspace{0.03cm}]\hspace{0.015cm}\}
=
-\hspace{0.02cm}\{\hspace{0.02cm}a^{+}_{1},a^{+}_{2}\}.
\hspace{0.2cm}
\]
Here, we have used the commutation rules (I.6.15). As a result, the first contribution in \eqref{ap:D2} takes the form
\begin{equation}
[\hspace{0.02cm}\{\hspace{0.03cm}L_{12}, M_{12}\hspace{0.02cm}\},\{\hspace{0.02cm}a^{+}_{1},a^{-}_{1}\}\hspace{0.02cm}]
=
-\hspace{0.02cm}\{L_{12},\{\hspace{0.02cm}a^{-}_{1},a^{-}_{2}\}\hspace{0.04cm}\!\}
-
\{M_{12},\{\hspace{0.02cm}a^{+}_{1},a^{+}_{2}\}\hspace{0.04cm}\!\}.
\label{ap:D3}
\end{equation}
For the second contribution in \eqref{ap:D2} a similar reasoning results in
\begin{equation}
[\hspace{0.02cm}\{\hspace{0.03cm}N_{12}, N_{21}\hspace{0.02cm}\},\{\hspace{0.02cm}a^{+}_{1},a^{-}_{1}\}\hspace{0.02cm}]
=
\{N_{12},\{\hspace{0.02cm}a^{-}_{1},a^{-}_{2}\}\hspace{0.04cm}\!\}
-
\{N_{21},\{\hspace{0.02cm}a^{+}_{1},a^{+}_{2}\}\hspace{0.04cm}\!\},
\label{ap:D4}
\end{equation}
and the third contribution in \eqref{ap:D2} vanishes.\\
\indent Further, we transform the right-hand side of \eqref{ap:D3}. For the first term on the right-hand side we make use of the operator identity (\ref{eq:9h})
\[
\{L_{12},\{\hspace{0.02cm}a^{-}_{1},a^{-}_{2}\}\hspace{0.04cm}\!\}
=
[\hspace{0.02cm}a^{-}_{1},[\hspace{0.02cm}a^{-}_{2},L_{12}\hspace{0.03cm}]\hspace{0.03cm}] +
\{\hspace{0.02cm}a^{-}_{2},\{L_{12},a^{-}_{1}\hspace{0.02cm}\}\!\hspace{0.04cm}\}
=
2\hspace{0.01cm}N_{1} + \{a^{-}_{2},\{L_{12},a^{-}_{1}\hspace{0.02cm}\}\!\hspace{0.03cm}\}.
\]
Here, we have used again the commutation rules (I.6.15). For the second term in \eqref{ap:D3} we have similarly
\[
\{M_{12},\{\hspace{0.02cm}a^{-}_{1},a^{-}_{2}\}\hspace{0.04cm}\!\}
=
-2\hspace{0.01cm}N_{1} + \{a^{+}_{2},\{M_{12},a^{+}_{1}\hspace{0.02cm}\}\!\hspace{0.03cm}\}.
\]
and then \eqref{ap:D3} goes into
\begin{equation}
[\hspace{0.02cm}\{\hspace{0.03cm}L_{12}, M_{12}\hspace{0.02cm}\},\{\hspace{0.02cm}a^{+}_{1},a^{-}_{1}\}\hspace{0.02cm}]
=
-\hspace{0.02cm}\{a^{-}_{2},\{L_{12},a^{-}_{1}\hspace{0.02cm}\}\!\hspace{0.03cm}\}
-
\hspace{0.02cm}\{a^{+}_{2},\{M_{12},a^{+}_{1}\hspace{0.02cm}\}\!\hspace{0.03cm}\}.
\label{ap:D5}
\end{equation}
We make another transformation of the right-hand side of the last expression with the aim that instead of the generators $L_{12}$ and $M_{12}$ the second pair of the generators $N_{12}$ and $N_{21}$ has appeared. By using the identity (\ref{eq:9j}) for the internal anticommutator in the first term in \eqref{ap:D5} and the definition of the generator $L_{12}$, we obtain
\[
\begin{split}
\{L_{12},a^{-}_{1}\hspace{0.02cm}\} = \frac{1}{2}\,&\{[\hspace{0.03cm}a^{+}_{1},a^{+}_{2}\hspace{0.03cm}],
a^{-}_{1}\hspace{0.01cm}\}
=
\frac{1}{2}\,\bigl(\hspace{0.02cm}\{\hspace{0.02cm}a^{+}_{1},
[\hspace{0.02cm}a^{+}_{2},a^{-}_{1}\hspace{0.02cm}]\}
-
[\hspace{0.04cm}a^{+}_{2},\{\hspace{0.02cm}a^{-}_{1},a^{+}_{1}\hspace{0.02cm}\}]
\hspace{0.03cm}\bigr) = \\[1ex]
=\,
&\{\hspace{0.02cm}a^{+}_{1},N_{21}\} - \frac{1}{2}\,[\hspace{0.04cm}a^{+}_{2},\{\hspace{0.02cm}a^{-}_{1},a^{+}_{1}\hspace{0.02cm}\}].
\end{split}
\]
A similar transformation for the internal anticommutator in the second term in \eqref{ap:D5} gives 
\[
\{M_{12},a^{+}_{1}\hspace{0.02cm}\}
=
-\hspace{0.02cm}\{\hspace{0.02cm}a^{-}_{1},N_{12}\} - \frac{1}{2}\,[\hspace{0.04cm}a^{-}_{2},\{\hspace{0.02cm}a^{+}_{1},a^{-}_{1}\hspace{0.02cm}\}]
\]
and the right-hand side of \eqref{ap:D5} becomes
\[
-\hspace{0.02cm}\{a^{-}_{2},\{\hspace{0.02cm}a^{+}_{1},N_{21}\}\!\hspace{0.04cm}\}
+
\hspace{0.02cm}\{a^{+}_{2},\{\hspace{0.02cm}a^{-}_{1},N_{12}\}\!\hspace{0.04cm}\}
+
\frac{1}{2}\,\bigl(\{a^{-}_{2},[\hspace{0.04cm}a^{+}_{2},\{\hspace{0.02cm}a^{-}_{1},a^{+}_{1}
\hspace{0.02cm}\}]\hspace{0.02cm}\}
-
\hspace{0.02cm}\{a^{+}_{2},[\hspace{0.04cm}a^{-}_{2},\{\hspace{0.02cm}a^{+}_{1},a^{-}_{1}
\hspace{0.02cm}\}]\hspace{0.01cm}\}\hspace{0.02cm}\bigr).
\]
The expression in parentheses here can be written in a more compact form:
\[
[\hspace{0.04cm}\{\hspace{0.02cm}a^{+}_{2},a^{-}_{2}\},
\{\hspace{0.02cm}a^{+}_{1},a^{-}_{1}\hspace{0.03cm}\}]
\equiv
[\hspace{0.04cm}\hat{\Lambda}_{2},\hat{\Lambda}_{1}\hspace{0.03cm}]
\]
and then the first contribution in \eqref{ap:D2} takes  its final form
\begin{equation}
[\hspace{0.03cm}\{\hspace{0.03cm}L_{12}, M_{12}\hspace{0.02cm}\},\{\hspace{0.02cm}a^{+}_{1},a^{-}_{1}\}\hspace{0.03cm}]
=
-\hspace{0.02cm}\{a^{-}_{2},\{\hspace{0.02cm}a^{+}_{1},N_{21}\}\!\hspace{0.04cm}\}
+
\hspace{0.02cm}\{a^{+}_{2},\{\hspace{0.02cm}a^{-}_{1},N_{12}\}\!\hspace{0.04cm}\}
+
\frac{1}{2}\;[\hspace{0.04cm}\hat{\Lambda}_{2},\hat{\Lambda}_{1}\hspace{0.03cm}].
\label{ap:D6}
\end{equation}
\indent It is sufficient to transform the right-hand side of the second contribution \eqref{ap:D4} to the form similar to \eqref{ap:D5}. Then instead of \eqref{ap:D4} we will have
\begin{equation}
[\hspace{0.03cm}\{\hspace{0.03cm}N_{12}, N_{21}\hspace{0.02cm}\},\{\hspace{0.02cm}a^{+}_{1},a^{-}_{1}\}\hspace{0.03cm}]
=
\{\hspace{0.02cm}a^{+}_{2},\{N_{12},a^{-}_{1}\hspace{0.02cm}\}\!\hspace{0.04cm}\}
-
\{\hspace{0.02cm}a^{-}_{2},\{N_{21},a^{+}_{1}\hspace{0.02cm}\}\!\hspace{0.04cm}\}.
\label{ap:D7}
\end{equation}
We should note that here the signs on the right-hand side {\it coincide} with the signs of the first two terms in \eqref{ap:D6} contrary to the expectations. Summing \eqref{ap:D6} and \eqref{ap:D7}, we get
\[
[\hspace{0.03cm}a_{0},\hat{\Lambda}_{1}\hspace{0.02cm}]
=
\frac{1}{2}\,\bigl(
\{a^{-}_{2},\{\hspace{0.02cm}a^{+}_{1},N_{21}\}\!\hspace{0.04cm}\}
-
\hspace{0.02cm}\{a^{+}_{2},\{\hspace{0.02cm}a^{-}_{1},N_{12}\}\!\hspace{0.04cm}\}
\hspace{0.01cm}\bigr)
+
\frac{1}{8}\;[\hspace{0.04cm}\hat{\Lambda}_{1},\hat{\Lambda}_{2}\hspace{0.03cm}].
\]
As we see, this expression does not vanish. The second term on the left-hand side of \eqref{ap:D1} can be obtained from the previous one by a simple replacement of indices $1 \rightleftarrows 2$ (the operator $a_{0}$ is invariant with respect to such a replacement, and $N_{12} \rightleftarrows N_{21}$):
\[
[\hspace{0.03cm}a_{0},\hat{\Lambda}_{2}\hspace{0.02cm}]
=
\frac{1}{2}\,\bigl(
\{a^{-}_{1},\{\hspace{0.02cm}a^{+}_{2},N_{12}\}\!\hspace{0.04cm}\}
-
\hspace{0.02cm}\{a^{+}_{1},\{\hspace{0.02cm}a^{-}_{2},N_{21}\}\!\hspace{0.04cm}\}
\hspace{0.01cm}\bigr)
-
\frac{1}{8}\;[\hspace{0.04cm}\hat{\Lambda}_{1},\hat{\Lambda}_{2}\hspace{0.03cm}].
\]
We sum the last two expressions
\[
[\hspace{0.04cm}a_{0},\hat{\Lambda}_{1} + \hat{\Lambda}_{2}\hspace{0.02cm}]
=
\]
\[
=
\frac{1}{2}\,\bigl(
\{a^{-}_{2},\{\hspace{0.02cm}a^{+}_{1},N_{21}\}\!\hspace{0.04cm}\}
-
\hspace{0.02cm}\{a^{+}_{1},\{\hspace{0.02cm}a^{-}_{2},N_{21}\}\!\hspace{0.04cm}\}
\hspace{0.01cm}\bigr)
+
\frac{1}{2}\,\bigl(
\{a^{-}_{1},\{\hspace{0.02cm}a^{+}_{2},N_{12}\}\!\hspace{0.04cm}\}
-
\hspace{0.02cm}\{a^{+}_{2},\{\hspace{0.02cm}a^{-}_{1},N_{12}\}\!\hspace{0.04cm}\}
\hspace{0.01cm}\bigr)
=
\]
\[
=
\frac{1}{2}\,\bigl(\hspace{0.02cm}
[\hspace{0.02cm}N_{21},[\hspace{0.02cm}a^{+}_{1},a^{-}_{2}\hspace{0.02cm}]\hspace{0.04cm}]
+
[\hspace{0.02cm}N_{12},[\hspace{0.02cm}a^{+}_{2},a^{-}_{1}\hspace{0.02cm}]\hspace{0.03cm}]
\hspace{0.02cm}\bigr)
\equiv
[\hspace{0.02cm}N_{21},N_{12}\hspace{0.03cm}]
+
[\hspace{0.02cm}N_{12},N_{21}\hspace{0.03cm}] = 0.
\]
By this mean, in contrast to (\ref{eq:9v}), the relation \eqref{ap:D1} is fulfilled only for a sum of two terms $\hat{\Lambda}_{1}$ and $\hat{\Lambda}_{2}$ and therefore the equalities (\ref{eq:9b}) are not the case.

\end{appendices}

%%%%%%%%%%%%%%%%%%%%%%%% section 14 %%%%%%%%%%%%%%%%%%%%%%%%%%%%

\end{document}